\documentclass[preprint]{elsarticle} 

\usepackage{epsfig,multicol,bbm}
\usepackage{latexsym}
\usepackage{amssymb,amsmath}
\usepackage{dsfont}

\newcommand\fverb{\setbox\pippobox=\hbox\bgroup\verb}
\newcommand\fverbdo{\egroup\medskip\noindent%
\fbox{\unhbox\pippobox}\ }
\newcommand\fverbit{\egroup\item[\fbox{\unhbox\pippobox}]}
\newbox\pippobox

\newcommand\eqn[1]{(\ref{#1})}      
\newcommand\Eqn[1]{Eq.~(\ref{#1})}  
\newcommand\Refer[1]{Ref.~\cite{#1}}  
\newcommand\Sec[1]{Sec.~\ref{#1}}  
\newcommand\Fig[1]{Fig.~\ref{#1}}  
\newcommand\App[1]{App.~\ref{#1}}  

\newcommand{\beq}{\begin{equation}}
\newcommand{\eeq}{\end{equation}}
\newcommand{\ba}{\begin{array}}
\newcommand{\bea}{\begin{eqnarray}}
\newcommand{\ea}{\end{array}}
\newcommand{\eea}{\end{eqnarray}}

\newcommand\comment[1]{ \hbox{[{\it Comment suppressed here.}\/]} }
\newcommand\hide[1]{}

\newcommand{\tr}{{\rm  tr}}

\newcommand{\Str}{\hbox{Str}}
\newcommand{\Ln}{\hbox{Ln}}

\newcommand{\Gammatpi}{\Gamma_{\mbox{\scriptsize 2PI}}}
\newcommand{\Gammaint}{\Gamma_{\!\mbox{\scriptsize int}}}
\newcommand{\bcG}{{\bar{\cal G}}}

\newcommand{\cG}{{\cal G}}
\newcommand{\cC}{{\cal C}}
\newcommand{\cM}{{\cal M}}
\newcommand{\cV}{{\cal V}}
\newcommand{\cL}{{\cal L}}
\newcommand{\cJ}{{\cal J}}
\newcommand{\cK}{{\cal K}}

\newcommand{\bG}{{\bar{G}}}
\newcommand{\bD}{{\bar{D}}}
\newcommand{\bV}{\bar{V}}
\newcommand{\bK}{\bar{K}}
\newcommand{\bM}{\bar{M}}
\newcommand{\bW}{\bar{W}}

\newcommand{\bSigma}{\bar{\Sigma}}

\newcommand{\bLambda}{\bar{\Lambda}}

\newcommand{\bPi}{\bar\Pi}

\newcommand{\bra}{\langle}
\newcommand{\ket}{\rangle}
\newcommand{\ketc}{\ket_{\!_c}}

\newcommand{\nn}{\nonumber \\}

\def\slashchar#1{\setbox0=\hbox{$#1$}           
   \dimen0=\wd0                                 
   \setbox1=\hbox{/} \dimen1=\wd1               
   \ifdim\dimen0>\dimen1                        
      \rlap{\hbox to \dimen0{\hfil/\hfil}}      
      #1                                        
   \else                                        
      \rlap{\hbox to \dimen1{\hfil$#1$\hfil}}   
      /                                         
\fi}

\journal{Annals of Physics}

\begin{document}

\begin{frontmatter}

\title{2PI functional techniques for gauge theories: QED}

\author{Urko Reinosa\fnref{CPHT}}
\ead{urko.reinosa@cpht.polytechnique.fr}
\address{Centre de Physique Th\'eorique \'Ecole Polytechnique, CNRS, 91128 Palaiseau, France}
\fntext[CPHT]{CPHT is unit\'e mixte de recherche UMR7644 (CNRS, \'Ecole Polytechnique).}

\author{Julien Serreau\fnref{APC}}
\ead{julien.serreau@apc.univ-paris7.fr}
\address{Astro-Particule et Cosmologie, Universit\'e Paris 7 - Denis Diderot,\\ 10, rue A.~Domon et L.~Duquet, 75205 Paris Cedex 13, France}
\fntext[APC]{APC is unit\'e mixte de recherche UMR7164 (CNRS, Universit\'e Paris 7, CEA, Observatoire de Paris).}

\begin{abstract}
We discuss the formulation of the prototype gauge field theory, QED, in the context of two-particle-irreducible (2PI) functional techniques with particular emphasis on the issues of renormalization and gauge symmetry. We show how to renormalize all $n$-point vertex functions of the (gauge-fixed) theory at any approximation order in the 2PI loop-expansion by properly adjusting a finite set of local counterterms consistent with the underlying gauge symmetry. The paper is divided in three parts: a self-contained presentation of the main results and their possible implementation for practical applications; a detailed analysis of ultraviolet divergences and their removal; a number of appendices collecting technical details.
\end{abstract}


\end{frontmatter}


\section{Introduction}
\label{sec:intro}

Two-particle-irreducible (2PI) functional techniques \cite{Luttinger:1960ua,Cornwall:1974vz} provide a powerful tool to generate systematic partial resummations of perturbative series in (quantum) field theory, of interest in numerous physical situations where the perturbative expansion breaks down. Topical examples include thermodynamic~\cite{Blaizot:1999ip,Andersen:2004re,Berges:2004hn,Blaizot:2005wr} and transport \cite{Aarts:2003bk} properties of bosonic fields at high temperatures or at a second order phase transition \cite{Alford:2004jj,Arrizabalaga:2006hj}, or genuine nonequilibrium situations~\cite{Berges:2004vw,Berges:2000ur,Aarts:2002dj}, with physical applications ranging from early-universe cosmology \cite{Berges:2002cz} to high-energy heavy-ion collisions experiments \cite{Berges:2002wr} and condensed matter physics \cite{Gasenzer:2005ze}. The application of such techniques to gauge theories is a nontrivial issue which requires one to understand the interplay between renormalization and symmetry. Important progress has been made recently concerning the issue of symmetries in the 2PI formalism, in particular, in the context of abelian gauge theories \cite{Reinosa:2007vi}. Moreover, the basic ideas concerning the renormalization of 2PI QED have been put forward in \Refer{Reinosa:2006cm}. We present here a complete renormalization theory of QED in covariant linear gauges in the 2PI framework. 

It is known that linearly realized global symmetries of a given theory are inherited by the corresponding 2PI-resummed effective action \cite{Aarts:2002dj,vanHees:2002bv}. This has recently been shown to be true also for abelian gauge theories in linear gauges \cite{Reinosa:2007vi}. In particular, it has been shown how to construct 2PI approximation schemes which systematically respect the corresponding Ward-Takahashi identities. Another important issue regarding gauge symmetry concerns the gauge-fixing independence of physical observables \cite{Arrizabalaga:2002hn,Mottola:2003vx}. The point is that, even though the latter, as defined in the exact theory, are gauge-fixing independent, the corresponding expressions obtained from a given 2PI approximation, contain residual gauge-fixing dependences due to the resummation of infinite subclasses of perturbative contributions inherent to 2PI techniques. It is to be emphasized that the fact that vertex functions obtained from a given 2PI approximation exactly satisfy the correct symmetry identities does not guarantee that the corresponding physical observables are gauge-invariant, i.e. gauge-fixing independent. 

For systematic approximation schemes, such as e.g. the 2PI loop-expansion, it can be shown that these gauge dependent contributions are parametrically suppressed in powers of the expansion parameter \cite{Arrizabalaga:2002hn} (see also~\cite{Andersen:2004re}). The observed good convergence properties of 2PI approximations schemes \cite{Berges:2004hn,Blaizot:2005wr} therefore suggests that they should be under control, at least in weakly coupled situations, provided that renormalization is well understood. This has recently been tested for QED in \Refer{Borsanyi:2007bf}, where the thermodynamic pressure has been computed at finite temperature from the two-loop approximation of the 2PI effective action in covariant gauge. The results indicate that gauge-fixing parameter dependences remain under control in a wide range of couplings and are comparable with the renormalization scale dependences, other sources of uncertainty inherent to such calculations. Moreover, the Landau gauge has been identified as a gauge minimizing both gauge-fixing parameter and renormalization scheme dependences.  

The second aspect mentioned above, renormalizability, is deeply related to that of symmetries. Renormalization in the 2PI framework has been understood only in recent years and has been extensively developed for theories with scalar~\cite{VanHees:2001pf,vanHees:2002bv,Berges:2004hn,Cooper:2004rs,Berges:2005hc,Arrizabalaga:2006hj}, but also fermionic~\cite{Reinosa:2005pj} and, recently, gauge~\cite{Reinosa:2006cm,Borsanyi:2007bf} degrees of freedom. Most studies have been concerned with the issue of renormalizing the basic equation of the 2PI formalism, that is the self-consistent equation for the two-point function. In \Refer{Berges:2005hc} a complete description of all $n$-point vertex functions in the 2PI framework has been put forward for scalar field theories. The present paper develops the renormalization theory of 2PI QED in covariant linear gauge, extending our previous work~\cite{Reinosa:2006cm} for the renormalization of the photon and fermion two-point functions. In particular, we show that all proper vertex functions obtained from the 2PI loop-expansion of QED can be made finite at any approximation order by adjusting a finite set of local counterterms consistent with the underlying gauge symmetry.

In Sec.~\ref{sec:generalities} we review the 2PI formulation of QED. In particular, we consider the field expansion of the so-called 2PI-resummed effective action, which defines the (2PI-resummed) proper vertex functions of the theory. These can be expressed in terms of a second class of vertex functions which we call 2PI vertices and which require independent renormalization. Sec.~\ref{sec:symmetry_counterterms} introduces the counterterms needed to renormalize both 2PI and 2PI-resummed vertex functions in a way that preserves the symmetries of the bare theory. The presence of two different classes of vertices and the fact that gauge symmetry constraints them differently \cite{Reinosa:2007vi} call for introducing additional counterterms as compared to the standard (i.e. perturbative) renormalization theory. Still, the number of independent renormalization conditions which define the theory is the same as usual, as we discuss in \Sec{sec:rencond}. Indeed, the extra-counterterms are fixed by imposing consistency conditions between 2PI and 2PI-resummed vertex functions at the renormalization point. This is a standard feature of 2PI renormalization theory \cite{Berges:2005hc}. Another important issue, which we discuss in Secs.~\ref{sec:WTI} and \ref{sec:rencond}, is to make sure that renormalization and consistency conditions respect the underlying gauge symmetry. In \Sec{sec:summary}, we provide a ready-to-use renormalization procedure which allows one to fix all the counterterms in a way that preserves the corresponding 2PI Ward-Takahashi identities. Readers interested in practical applications will find everything they need in those sections. 

Finally, Secs.~\ref{sec:renormalization_2PI_vertices} and \ref{sec:renormalization_2PI_resummed_vertices} give a detailed analysis of ultraviolet divergences present in 2PI and 2PI-resummed vertex functions and describes how they can be systematically absorbed in the above mentioned counterterms. In particular, we show that the structure of the (sub-)divergences and, consequently, of the corresponding counterterms, is constrained by 2PI Ward identities in precisely such a way that no divergence appears which would call for symmetry-breaking couterterms. In turn, this guarantees that gauge symmetry is preserved by the renormalization procedure, provided one imposes suitable (i.e. gauge-symmetric) renormalization conditions. Some technical issues are presented in the appendices. 

\section{The 2PI-resummed effective action for QED}
\label{sec:generalities}

\subsection{Generalities}

We consider QED in the linear covariant gauge. The gauge-fixed classical action reads 
\begin{equation}
\label{eq:classact}
S[A,\psi,\bar\psi]=\int_x \left\{\bar\psi\Big[i\slashchar{\partial}-e\slashchar{A}-m\Big]\psi+\frac{1}{2}A^\mu\Big[g_{\mu\nu}\partial^2-(1-\lambda)\partial_\mu\partial_{\nu}\Big]A^\nu\right\}\,,
\end{equation}where $\int_x\equiv\int d^4x$. Aside from the gauge-fixing term, proportional to the parameter $\lambda$, the classical action is invariant under the gauge transformation
\begin{equation}\label{eq:gauge}
\psi(x)\rightarrow e^{i\alpha(x)}\,\psi(x)\,,\,\,
\bar\psi(x)\rightarrow e^{-i\alpha(x)}\,\bar\psi(x)\,,\,\,
A_\mu(x)\rightarrow A_\mu(x)-\frac{1}{e}\,\partial_\mu\alpha(x)\,,
\end{equation}
where $\alpha(x)$ denotes an arbitrary real function. The free inverse fermion and photon propagators are given by
\begin{eqnarray}
\label{eq:freepf}
iD_{0,\bar\alpha\alpha}^{-1}(x,y) & = & \left[\,i\slashchar{\partial}_x-m\,\right]_{\bar\alpha\alpha}\delta^{(4)}(x-y)\,,\\
\label{eq:freepb}
iG_{0,\mu\nu}^{-1}(x,y) & = & \left[\,g_{\mu\nu}\partial_x^2-
(1-\lambda)\partial^x_\mu\partial^x_\nu\,\right]\delta^{(4)}(x-y)\,.
\end{eqnarray}

A derivation of the 2PI effective action for QED can be found in Ref.~\cite{Reinosa:2007vi}. We briefly review the main formalism here. In order to define the 2PI effective action in full generality, one needs to introduce the usual connected correlators
for time-ordered Maxwell and Dirac fields
\beq
\label{eq:cor1}
 G_{\mu\nu}(x,y)\equiv\bra T A_\mu(x) A_\nu(y)\ketc\quad,\quad D_{\alpha\bar\beta}(x,y)\equiv\bra T
 \psi_{\alpha}(x)\bar\psi_{\bar\beta}(y)\ketc\,,
\eeq
as well as the mixed connected correlators\footnote{Note that $G$, $D$, $F$ and $\bar F$ are commuting functions, whereas $K$ and $\bar K$ are anticommuting.}
\beq
\label{eq:cor2}
 K_{\alpha\nu}(x,y)\equiv\bra T \psi_\alpha(x)A_\nu(y)\ketc\quad,\quad\bar K_{\mu\bar\beta}(x,y)\equiv\bra T
 A_\mu(x)\bar\psi_{\bar\beta}(y)\ketc
\eeq
and
\beq
\label{eq:cor3}
 F_{\alpha\beta}(x,y)\equiv\bra T \psi_\alpha(x)\psi_\beta(y)\ketc\quad,\quad\bar F_{\bar\alpha\bar\beta}(x,y)\equiv\bra T \bar\psi_{\bar\alpha}(x)\bar\psi_{\bar\beta}(y)\ketc\,.
\eeq
It proves convenient to grab the bosonic and fermionic fields $A$, $\psi$ and $\bar\psi$ in a $12$-component superfield $\varphi$ and the various
correlators \eqn{eq:cor1}-\eqn{eq:cor3} in a corresponding supercorrelator $\cG\equiv\bra T \varphi\varphi^t\ketc$:
\beq
\label{eq:sfield}
 \varphi\equiv\left(
 \begin{tabular}{c}
  $A$\\
  $\psi$\\
  $\bar\psi^t$
  \end{tabular}
  \right)
  \qquad,\qquad
  \cG\equiv \left(
 \begin{tabular}{ccc}
 $G$&$K^t$&$\bar K$\\
 $K$&$F$&$D$\\
 $\bar K^t$&$-D^t$&$\bar F$
 \end{tabular}
 \right)\,,
\eeq
where the upperscript $t$ means transposition, of Dirac or Lorentz indices and of space-time variables.\footnote{We employ a notation where space-time variables, if not written explicitly, are put together with Lorentz, Dirac, or superfield indices. Repeated indices are implicitly summed over, which includes an integration over space-time variables.} We assign a fermion number $q_m$ to the component $\varphi_m$ of the superfield: $\smash{q_m=0}$ for
bosonic $A$-like components, $\smash{q_m=1}$ for fermionic $\psi$-like components and $\smash{q_m=-1}$ for fermionic
$\bar\psi$-like components. Note the following symmetry property~\cite{Reinosa:2007vi}: 
\beq\label{eq:sympropG}
 \cG_{mn}=(-1)^{q_mq_n}\cG_{nm}\,.
\eeq
Written as a functional of $\varphi$ and $\cG$, the 2PI effective action reads \cite{Reinosa:2007vi},
\beq 
\label{eq:2PI}
\Gammatpi[\varphi,\cG]=S_0[\varphi]+\frac{i}{2}\,\Str\,\Ln\,\cG^{-1}+\frac{i}{2}\,\Str\,\cG_0^{-1}\cG+\Gammaint[\varphi,\cG]\,,
\eeq
where $\Str$ denotes the functional supertrace\footnote{For a supermatrix $\cal M$ with bosonic and fermionic elements as in \Eqn{eq:sfield}, one has ${\rm str}\,{\cal M}\equiv\sum_n (-1)^{q_n}{\cal M}_{nn}$. 
The functional
supertrace ${\rm Str}$ involves, in addition, an integration over space-time variables.} and $S_0[\varphi]$ is the free action, defined as
\beq
\label{eq:freeact}
 S_0[\varphi]\equiv\frac{1}{2}\,\varphi_m i\cG_{0,mn}^{-1}\,\varphi_n\,,
\eeq
with the free inverse propagator\footnote{Throughout this paper, functional derivatives are understood as being right derivatives and
successive derivatives are noted such that the rightmost derivative acts first. For any functional $F$ of
$\chi\equiv\varphi$ or $\cG$, one has
$$
 F[\chi+\delta\chi]=\sum_{n\ge
 0}\frac{1}{n!}\,\delta\chi_1\ldots\delta\chi_n\,\frac{\delta^{n}F[\chi]}{\delta\chi_n\ldots\delta\chi_1}\,.
$$}
\beq
 i\cG_{0,mn}^{-1}\equiv\left.(-1)^{q_n}\frac{\delta^{2} S[\varphi]}{\delta\varphi_m\delta\varphi_n}\right|_{\varphi=0}\,.
\eeq
The only nonvanishing components of $\cG_0^{-1}$ are given by Eqs.~\eqn{eq:freepf}-\eqn{eq:freepb}:
\bea
\label{eq:scor0}
 i\cG_0^{-1}=\left(
 \begin{tabular}{ccc}
 $iG_0^{-1}$&$0$&$0$\\
 $0$&$0$&$\left(-iD_0^{-1}\right)^t$\\
 $0$&$iD_0^{-1}$&$0$
 \end{tabular}
 \right)\,.
\eea

The functional $i\Gamma_{\rm int}[\varphi,\cG]$ is the sum of closed
two-particle-irreducible (2PI) diagrams with lines corresponding to the components of $\cG$ and
vertices obtained from the shifted action $S_{\rm int}[\varphi +\delta\varphi]$. In QED, and more generally in theories with only cubic interactions, the complete field dependence of $\Gammaint[\varphi,\cG]$ is contained in its zero (classical) and one-loop contributions~\cite{Reinosa:2007vi}. We may thus write 
\beq\label{eq:2PI_loops}
\Gammaint[\varphi,\cG]\equiv S_{\rm int}[\varphi]+\Gammaint^{\rm 1loop}[\varphi,\cG]+\Gamma_2[\cG]\,,
\eeq
where, in term of the components of the superfields $\varphi$ and $\cG$,  
\bea
S_{\rm int}[\varphi] \!\!& = &\!\! -\,e\int_x\bar\psi(x)\slashchar{A}\psi(x)\,,\\
\Gammaint^{\rm 1loop}[\varphi,\cG] \!\!& = &\!\! e\int_x
\Big(\tr\left[\slashchar{A}(x)D(x,x)\right]-\bar\psi(x)\slashchar{K}(x,x)-\bar{\slashchar{K}}(x,x)\psi(x)\Big).
\eea

The effective action $\Gamma[\varphi]$, i.e. the generating functional for 1PI $n$-point functions, can be obtained from the 2PI effective
action (\ref{eq:2PI}) as \cite{Luttinger:1960ua,Cornwall:1974vz,vanHees:2002bv,Reinosa:2007vi}
\beq\label{eq:1PI}
\Gamma[\varphi]=\Gamma_{\rm 2PI}[\varphi,\bcG[\varphi]]\,,
\eeq
where the physical correlator $\bcG[\varphi]$ in presence of a nonvanishing field $\varphi$ is obtained from the stationarity condition
\beq
\label{eq:stat}
 \left.\frac{\delta\Gammatpi[\varphi,\cG]}{\delta \cG}\right|_{\bcG[\varphi]}=0\qquad\Longleftrightarrow\qquad\bar\cG^{-1}[\varphi]=\bar\cG^{-1}_{0}-\bar\Sigma[\varphi]
\eeq
where we introduced the self-energy
\beq\label{eq:sig}
 \bar\Sigma_{nm}[\varphi]=\left.(-1)^{q_m}\,2i\,\frac{\delta\Gammaint[\varphi,\cG]}{\delta
 \cG_{mn}}\right|_{\bcG[\varphi]}.
\eeq
The independent components of the latter are given by
\bea
 \bSigma\equiv\left(
 \begin{tabular}{ccc}
 $\bSigma_{AA}$&$\bSigma_{A\psi}$&$-\bSigma^t_{\bar\psi A}$\\
 $-\bSigma^t_{A\psi}$&$\bSigma_{\psi\psi}$&$-\bSigma^t_{\bar\psi\psi}$\\
 $\bSigma_{\bar\psi A}$&$\bSigma_{\bar\psi\psi}$&$\bSigma_{\bar\psi\bar\psi}$
 \end{tabular}
 \right).
\eea
Although it is a trivial identity in the exact theory, Eq.~\eqn{eq:1PI} provides an efficient starting point to devise systematic nonperturbative approximations  of the effective action $\Gamma[\varphi]$ through systematic expansions of the 2PI functional $\Gammatpi[\varphi,\cG]$.\footnote{The nonperturbative character of the approximation scheme is encoded in the self-consistent definition of $\bcG[\varphi]$, \Eqn{eq:stat}.} This is the so-called 2PI resummation scheme \cite{Berges:2005hc}. For any given approximation of the 2PI effective action, Eq.~\eqn{eq:1PI} defines the corresponding 2PI-resummed effective action.

\subsection{Vertex functions}\label{sec:barevertex}

Proper vertex functions of the theory can be defined as field derivatives of the (2PI-resummed) effective action
\beq
\label{eq:npoint}
 \Gamma^{(n)}_{1\ldots n}\equiv\left.\frac{\delta^n\Gamma[\varphi]}{\delta\varphi_n\cdots\delta\varphi_1}\right|_{\bar\varphi}
\eeq
evaluated at the physical value of the field $\smash{\varphi=\bar\varphi}$, defined as\footnote{For gauge-fixed QED in C-invariant physical states, which we assume in this paper, $\smash{\bar\varphi=0}$. All equations in the main text assume that the global $U(1)$ symmetry of the theory is not broken, in particular, this implies that the fermionic components of $\bar\varphi$ vanish.}
\beq
\label{eq:barphi}
 \left.\frac{\delta\Gamma[\varphi]}{\delta\varphi_1}\right|_{\bar\varphi}=0\,.
\eeq
We shall refer to the functions \eqn{eq:npoint} as 2PI-resummed vertex functions. Using Eqs.~(\ref{eq:2PI}) and (\ref{eq:1PI}), the latter can be expressed in terms of derivatives of $\Gammaint[\varphi,\cG]$ with respect to $\varphi$ or $\cG$, so-called 2PI kernels~\cite{Berges:2005hc,Reinosa:2007vi}. One has for instance
\beq
\label{eq:first} \frac{\delta\Gamma}{\delta\varphi_1}=\left.\frac{\delta\Gammatpi}{\delta\varphi_1}\right|_\bcG=i\cG_{0,1n}^{-1}\varphi_n+\left.\frac{\delta\Gammaint}{\delta\varphi_1}\right|_\bcG\,,
\eeq
where we have used the stationarity condition \eqn{eq:stat}. Similar expressions can be obtained for higher 2PI-resummed vertex functions. This has been worked out in detail in the context of scalar field theories in \Refer{Berges:2005hc}. In the present case, one has to pay particular attention to additional signs which arise due to the presence of Grassman field components. This is detailed in 
App.~\ref{appsec:2PI_and_2PI_resummed_vertices}. We
obtain, for the 2PI-resummed two-point vertex function,\footnote{We specialize to the case of a purely cubic interaction, see \Eqn{eq:2PI_loops}.} 
\beq\label{eq:g2}
\Gamma^{(2)}_{12}=(-1)^{q_1}i\cG_{0,21}^{-1}+\left.\frac{\delta^{2}\Gammaint}{\delta\varphi_2\delta\varphi_1}\right|_\bcG+\left(\bcG\,\frac{\delta\bSigma}{\delta\varphi_2}\,\bcG\right)_{\!\!\!mn}\left.\frac{\delta^{2}\Gammaint}{\delta\cG_{mn}\delta\varphi_1}\right|_\bcG\,,
\eeq
and for the 2PI-resummed three-point vertex function,
\beq\label{eq:g3}
\Gamma^{(3)}_{123}=\left.\frac{\delta^{3}\Gammaint}{\delta\varphi_3\delta\varphi_2\delta\varphi_1}\right|_\bcG+\left(2\,\bcG\,\frac{\delta\bSigma}{\delta\varphi_2}\,\bcG\,\frac{\delta\bSigma}{\delta\varphi_3}\,\bcG+\bcG\,\frac{\delta^{2}\bSigma}{\delta\varphi_3\delta\varphi_2}\,\bcG\right)_{\!\!\!mn}\left.\frac{\delta^{2}\Gammaint}{\delta\cG_{mn}\delta\varphi_1}\right|_\bcG\,,
\eeq
where it is understood that the field assumes its physical value $\smash{\varphi=\bar\varphi}$.

It is clear from the previous formulae (see also App.~\ref{appsec:2PI_and_2PI_resummed_vertices}) that 2PI-resummed vertex functions can be conveniently expressed in terms of field derivatives of the self-energy $\bar\Sigma[\varphi]$ taken at $\smash{\varphi=\bar\varphi}$. The latter implicitly contains all the nonperturbative resummations intrinsic to the 2PI resummation scheme, through \Eqn{eq:sig}. This is made explicit by the fact that its derivatives satisfy linear integral, Bethe-Salpeter--like equations, see App.~\ref{appsec:2PI_and_2PI_resummed_vertices}. One has for instance
\beq\label{eq:a1}
\frac{\delta\bar\Sigma_{nm}}{\delta\varphi_1}=(-1)^{q_m}\!\left.\frac{2i\,\delta^2\Gammaint}{\delta\varphi_1\delta\cG_{mn}}\right|_{\bcG}+\left(\bcG\frac{\delta\bar\Sigma}{\delta\varphi_1}\bcG\right)_{\!\!\!ab}(-1)^{q_m}\!\left.\frac{2i\,\delta^2\Gammaint}{\delta\cG_{ab}\delta\cG_{mn}}\right|_{\bcG}\,, 
\eeq
where, again it is understood that all derivatives are evaluated at $\varphi=\bar\varphi$.
Similarly, the second derivative, needed e.g. in \Eqn{eq:g3}, reads
\bea\label{eq:a2}
\frac{\delta^2\bar\Sigma_{nm}}{\delta\varphi_2\delta\varphi_1} & = & \left(\cG\frac{\delta\bSigma}{\delta\varphi_1}\cG\right)_{\!\!ab}\!\!\left(\cG\frac{\delta\bSigma}{\delta\varphi_2}\cG\right)_{\!\!cd}(-1)^{q_m}\!\left.\frac{2i\,\delta^3\Gammaint}{\delta\cG_{cd}\delta\cG_{ab}\delta\cG_{mn}}\right|_{\bcG}\nonumber\\
& + &
\left(2\,\bcG\,\frac{\delta\bSigma}{\delta\varphi_1}\,\bcG\,\frac{\delta\bSigma}{\delta\varphi_2}\,\bcG+\bcG\,\frac{\delta^{2}\bSigma}{\delta\varphi_2\delta\varphi_1}\,\bcG\right)_{\!\!\!ab}(-1)^{q_m}\!\left.\frac{2i\,\delta^2\Gammaint}{\delta\cG_{ab}\delta\cG_{mn}}\right|_{\bcG}.
\eea

It is to be emphasized that field-derivatives of the self-energy, or more precisely, the functions
\beq
\label{eq:V}
V^{(p+2)}_{mn;1\cdots p}\equiv -i(-1)^{q_m}\left.\frac{\delta^p\bar\Sigma_{nm}[\varphi]}{\delta\varphi_p\cdots\delta\varphi_1}\right|_{\bar\varphi},
\eeq
provide an alternative definition of the proper vertex functions of the theory. We shall refer to the latter as 2PI vertex functions~\cite{Reinosa:2007vi,Berges:2005hc}. Although both definitions coincide in the exact theory, they differ at finite approximation order. 2PI vertex functions can be computed -- and renormalized -- independently from 2PI-resummed ones. In contrast, 2PI-resummed vertex functions implicitly involve 2PI vertex functions and their renormalization requires a prior renormalization of the latter.
 
We end this section by writing down the explicit equations to be discussed in the following, namely Eqs.~\eqn{eq:g2}-\eqn{eq:a2} written in terms of the components of the superfields $\varphi$ and $\cG$. To this purpose, some care must be taken concerning the fact that the various components of $\cG$ are not independent. One has, for instance,
\beq
 \frac{\delta\Gammaint}{\delta D_{\alpha\bar\alpha}}=\frac{\delta\cG_{mn}}{\delta{D}_{\alpha\bar\alpha}}\frac{\delta\Gammaint}{\delta\cG_{mn}}=\frac{\delta\Gammaint}{\delta\cG_{\psi\bar\psi}^{\alpha\bar\alpha}}-\frac{\delta\Gammaint}{\delta\cG_{\bar\psi\psi}^{\bar\alpha\alpha}}=2\frac{\delta\Gammaint}{\delta\cG_{\psi\bar\psi}^{\alpha\bar\alpha}}
\eeq
and similarly for other functional derivatives. 

We shall also take advantage of the global symmetries of the theory which we assume to be preserved by the approximation at hand.\footnote{Some global symmetries might be broken by the particular physical (initial) state one considers, without affecting the discussion of renormalization. For instance, the presence of a thermal bath explicitly breaks Lorentz symmetry but does not affect the structure of UV divergences.} To be specific, we shall focus on the 2PI loop-expansion. The global $U(1)$ symmetry implies that any quantity having a nonvanishing fermion number vanishes in the absence of external sources (i.e. ``on-shell"). For instance all mixed correlators $K$, $\bar K$, $F$ and $\bar F$ vanish on-shell; also
$\delta\bSigma_{mn}/\delta\varphi_p=0$ unless $\smash{q_m+q_n+q_p=0}$ etc., see \App{appsec:charge_conjugation_invariance}. Parity and charge conjugation symmetry lead to further simplifications of the equations, see App.~\ref{appsec:charge_conjugation_invariance} for details. For instance, one deduces from charge conjugation symmetry that 2PI $n$-photon functions with $n$ odd vanish identically, a generalization of Furry's theorem for 2PI vertices ($\varphi=\bar\varphi=0$ implicit):
\beq
\frac{\delta^{2p+1}\bSigma_{AA}}{\delta A_{2p+1}\cdots\delta A_1}=0\,.
\eeq
One also obtains that the 2PI three-point functions $\delta\bSigma_{A\psi}/\delta\bar\psi$ and $\delta\bar\Sigma_{\bar\psi A}/\delta\psi$ are linearly related to each other by means of the charge conjugation matrix $C$ ($\varphi=\bar\varphi=0$ implicit):
\beq\label{eq:3p}
\frac{\delta\bSigma_{A\psi}^{\mu\alpha}(x,y)}{\delta\bar\psi_{\bar\alpha}(z)}=-C_{\alpha\bar\beta}\frac{\delta\bSigma_{\bar\psi A}^{\bar\beta\mu}(y,x)}{\delta\psi_\beta(z)}C^{-1}_{\beta\bar\alpha}\,.
\eeq

The only two independent components of the 2PI three-point vertex $V^{(3)}\propto\delta\bSigma/\delta\varphi$, namely $\delta\bSigma_{\bar\psi\psi}/\delta A$ and $\delta\bSigma_{\bar\psi A}/\delta \psi$ fulfill the following self-consistent equations ($\varphi=\bar\varphi=0$ implicit)
\beq\label{eq:2PI3_bpap}
\frac{\delta\bSigma_{\bar\psi
A}^{\bar\alpha\mu}}{\delta\psi_\alpha}=\frac{i\delta^2\Gammaint}{\delta\psi_\alpha\delta\bK_{\mu\bar\alpha}}+\bD_{\beta\bar\beta}\frac{\delta\bSigma_{\bar\psi
A}^{\bar\beta\rho}}{\delta\psi_\alpha}\bG_{\rho\nu}\left.\frac{i\delta^2\Gammaint}{\delta K_{\beta\nu}\delta\bK_{\mu\bar\alpha}}\right|_\bcG
\eeq
and
\beq\label{eq:2PI3_bppa}
\frac{\delta\bSigma_{\bar\psi\psi}^{\bar\alpha\alpha}}{\delta A_\mu}=\frac{-i\delta^2\Gammaint}{\delta A_\mu\delta D_{\alpha\bar\alpha}}+\bar D_{\beta\bar\gamma}\,\frac{\delta\bSigma_{\bar\psi\psi}^{\bar\gamma\gamma}}{\delta A_\mu}\,\bar D_{\gamma\bar\beta}\left.\frac{-i\delta^2\Gammaint}{\delta D_{\beta\bar\beta}\delta D_{\alpha\bar\alpha}}\right|_\bcG\,.
\eeq
We will also have to discuss the 2PI four-point function $\delta^2\bSigma_{\bar\psi A}/\delta A\delta\psi$ which reads, explicitly,
\bea
\frac{\delta^2\bar\Sigma_{\bar\psi A}^{\bar\beta\nu}}{\delta A_\mu\delta\psi_\alpha} & = & \left(\bD\frac{\delta\bSigma_{\bar\psi
A}}{\delta\psi_\alpha}\bG\right)_{\!\!\beta\rho}\!\!\left(\bD\frac{\delta\bSigma_{\bar\psi\psi}}{\delta
A_\mu}\bD\right)_{\!\!\gamma\bar\gamma}\left.\frac{i\delta^3\Gammaint}{\delta D_{\gamma\bar\gamma}\delta K_{\beta\rho}\delta
\bK_{\nu\bar \beta}}\right|_\bcG\nonumber\\
\label{eq:explicitely}
& + & \left(\bD \frac{\delta\bSigma_{\bar\psi\psi}}{\delta A_\mu}\bD\frac{\delta\bSigma_{\bar\psi
A}}{\delta\psi_\alpha}\bG+\bar D\frac{\delta^2\bSigma_{\bar\psi A}}{\delta A_\mu\delta\psi_\alpha}\bar G\right)_{\!\!\beta\rho}\!\!\left.\frac{i\delta^2\Gammaint}{\delta K_{\beta\rho}\delta\bar K_{\nu\bar\beta}}\right|_\bcG.
\eea
Finally, the 2PI-resummed two- and three-point vertex functions read ($\varphi=\bar\varphi=0$ implicit)
\beq\label{eq:2PIr2_pbp}
\frac{-i\delta^2\Gamma}{\delta \psi_\alpha\delta
\bar\psi_{\bar\alpha}}=D_{0,\,\bar\alpha\alpha}^{-1}+\frac{-i\delta^2\Gammaint}{\delta
\psi_\alpha\delta\bar\psi_{\bar\alpha}}+\bar D_{\beta\bar\beta}\frac{\delta\bSigma_{\bar\psi A}^{\bar\beta\nu}}{\delta\psi_\alpha}\bar G_{\nu\mu}\frac{-i\delta^2\Gammaint}{\delta K_{\beta\mu}\delta \bar\psi_{\bar\alpha}}\,,
\eeq
\beq\label{eq:2PIr2}
\frac{-i\delta^2\Gamma}{\delta A_\mu\delta A_\nu}=G_{0,\,\mu\nu}^{-1}+\frac{-i\delta^2\Gammaint}{\delta A_\mu\delta
A_\nu}+\bar D_{\alpha\bar\beta}\frac{\delta\bSigma_{\bar\psi\psi}^{\bar\beta\beta}}{\delta A_\mu}\bar D_{\beta\bar\alpha}\frac{-i\delta^2\Gammaint}{\delta D_{\alpha\bar\alpha}\delta A_\nu} 
\eeq
and
\beq\label{eq:2PIr34}
\frac{i\delta^3\Gamma}{\delta A_\mu\delta\psi_\alpha\delta\bar\psi_{\bar\alpha}}=\frac{i\delta^3\Gammaint}{\delta
A_\mu\delta\psi_\alpha\delta\bar\psi_{\bar\alpha}}\!+\!\left(\bar
D\frac{\delta\bSigma_{\bar\psi \psi}}{\delta A_\mu}\bar D\frac{\delta\bSigma_{\bar\psi
A}}{\delta\psi_\alpha}\bar G+\bar D\frac{\delta^2\bSigma_{\bar\psi A}}{\delta A_\mu\delta\psi_\alpha}\bar G\right)_{\!\!\beta\nu}\!\!\frac{i\delta^2\Gammaint}{\delta
K_{\beta\nu}\delta \bar\psi_{\bar\alpha}}.
\eeq
Notice that \Eqn{eq:2PIr34} can be written in a different way (depending on which field derivative is taken first), involving the four-point 2PI vertex $\delta\bSigma_{AA}/\delta\psi\delta\bar\psi$, see e.g. \cite{Reinosa:2007vi}. The above expression is more suited to discuss renormalization.

\section{Renormalized 2PI effective action}
\label{sec:symmetry_counterterms}
The bare 2PI and 2PI-resummed vertex functions introduced in the previous section are plagued by UV divergences and require regularization, see \Sec{sec:regularization} and renormalization. To this purpose, we introduce the corresponding renormalized vertex functions, which can be conveniently defined from the renormalized 2PI effective action, defined in \Sec{sec:definition}. In particular, the latter includes a counterterm part consistent with the symmetries of the theory, see \Sec{sec:counterterms}. This allows one to renormalize all vertex functions in a gauge symmetric way, provided one imposes suitable (i.e. gauge symmetric) renormalization conditions, as discussed in Secs.~\ref{sec:WTI} and \ref{sec:rencond}. \Sec{sec:summary} summarizes the renormalization procedure.

\subsection{Regularization}\label{sec:regularization}

Before starting our discussion of renormalization, a point of caution is in order. In what follows, we shall not actually perform a nonperturbative proof of renormalization in the 2PI framework. This is beyond the scope of the present paper. Instead, we shall employ a formal coupling expansion of the various renormalized vertex functions obtained from an arbitrary 2PI approximation and show that the coefficients of such formal series are all finite. 

Since each of these coefficients is given by a finite sum of standard perturbative diagrams (those which are summed up in the 2PI approximation at hand), we are free to use standard gauge-symmetric regularization schemes, such as Pauli-Villars, lattice, or dimensional regularizations. We choose to work with the latter, the simplest one, which respects all symmetries of the theory (unlike the lattice) and does not require any modification of the classical action (unlike Pauli-Villars or the lattice). All four-dimensional momentum space integrals are replaced by $d$-dimensional ones and expressed as analytic functions of $\epsilon=4-d$ in the usual manner. In Secs.~\ref{sec:renormalization_2PI_vertices} and \ref{sec:renormalization_2PI_resummed_vertices}, we show that the coefficients of the formal coupling expansions of any vertex functions in the 2PI formalism all have a finite limit when $\epsilon$ is taken to zero.\footnote{In Secs.~\ref{sec:renormalization_2PI_vertices} and \ref{sec:renormalization_2PI_resummed_vertices}, we also employ a ladder-rung expansion for some equations, which greatly simplifies the analysis. In principle, dimensional regularization does not apply to the diagram of such an expansion, which involve resummed propagator lines. Strictly speaking, all the expression we shall write are to be understood in the sense of their formal coupling expansions.}
 
The limitations of such an analysis are the same as those of standard perturbative renormalization proofs. In particular, the obtained renormalized series are at best asymptotic ones and one has to wonder how to make sense out of them. Of course, the whole point of 2PI techniques is to directly solve the nonperturbative equations for resummed propagators and vertices without resorting to such perturbative expansion, which requires a truly nonperturbative renormalization.\footnote{Since QED is expected to exhibit a Landau pole at a nonperturbatively high momentum scale, see e.g. \cite{Gies:2004hy}, renormalization has to be understood in the usual sense of minimizing the cutoff-sensitivity of physical observables in a range of cutoff well above any physical momentum scale in the problem and well below the Landau pole.} Nonperturbative proofs of 2PI renormalization are accessible to analytical analysis under the (conservative) assumption that the asymptotic large momentum behavior of resummed propagator lines $\bcG(p)$ at any given 2PI approximation order is bounded -- up to logarithms -- by that of the corresponding free propagators so that the standard power counting analysis of divergent loop integrals applies\footnote{\label{ft:ft} We stress that this is not guaranteed to be true a priori for arbitrary 2PI approximations. Partial resummations of perturbative logarithms can modify the convergence properties of diagrams with resummed propagator lines as compared to naive power counting. Such modifications occur, for instance, in the standard $1/N$ expansion of $0(N)$ scalar field theory, see e.g. \cite{Fejos:2009dm}.}, see e.g.~\cite{VanHees:2001pf,vanHees:2002bv,Berges:2004hn,Cooper:2004rs,Reinosa:2005pj,Reinosa:2006cm}. 
This assumption must ultimately be checked by actual (numerical) calculations, see e.g.~Refs.~\cite{VanHees:2001pf,Berges:2004hn,Arrizabalaga:2006hj,Borsanyi:2007bf}. 

Such an analysis, not to be performed here, requires a nonperturbative -- in the present context gauge-invariant -- regularization, such as the lattice. The latter involves specific lattice vertices in the discretized classical action. However, it can be checked that these do not modify the usual power counting of divergent integrals \cite{Borsanyi:2007bf} and do not require additional renormalization/consistency conditions. Therefore, we expect that no counterterms other than those introduced in dimensional regularization are needed (see however footnote \ref{ft:ft}) and that the renormalization procedure presented below is complete. A recent numerical calculation of the QED pressure, using the (isotropic) lattice formulation of the 2PI effective action to two-loop order, seems to support this idea \cite{Borsanyi:2007bf}.

\subsection{Renormalized 2PI effective action: definition}\label{sec:definition}
The renormalized 2PI effective action is defined as:\footnote{The term ``renormalized 2PI effective action'' means that all (renormalized) vertex functions derived from it, which only involve the physical $\varphi_R$ and $\cG_R$, are UV convergent once a proper renormalization procedure is applied. Notice, however, that we do not discuss the renormalization of composite operators in this paper, which would involve further independent counterterms.}
\beq\label{eq:bar_ren_2PI}
\Gamma_{\rm 2PI}^R[\varphi_R,\cG_R]\equiv\Gamma_{\rm 2PI}[\varphi,\cG]\,,
\eeq
where the renormalized superfield $\varphi_R$ and superpropagator $\cG_R$ are given by
\beq\label{eq:sfield_redef}
\varphi_R\equiv Z^{-1/2}\varphi \quad {\rm and} \quad \cG_R\equiv Z^{-1/2}\,\cG\,Z^{-1/2}\,,
\eeq
with the supermatrix $\smash{Z\equiv {\rm diag}\,(Z_3\mathds{1}_{4\times 4},Z_2\mathds{1}_{4\times 4},Z_2\mathds{1}_{4\times 4})}$. We also introduce the standard renormalized mass, coupling and gauge-fixing parameters:
\beq\label{eq:param_redef}
Z_0m_R\equiv Z_2m\,,\;\; Z_1e_R\equiv Z_2Z_3^{1/2}e \,,\;\;Z_4\lambda_R\equiv Z_3\lambda\,.
\eeq
Let us here recall that, in the exact theory, the Ward-Takahashi identities associated with the underlying $U(1)$ gauge symmetry ensure that $Z_1/Z_2$ and $Z_4$ are UV finite. It follows that, in the exact theory, one can choose $\smash{Z_1/Z_2=1}$ and $\smash{Z_4=1}$, up to finite redefinitions of the coupling constant $e_R$ and gauge-fixing parameter $\lambda_R$ respectively.

Renormalized vertex functions are obtained from the renormalized 2PI effective action in the very same way as bare vertex functions were obtained in Sec.~\ref{sec:barevertex} from the bare 2PI effective action. In particular, the renormalized 2PI-resummed effective action $\Gamma_R[\varphi_R]\equiv\Gamma[\varphi]$, which field expansion defines the renormalized 2PI-resummed vertex functions, can be obtained as, see \Eqn{eq:1PI},
\beq\label{eq:1PIRen}
\Gamma_R[\varphi_R]=\Gamma^R_{\rm 2PI}[\varphi_R,\bcG_R[\varphi_R]]\,,
\eeq
where the renormalized physical correlator $\bcG_R[\varphi_R]$ is obtained from, see \Eqn{eq:stat},
\beq
\label{eq:statRen}
 \left.\frac{\delta\Gamma^R_{\rm 2PI}[\varphi_R,\cG_R]}{\delta \cG_R}\right|_{\bcG_R[\varphi_R]}=0\,.
\eeq

It is convenient to write the renormalized 2PI effective action in a form similar to (\ref{eq:2PI}) but involving exclusively renormalized superfields, superpropagators and parameters:
\beq\label{eq:ren_2PI}
\Gamma_{\rm
2PI}^R[\varphi_R,\cG_R]\equiv S_{0,R}[\varphi_R]+\frac{i}{2}\,\Str\,\Ln\,\cG_R^{-1}+\frac{i}{2}\,\Str\,\cG_{0,R}^{-1}\,\cG_R+\Gammaint^R[\varphi_R,\cG_R]\,,
\eeq
which defines $\Gammaint^R[\varphi_R,\cG_R]$, see below. The renormalized free action and propagator $S_{0,R}$ and ${\cal G}_{0,R}$ are obtained from $S_0$ and ${\cal G}_0$ after replacing the original bare parameters $m$ and $\lambda$ by the renormalized ones $m_R$ and $\lambda_R$. The decomposition (\ref{eq:ren_2PI}) allows one to easily obtain the equations for renormalized vertex functions from the corresponding equations for bare vertex functions by simply replacing bare superfields and (free) superpropagators by renormalized ones, and $\Gammaint[\varphi,\cG]$ by $\Gammaint^R[\varphi_R,\cG_R]$. In what follows, we shall refer to the equations at the end of Sec.~2.2, as equations for renormalized vertex functions, assuming implicitly that the relevant replacements have been made.

\subsection{Counterterms}\label{sec:counterterms}
Using Eqs.~(\ref{eq:2PI}), (\ref{eq:bar_ren_2PI}), (\ref{eq:sfield_redef}) and (\ref{eq:ren_2PI}), the functional $\Gammaint^R[\varphi_R,\cG_R]$ can be written, up to a constant contribution, as
\beq
\label{eq:deltaGammaint} \smash{\Gammaint^R[\varphi_R,\cG_R]=\Gammaint[\varphi_R,\cG_R;e_R]+\delta\Gammaint[\varphi_R,\cG_R]}
\eeq
where the first term on the RHS is obtained from $\Gammaint[\varphi,\cG;e]$ simply by replacing the bare superfield, superpropagator and coupling by their renormalized counterparts (we made the coupling-dependence explicit here) and the second term contains all explicit reference to the counterterms $\smash{\delta Z_i\equiv Z_i-1}$, $i=0,\dots,3$. It can be expressed as\footnote{We use the fact that $\smash{\Gammaint[\varphi,\cG;e]=\Gammaint[\varphi_R,\cG_R;Z_1e_R]}$.}
\beq\label{eq:dGammaint}
\delta\Gammaint[\varphi_R,\cG_R]\!=\!\frac{i}{2}\varphi_R^t \delta\cG_{0}^{-1}\varphi_R+\frac{i}{2}\Str\delta\cG_{0}^{-1}\cG_R+\Gammaint[\varphi_R,\cG_R;Z_1 e_R]-\Gammaint[\varphi_R,\cG_R;e_R],
\eeq
where $\smash{\delta\cG_{0}^{-1}\equiv Z\cG_{0}^{-1}-\cG_{0,R}^{-1}}$. By construction, two-point -- i.e. mass and field-strength -- counterterms $\smash{\delta m\equiv\delta Z_0 m_R}$, $\delta Z_2$ and $\delta Z_3$ (recall that the gauge-fixing parameter counterterm $\smash{\delta Z_4=0}$), only appear in the first two terms of the RHS of \Eqn{eq:dGammaint}; the last two terms contain the dependence on the coupling counterterm $\delta Z_1$.

As we did previously in \Eqn{eq:2PI_loops}, it is convenient to separate the classical (i.e. $\cG_R$-independent), one-loop (i.e.  linear in $\cG_R$), and higher-loop contributions in the counterterm contribution $\delta\Gammaint[\varphi_R,\cG_R]$:
\beq
\label{eq:dGdecomp}
\delta\Gammaint[\varphi_R,\cG_R]\equiv\delta S_{\rm int}[\varphi_R]+\delta\Gammaint^{\rm 1loop}[\varphi_R,\cG_R]+\delta\Gamma_2[\cG_R]\,,
\eeq
where, in terms of the superfield and superpropagator components,
\bea\label{eq:Sint}
\delta S_{\rm int}[\varphi_R] & = & \frac{1}{2}\int_x \delta Z_3\,A_R^{\mu}(x)\Big(g_{\mu\nu}\partial_x^2-\partial^x_\mu\partial^x_\nu\Big)A_R^\nu(x)\nn
&+&\int_x\bar\psi_R(x)\Big(i\delta Z_2\slashchar{\partial}_x-\delta
m-\delta Z_1e_R\,\slashchar{A}_R(x)\Big)\psi_R(x)\,,
\eea
\bea\label{eq:dGammaint1loop2}
\delta\Gamma_{\rm int}^{\rm 1loop}[\varphi_R,\cG_R]  &=&
\frac{1}{2}\int_x\delta Z_3\Big(g_{\mu\nu}\partial_x^2-\partial^x_\mu\partial^x_\nu\Big) G_R^{\mu\nu}(x,y)\Big|_{y=x}\nn
&-&\int_x\tr\Big(i\delta Z_2\slashchar{\partial}_x-\delta m-\delta Z_1e_R\,\slashchar{A}_R(x)\Big) D_R(x,y)\Big|_{y=x}\nn
&-&\int_x
\delta Z_1e_R\Big(\bar\psi_R(x)\slashchar{K}_R(x,x)+\bar{\slashchar{K}}_R(x,x)\psi_R(x)\Big)\,,
\eea
and
\beq\label{eq:dGamma2}
\delta\Gamma_2[\cG_R]=\Gamma_2[\cG_R;Z_1e_R]-\Gamma_2[\cG_R;e_R]\,.
\eeq
The functional $i\delta\Gamma_2[\cG_R]$ is the series of closed 2PI diagrams having two or more loops with lines corresponding to the components of $\cG_R$ and classical QED vertices with either the renormalized coupling $e_R$ or the coupling counterterm $\delta Z_1e_R$ and containing at least one vertex counterterm. It is $\varphi_R$-independent, just as $\Gamma_2[\cG_R]$.

At this stage, it is important to realize that, although the same counterterms $\delta Z_{0,\ldots,3}$ appear in $\delta S_{\rm int}[\varphi_R]$, $\delta\Gamma_{\rm int}^{\rm 1loop}[\varphi_R,\cG_R]$ and $\delta\Gamma_2[\cG_R]$, they play in fact very different roles.
For instance, the last term of \Eqn{eq:Sint} leads to a tree level contribution to the (2PI-resummed) three-point function $i\delta^3\Gamma_R/\delta A_R\delta\psi_R\delta\bar\psi_R$, see \Eqn{eq:2PIr34}, proportional to $\delta Z_1$, which, as we shall see later, cancels a global divergence in this function. Similarly, the last three terms of \Eqn{eq:dGammaint1loop2}
respectively lead to tree level contributions to the (2PI) three-point functions $\delta\bar\Sigma_R^{\bar\psi\psi}/\delta A_R$, $\delta\bar\Sigma_R^{A\psi}/\delta\bar\psi_R$ and $\delta\bar\Sigma_R^{\bar\psi A}/\delta\psi_R$, see Eqs.~\eqn{eq:2PI3_bpap}-\eqn{eq:2PI3_bppa}, again proportional to $\delta Z_1$, which absorb global divergences in these functions. Of course, in the exact theory, these four functions coincide and a single coupling counterterm is sufficient to absorb their global divergences. However, at any finite approximation order, these functions differ and one needs to allow for a priori different coupling counterterms in the various contributions to $\delta S_{\rm int}[\varphi_R]$, $\delta\Gamma_{\rm int}^{\rm 1loop}[\varphi_R,\cG_R]$ and $\delta\Gamma_2[\cG_R]$. 

This is in fact a general feature of 2PI renormalization theory \cite{Berges:2005hc}: since 2PI and 2PI-resummed vertex functions do not coincide in general at finite approximation order, their renormalization calls for introducing different counterterms.\footnote{These are actually different approximations to the various QED counterterms $\delta Z_{0,\ldots,4}$. For systematic approximation schemes though, 2PI-resummed and 2PI vertex functions only differ beyond the accuracy of the approximation and so do the corresponding counterterms, provided one imposes suitable renormalization conditions \cite{Berges:2005hc,Reinosa:2007vi}, see Sec.~\ref{sec:rencond}.} Previous analysis of (renormalizable) scalar and fermionic field theories \cite{VanHees:2001pf,vanHees:2002bv,Berges:2004hn,Cooper:2004rs,Berges:2005hc,Arrizabalaga:2006hj} teach us that one should actually include all local counterterms of mass dimension less or equal to four\footnote{More precisely, one should include in the 2PI diagrammatic expansion all diagrams made of lines $\cG_R$ and local counterterms vertices having a number of legs $n$ such that $\sum_{i=1}^n (d_i+k_i)\le d=4$, where the sum runs over the legs of the vertex and $d_i$ is the canonical mass dimension of the field operator corresponding to the $i$-th leg ($d_i=(d-2)/2$ for scalars or gauge fields, $d_i=(d-1)/2$ for spin 1/2 fermions fields etc) and $k_i$ accounts for possible derivative couplings.} consistent with the symmetries of the theory \cite{Berges:2005hc}. The aim of the present paper is to show that this generalizes to abelian gauge theories.

It is important to notice that, in the present context, symmetry constraints may allow for new type of counterterms, which have no analog in standard renormalization theory, simply because one has, in general, more possibilities to construct symmetry invariants with both $\varphi_R$ and $\cG_R$ than with $\varphi_R$ alone \cite{Reinosa:2006cm}. This is the case for gauge theories where, because the gauge transformation of the field $\varphi_R$ involves an affine contribution, the gauge transformation of the bilinear $\varphi_R\varphi_R$ differs from that of the connected correlator $\cG_R$ \cite{Reinosa:2007vi}. The most general counterterm contribution $\delta\Gamma_{\rm int}[\varphi_R,\cG_R]$ consistent with the above requirements is given by \Eqn{eq:dGdecomp}, with 
\bea\label{eq:Sintapp}
\delta S_{\rm int}[\varphi_R] & = & \frac{1}{2}\int_x \delta Z_3\,A_R^{\mu}(x)\Big(g_{\mu\nu}\partial_x^2-\partial^x_\mu\partial^x_\nu\Big)A_R^\nu(x)\nn
&+&\int_x\bar\psi_R(x)\Big(i\delta Z_2\slashchar{\partial}_x-\delta
m-\delta Z_1e_R\,\slashchar{A}_R(x)\Big)\psi_R(x)\,,
\eea
\bea\label{eq:dGammaint1loop2app}
\delta\Gamma_{\rm int}^{\rm 1loop}[\varphi_R,\cG_R] \!\!\! &=& \!\!\!
\frac{1}{2}\int_x\!\Big(\delta\bar Z_3\!\left(g_{\mu\nu}\partial_x^2-\partial^x_\mu\partial^x_\nu\right)\!+\!\delta\bar M^2\!g_{\mu\nu}\!\!+\!\delta\bar\lambda\,\partial^x_\mu\partial^x_\nu\Big) G_R^{\mu\nu}(x,y)\Big|_{y=x}\nn
&-&\int_x\tr\Big(i\delta \bar Z_2\slashchar{\partial}_x-\delta\bar m-\delta \bar Z_1e_R\,\slashchar{A}_R(x)\Big) D_R(x,y)\Big|_{y=x}\nn
&-&\int_x
\delta\tilde Z_1e_R\Big(\bar\psi_R(x)\slashchar{K}_R(x,x)+\bar{\slashchar{K}}_R(x,x)\psi_R(x)\Big)\,,
\eea
and 
\beq\label{eq:dGamma2app}
\delta\Gamma_2[\cG_R]=\!\int_x \!\left(\frac{\delta\bar g_1}{8}G_{R\,\mu}^\mu(x,x)G_{R\nu}^\nu(x,x)\!+\!\frac{\delta\bar
g_2}{4}G_R^{\mu\nu}(x,x)G^R_{\mu\nu}(x,x)\right)+\delta\Gamma_2^{\rm BPHZ}[\cG_R].
\eeq

In the classical contribution $\delta S_{\rm int}[\varphi_R]$, terms $A_R^2$, $(\partial A_R)^2$, $(\partial A_R) A_R^2$ and $A_R^4$, which would be allowed by power counting are forbidden by gauge symmetry. Similarly, gauge invariance forbids one-loop terms of the form $(\partial A_R) G_R$ and $A_R^2 G_R$. In Secs.~4 and 5, we shall actually prove that, in a given truncation of the 2PI effective action, no diagrams generate divergences which would call for introducing such counterterm contributions. In contrast, since the gauge transformation of the photon connected correlator $G_R$ is trivial (see \cite{Reinosa:2007vi} and below), one-loop terms of the form $G_R$ or $\partial\partial G_R$ are allowed, giving rise to mass-like and gauge-fixing-parameter--like counterterms $\delta\bar M^2$ and $\delta\bar\lambda$. The same is true for two-loop terms $\sim G_R^2$ in  $\delta\Gamma_2[\cG_R]$, which give rise to four-photon--like counterterms $\delta\bar g_1$ and $\delta\bar g_2$ \cite{Reinosa:2006cm}. 
As discussed in detail in \Refer{Reinosa:2006cm}, see also Sec.~\ref{subsec:renormalization_2PI_vertices_2} below, these new counterterms actually absorb (sub)divergences in the longitudinal (in momentum space) part of the inverse photon correlator $\bar G_R^{-1}$. We stress that such divergences are a pure artifact of the (2PI) approximation. For systematic approximation schemes, they -- and the corresponding counterterms -- are systematically of higher order as compared to the order of approximation \cite{Reinosa:2006cm}. 

As for QED coupling counterterms in Eqs.~\eqn{eq:Sintapp}-\eqn{eq:dGamma2app}, charge-conjugation invariance implies that the terms $\bar\psi_R \slashchar{K}_R$ and $\bar{\slashchar{K}}_R\psi_R$  come with the same coefficient $\delta \tilde Z_1 e_R$. However, unlike what happens in the exact theory, these terms, which are actually gauge-symmetric (see below), are not related to the third coupling counterterm contribution $\slashchar{A}_R D_R$ and, consequently, the latter comes with an independent coefficient $\delta \bar Z_1 e_R$. We show in the next section that gauge symmetry further implies that the counterterms $\delta Z_1$, $\delta Z_2$, $\delta \bar Z_1$ and $\delta\bar Z_2$ are not independent.

Finally, we mention that the term $\delta\Gamma_2^{\rm BPHZ}[\cG_R]$ in \Eqn{eq:dGamma2app} only starts at three-loop order. It involves QED-coupling counterterms and, beyond five-loop order, (gauge-symmetric) four-photon counterterms. Its construction has been discussed in detail in \Refer{Reinosa:2006cm}. We shall not discuss it any further here.\\


\subsection{Renormalized 2PI Ward-Takahashi identities}\label{sec:WTI}

The issue of gauge symmetries in the 2PI formalism has been discussed in \cite{Arrizabalaga:2002hn,Calzetta:2004sh,Reinosa:2007vi}. In \Refer{Reinosa:2007vi} the generalization of Ward-Takahashi identities for the bare 2PI effective action were derived and explicitly solved for abelian theories. In particular, it was shown that, for linear gauges, the bare functional 
\begin{equation}\label{eq:GammaBaresym}
\Gamma_{\rm sym}[\varphi,\cG]\equiv\Gamma_{\rm 2PI}[\varphi,\cG]-S_{\rm gf}[\varphi]\,,
\end{equation} 
with $S_{\rm gf}[\varphi]\equiv-(\lambda/2)\int_x [\partial_\mu A^\mu(x)]^2$ the classical gauge-fixing term, is invariant:
\beq\label{eq:GammaBareinv}
 \Gamma_{\rm sym}[\varphi,\cG]=\Gamma_{\rm sym}[\varphi^{(\alpha)},\cG^{(\alpha)}]
\eeq
under the gauge transformation of the bare fields:
\beq
\label{eq:field_tr_bare}
\psi^{(\alpha)}(x)= e^{i\alpha(x)}\psi(x)\,,\,\,\bar\psi^{(\alpha)}(x)= e^{-i\alpha(x)}\bar\psi(x)\,,\,\, A^{(\alpha)}_\mu(x)= A_\mu(x)-\frac{1}{e}\partial_\mu\alpha(x)
\eeq
and bare superpropagator:
\beq
\label{eq:prop_tr_bare}
\cG^{(\alpha)}_{mn}(x,y) = e^{iq_m\alpha(x)}\cG_{mn}(x,y)\,e^{iq_n\alpha(y)}\,,
\eeq
where $\alpha(x)$ is an arbitrary real function. From \Eqn{eq:GammaBaresym}, one derives symmetry identities for the bare (2PI, 2PI-resummed, etc) vertex functions of the theory, which generalize the standard Ward-Takahashi identities. \Refer{Reinosa:2007vi} shows that these 2PI Ward-Takahashi identities are exactly satisfied at any order of approximation in the 2PI loop-expansion of QED.

Let us now see how Eqs.~\eqn{eq:GammaBaresym}-\eqn{eq:prop_tr_bare} are modified by renormalization at finite approximation order, i.e. by the inclusion of the counterterm contribution $\delta\Gamma_{\rm int}[\varphi_R,\cG_R]$, Eqs.~\eqn{eq:Sintapp}-\eqn{eq:dGamma2app}. Defining $\smash{\Delta_i\equiv\bar Z_i-Z_i=\delta\bar Z_i-\delta Z_i}$, for $i=1,2$, it is an easy exercise to check that the functional\footnote{The functional $\Gamma_{\rm sym}^R[\varphi_R,\cG_R]$ is obtained from $\Gamma_{\rm 2PI}^R[\varphi_R,\cG_R]$ by subtracting the renormalized classical gauge fixing action and replacing $\delta Z_{1,2}\to\delta\bar Z_{1,2}$ in \Eqn{eq:Sintapp}.}
\beq\label{eq:GammaRsym}
\Gamma_{\rm sym}^R[\varphi_R,\cG_R]\equiv\Gamma_{\rm 2PI}^R[\varphi_R,\cG_R]-S^R_{\rm gf}[\varphi_R]+\int_x\bar\psi_R(x)\Big(i\Delta_2\slashchar{\partial}_x-\Delta_1 e_R\,\slashchar{A}_R(x)\Big)\psi_R(x)\,,
\eeq
where $S^R_{\rm gf}[\varphi_R]\equiv-(\lambda_R/2)\int_x [\partial_\mu A^\mu_R(x)]^2$, is invariant:
\beq\label{eq:GammaRinv}
 \Gamma^R_{\rm sym}[\varphi_R,\cG_R]=\Gamma^R_{\rm sym}[\varphi^{(\alpha)}_R,\cG^{(\alpha)}_R]
\eeq
under the following gauge transformations of the renormalized fields:
\beq
\label{eq:field_tr}
\psi^{(\alpha)}_R(x)=e^{i\alpha(x)}\psi_R(x)\,,\,\bar\psi^{(\alpha)}_R(x)=e^{-i\alpha(x)}\bar\psi_R(x)\,,\, A^{(\alpha)}_{R\mu}(x)=A_{R\mu}(x)-\frac{1}{\tilde e}\partial_\mu\alpha(x)
\eeq
and renormalized superpropagator:
\beq
\label{eq:prop_tr}
\cG^{(\alpha)}_{R,mn}(x,y)=e^{iq_m\alpha(x)}\,\cG_{R,mn}(x,y)\,e^{iq_n\alpha(y)}\,,
\eeq
where $\tilde e/e_R=\bar Z_1/\bar Z_2$ and  $\alpha(x)$ is an arbitrary real function. This uses the fact that $\delta\Gamma_2[\cG_R]$ is gauge-invariant since it only depends on the components of $\cG_R$, whose gauge transformations \eqn{eq:prop_tr} are purely linear: in the diagrammatic loop expansion, one easily sees that the local phase factors associated with the fermionic end of each propagator line cancel each other at the QED vertex they are attached to\footnote{Diagrams containing possible four-photon vertex counterterms do not pose any problem since only photon lines are attached to these vertices and the photon propagator is invariant under the gauge transformation \eqn{eq:prop_tr} \cite{Reinosa:2006cm}.} \cite{Reinosa:2007vi}. 
It also follows that the functional $\Gamma^R_{\rm sym}[\varphi_R]\equiv\Gamma_{\rm sym}^R[\varphi_R,\bcG_R[\varphi_R]]$, that is,
\beq\label{eq:ren2piresumsym}
 \Gamma_{\rm sym}^R[\varphi_R]\equiv\Gamma_R[\varphi_R]-S^R_{\rm gf}[\varphi_R]+\int_x\bar\psi_R(x)\left[i\Delta_2\slashchar{\partial}_x-\Delta_1 e_R\,\slashchar{A}_R(x)\right]\psi_R(x)\,,
\eeq
where $\smash{\Gamma_R[\varphi_R]\equiv\Gamma_{\rm 2PI}^R[\varphi_R,\bcG_R[\varphi_R]]}$ is the renormalized 2PI-resummed effective action, is symmetric:
\beq\label{eq:ren2piresumsym2}
 \Gamma_{\rm sym}^R[\varphi_R]=\Gamma_{\rm sym}^R[\varphi_R^{(\alpha)}]\,.
\eeq 

The gauge transformation of the renormalized gauge field, see Eq.~(\ref{eq:field_tr}), involves the ratio $\bar Z_2/\bar Z_1$, which must, therefore, be finite. Using 2PI Ward-Takahashi identities for renormalized vertices, we show below that this is indeed the case. Moreover, the symmetry property \eqn{eq:GammaBareinv} of the bare 2PI effective action is modified by a term involving (a priori infinite) counterterms, \Eqn{eq:GammaRsym}. The variation of this term under the gauge transformation \eqn{eq:field_tr} can be written as
\begin{equation}\label{eq:gaugevarrr}
\delta_\alpha\!\!\int_x\!\bar\psi_R(x)\Big(i\Delta_2\slashchar{\partial}_x-\Delta_1 e_R\,\slashchar{A}_R(x)\Big)\psi_R(x)=i\!\left(\Delta_2-\Delta_1\frac{\bar Z_2}{\bar Z_1}\right)\!\delta_\alpha\!\!\int_x\!\bar\psi_R(x)\slashchar{\partial}_x\psi_R(x).
\end{equation}
where, for a given functional $F$, $\delta_\alpha F[\varphi_R,\cG_R]\equiv F[\varphi_R^{(\alpha)},\cG_R^{(\alpha)}]-F[\varphi_R,\cG_R]$.
As we discuss below, this only slightly modifies the (renormalized) 2PI Ward-Takahashi identities. Using these modified identities, one can show that the combination of counterterms on the RHS of \Eqn{eq:gaugevarrr} is actually finite and can be set to zero by an appropriate choice of renormalization conditions. This ensures that gauge symmetry (i.e. 2PI Ward-Takahashi identities) is exactly preserved by the renormalization procedure at any approximation order.\footnote{Note that this alone does not guarantee gauge-invariance in the sense of gauge-fixing independence.} In particular: 
\beq
 \delta_\alpha\!\Big(\Gamma^R_{\rm 2PI}[\varphi_R,\cG_R]-S^R_{\rm gf}[\varphi_R]\Big)=0\,,
\eeq
and similarly for the 2PI-resummed effective action 
\beq
 \delta_\alpha\!\Big(\Gamma_R[\varphi_R]-S^R_{\rm gf}[\varphi_R]\Big)=0\,.
\eeq 

To prove this, it is convenient (but not necessary) to switch to 4-momentum space\footnote{To discuss renormalization, it is sufficient to assume space-time translation invariance.}. For later convenience, we introduce the following notations (Dirac and Lorentz indices implicit) for two- and three-point vertex functions in momentum space:
\beq\label{eq:mom1}
 \bar D_R\to\bar D_R(p) \,,\,\, \bar G_R\to \bar G_R(p)\,,\,\, \frac{\delta^{2}\Gamma_R}{\delta\psi_R\delta\bar\psi_R}\to\Gamma_{R}^{(2,0)}(p)\,,\,\,\frac{\delta^{2}\Gamma_R}{\delta A_R\delta A_R}\to\Gamma_{R}^{(0,2)}(p)
\eeq
and
\beq\label{eq:mom3}
\frac{\delta\bSigma_R^{\bar\psi\psi}}{\delta A_R^\mu}\to iV_{R\,\mu}^{(2,1)}(p',p)\,,\,\,
\frac{\delta\bSigma_{R\,\mu}^{\bar\psi A}}{\delta \psi_R}\to i\tilde V_{R\,\mu}^{(2,1)}(p',p)\,,\,\,
 \frac{\delta^{3}\Gamma_R}{\delta A_R^\mu\delta\psi_R\delta\bar\psi_R}\to\Gamma_{R\,\mu}^{(2,1)}(p',p) \,.
\eeq
For three-point functions, we use the convention that the first and second momenta (here $p'$ and $p$) are associated with the outgoing and ingoing fermion legs respectively. 
Ward-Takahashi identities for the renormalized 2PI vertex functions are obtained by noticing that, since the last term of \Eqn{eq:GammaRsym} is independent of $\cG_R$, one has
\beq
 \frac{\delta\Gamma_{\rm 2PI}^R[\varphi_R,\cG_R]}{\delta\cG_R}=\frac{\delta\Gamma_{\rm sym}^R[\varphi_R,\cG_R]}{\delta\cG_R}
\eeq
which implies that the function $\bcG_R[\varphi_R]$, defined as the extremum of the renormalized 2PI effective action $\Gamma_{\rm 2PI}^R[\varphi_R,\cG_R]$ for given $\varphi_R$, is actually the extremum of a gauge-invariant functional. It follows that it transforms covariantly under a gauge transformation of its argument \cite{Reinosa:2007vi}: 
\beq\label{eq:Rcov}
 \bcG_R[\varphi_R^{(\alpha)}]=\bcG_R^{(\alpha)}[\varphi_R]\,.
\eeq
This generalizes a result derived in \Refer{Reinosa:2007vi} for the bare supercorrelator. In particular, this implies that the symmetry identities for bare 2PI vertices are not modified by the presence of counterterms and can be directly transposed to renormalized 2PI vertices. 
Of particular interest for our present purposes is the identity relating the renormalized three-point vertex $\delta\bSigma^{\bar\psi\psi}_R/\delta A_R\equiv iV_R^{(2,1)}$ to the inverse fermion correlator $D_R^{-1}$ \cite{Reinosa:2007vi}:
\beq
\label{eq:2PIWIFourier1}
 -\frac{\bar Z_2}{\bar Z_1}\frac{1}{e_R}k^\mu \,V_{R\,\mu}^{(2,1)}(p+k,p)=i\bar D_R^{-1}(p+k)-i\bar D_R^{-1}(p)\,.
\eeq
Another important consequence of Eq.~(\ref{eq:Rcov}), which we shall use extensively in the next sections to show how UV divergences are constrained by gauge symmetry, is the fact that the 2PI four-photon vertex $\delta^2\bSigma^{AA}_R/\delta A_R\delta A_R$ is transverse with respect to the two external legs corresponding to field derivatives: One has, schematically, 
\beq\label{eq:tata}
k^\rho\frac{\delta^2\bSigma_{R\,\mu\nu}^{AA}}{\delta A^\rho_R\delta A^\sigma_R}=0\,.
\eeq

Similarly, one can deduce from Eqs.~(\ref{eq:ren2piresumsym}) and (\ref{eq:ren2piresumsym2}) symmetry identities for the renormalized 2PI-resummed vertex functions \cite{Reinosa:2007vi}. It is easy to see that, since the last term in the RHS of \Eqn{eq:ren2piresumsym} is at most trilinear in the fields, the only modification it brings as compared to the usual Ward-Takahashi identities concerns the relation between the three-point function and the fermion two-point function. We find:
\beq
\label{eq:2PIWIFourier2}
 -\frac{\bar Z_2}{\bar Z_1}\frac{1}{e_R}k^\mu\Gamma_{R\,\mu}^{(2,1)}(p+k,p)=\Gamma_R^{(2,0)}(p+k)-\Gamma_R^{(2,0)}(p)-\slashchar{k}\Big(\Delta_2-\Delta_1\frac{\bar Z_2}{\bar Z_1}\Big)\,.
\eeq
Any other 2PI-resummed vertex function fulfills the usual Ward-Takahashi identities. In particular, the 2PI-resummed photon polarization tensor, defined through $\Gamma^{(0,2)}_R=iG_{0,R}^{-1}-i\Pi_R$, is transverse:
\beq
\label{eq:transmom}
 k_\mu\Pi_R^{\mu\nu}(k)=0\,.
\eeq

We show in Secs.~\ref{sec:renormalization_2PI_vertices} and \ref{sec:renormalization_2PI_resummed_vertices} that all renormalized vertex functions can actually be made finite by a proper choice of the counterterms introduced above. It then follows from \Eqn{eq:2PIWIFourier1} that the ratio $\bar Z_2/\bar Z_1$ is finite and that one can set 
\beq\label{eq:equalWard}
 \bar Z_1=\bar Z_2
\eeq
by a suitable definition of the renormalized charge $e_R$. This generalizes the usual textbook result in the 1PI formalism. Similarly, Eq.~(\ref{eq:2PIWIFourier2}) proves that the combination $\smash{\Delta_2-\Delta_1(\bar Z_2/\bar Z_1)}$ is finite and can, therefore, be set to zero by adjusting the finite part of some counterterm, that is by a suitable (i.e. gauge-symmetric) choice of renormalization conditions as discussed below. This implies that 
\beq\label{eq:cafaitzero}
 \frac{Z_1}{Z_2}=\frac{\bar Z_1}{\bar Z_2}\,.
\eeq
As already mentioned, this guarantees that the renormalized 2PI effective action satisfies the (2PI) Ward-Takahashi identities. We stress again that this is true at any approximation order in the 2PI loop expansion.

\subsection{Renormalization conditions and gauge symmetry}
\label{sec:rencond}

The QED counterterms $\delta Z_0,\dots,\delta Z_3$ are fixed by imposing four independent renormalization conditions for the (2PI-resummed) vertex functions of the theory with non-negative superficial degree of divergence, namely the two- and three-point functions $\Gamma^{(0,2)}_R$, $\Gamma^{(2,0)}_R$ and $\Gamma^{(2,1)}_R$. The Ward-Takahashi identity \eqn{eq:transmom} guarantees that the longitudinal part of the photon two-point function $\Gamma^{(0,2)}_R$ is not modified by loop corrections, which implies that $\delta Z_4=0$. The Lorentz decomposition of the 2PI-resummed photon polarization tensor $\Pi_R$ reads
\beq\label{eq:lolorentz}
 \Pi_{R\,\mu\nu}(k)=\Pi_R^T(k^2)P^T_{\mu\nu}(k)\,,
\eeq
with the usual projectors $P^L_{\mu\nu}(k)= k_\mu k_\nu/k^2$ and $P^T_{\mu\nu}(k)+P^L_{\mu\nu}(k)=g_{\mu\nu}$. Because it is one-particle-irreducible, $\Pi_{R\,\mu\nu}(k)$ is expected not to have a pole at $k^2=0$, which implies that $\Pi_R^T(0)=0$ (i.e. the photon is massless). The global divergence of $\Pi_R^T$ is absorbed by the photon field-strength counterterm $\delta Z_3$ in \Eqn{eq:Sintapp} which can be fixed by requiring that, at a given renormalization point $k=k_1^*$,
\beq\label{eq:2PIresrenG}
\left.\frac{d\Pi_{R}^T(k^2)}{dk^2}\right|_{k=k^*_1}=0\,.
\eeq
Similarly, the mass and fermion field strength counterterms $\delta Z_0 =\delta m/m_R$ and $\delta Z_2$ are fixed by imposing renormalization conditions on the fermion two-point function or, equivalently, on the (2PI-resummed) fermion self-energy $\Sigma_R$, defined through $\Gamma^{(2,0)}_R=iD_{0,R}^{-1}-i\Sigma_R$. The latter has the following structure in Dirac space:
\beq\label{eq:structstruct}
 \Sigma_{R}(p)=\sigma_0(p^2)\mathds{1}+ \sigma_1(p^2)\slashchar{p}\,,
\eeq
where $\sigma_0$ and $\sigma_1$ are globally divergent real functions, see \App{appsec:lorstruct}. The mass counterterm $\delta m$ absorbs the global divergence of $\sigma_0(p^2)$ and can be fixed by demanding that, at a given renormalization point $p=p_1^*$,
\beq\label{eq:ren_ferm}
 {\rm tr}\Big[\Sigma_R(p)\Big]_{p=p^*_1}=0\,,
\eeq
i.e. $\sigma_0(p_1^{*2})=0$. The global divergence of $\sigma_1(p^2)$ is absorbed in the fermion field strength counterterm $\delta Z_2$. To fix the latter, it proves convenient for later use to consider, instead of the standard renormalization condition involving the momentum derivative of $\Sigma_R(p)$, the finite difference at the renormalization point $p=p_2^*$, $k=k_2^*$:
\beq
\label{eq:rendiff2}
 \Big\{\Sigma_R(p+k)-\Sigma_R(p)\Big\}_{(p^*_2,k^*_2)}=0\,,
\eeq
where we introduced the following notation:
\beq\label{eq:crochet}
 \Big\{F(p,k)\Big\}_{(p^*,k^*)}\equiv\frac{1}{4}{\rm tr}\left[\frac{\slashchar{k}}{k^2}\,F(p,k)\right]_{p=p^*,k=k^*}\,.
\eeq
Notice, in particular, that $\{\slashchar{k}\}_{(p,k)}=1$. Here the Dirac trace ${\rm tr}[\gamma_\mu\Sigma_R]$ isolates the vector component  $\sigma_1$ of $\Sigma_R$ and the contraction with $k^\mu$ is chosen such that, for on-shell fermion momenta $(p_2^*+k_2^*)^2=p_2^{*2}$,
\Eqn{eq:rendiff2} reduces to the simpler condition $\sigma_1(p_2^{*2})=0$.

Finally, the coupling counterterm $\delta Z_1$ absorbs the global divergence of the 2PI-resummed three-point vertex $\Gamma^{(2,1)}_{R,\mu}$. The latter admits the following Dirac decomposition, see \App{appsec:lorstruct}:
\beq\label{eq:lorstructvertex}
 \Gamma^{(2,1)}_{R,\mu}(p',p)=V^S_\mu(p',p)\mathds{1}+V^V_{\mu,\nu}(p',p)\gamma^\nu+iV^A_{\mu,\nu}(p',p)\gamma_5\gamma^\nu+\frac{i}{2}V^T_{\mu,\nu\rho}(p',p)\sigma^{\nu\rho}\,,
\eeq
where $V^S,\dots,V^T$ are real functions, among which only the vector component $V^V$ presents a global divergence, see \App{appsec:lorstruct}. To isolate the latter, we impose the following condition at a renormalization point $p=p_3^*$, $k=k_3^*$
\beq\label{eq:ren_coup}
 \left\{k^\mu\Gamma^{(2,1)}_{R,\mu}(p+k,p)\right\}_{(p^*_3,k^*_3)}=-e_R\,.
\eeq
Here the contraction with $k^\mu$ is introduced in prevision of a later use of the symmetry identity \eqn{eq:2PIWIFourier2}. 

Equations \eqn{eq:2PIresrenG}, \eqn{eq:ren_ferm}, \eqn{eq:rendiff2} and \eqn{eq:ren_coup} define the theory. However, at any finite approximation order in the 2PI expansion, these are not sufficient to fix all the counterterms in Eqs.~\eqn{eq:Sintapp}-\eqn{eq:dGamma2app}. 
The remaining counterterms must be fixed without introducing neither extra renormalization condition nor new parameters other than the two QED parameters $m_R$ and $e_R$, and the gauge-fixing parameter $\lambda_R$. To this aim, recall that the occurrence of extra counterterms in the 2PI formalism originates from the various possible definitions of vertex functions, which need a priori independent renormalization at finite approximation order. However, renormalization conditions are an aspect of the exact theory, where e.g. experimental input is introduced, and, as such, should not differ for various definitions of vertex functions. It is thus natural to require the consistent condition that various (e.g. 2PI and 2PI-resummed) vertex functions with non-negative superficial degree of divergence agree at the renormalization point. This is a standard feature of 2PI renormalization theory, see \Refer{Berges:2005hc} for the case of scalar field theories.

The only independent, globally divergent 2PI vertex functions are the photon and fermion self-energies $\bSigma_R^{AA}\equiv\bPi_R$ and $\bSigma_R^{\bar\psi\psi}\equiv\bSigma_R$ and the three-point vertices $\delta\bSigma_R^{\bar\psi\psi}/\delta A_R=iV^{(2,1)}_R$ and $\delta\bSigma_R^{\bar\psi A}/\delta \psi_R=i\tilde V^{(2,1)}_R$. The photon self-energy admits the following Lorentz decomposition in momentum space:
\beq
\label{eq:projectors} \bSigma_{R\,\mu\nu}^{AA}(k)\equiv\bar\Pi_{R\,\mu\nu}(k)=\bar\Pi_R^T(k^2)P^T_{\mu\nu}(k)+\bar\Pi_R^L(k^2)P^L_{\mu\nu}(k)\,,
\eeq
where both transverse and longitudinal components $\bar\Pi_R^T(k^2)$ and $\bar\Pi_R^L(k^2)$ are UV-divergent \cite{Reinosa:2006cm}, see \App{appsec:lorstruct}. In particular, their global divergences are absorbed in the counterterms $\delta\bar Z_3$, $\delta\bar M^2$ and $\delta\bar\lambda$ in \Eqn{eq:dGammaint1loop2app}. To fix the latter we impose the consistency conditions \cite{Reinosa:2006cm}
\beq\label{eq:renG1}
 \left.\frac{d\bar\Pi_{R}^T(k^2)}{dk^2}\right|_{k=k^*_1}=\left.\frac{d\Pi_R^T(k^2)}{dk^2}\right|_{k=k^*_1}\,,
\eeq
and, using the fact that the longitudinal part of the 2PI-resummed photon polarization tensor vanishes identically,  $\Pi_R^L(k^2)=P^L_{\mu\nu}\Pi_R^{\mu\nu}(k^2)=0$, 
\beq\label{eq:renG2}
 \bar\Pi_{R}^L(k^2)|_{k=k^*_1}=\Pi_R^L(k^2)|_{k=k^*_1}=0\quad , \quad\left.\frac{d\bar\Pi_{R}^L(k^2)}{dk^2}\right|_{k=k^*_1}=\left.\frac{d\Pi_{R}^L(k^2)}{dk^2}\right|_{k=k^*_1}=0\,.
\eeq

A related aspect is the presence of divergent four-photon subdiagrams in $\bar\Pi_R$ \cite{Reinosa:2006cm}. The latter actually add up to the complete four-photon vertex function $\bar V_{R\,\mu\nu,\rho\sigma}(p,q,k)$ obtained from the 2PI kernel $\delta^2\Gammaint^R/\delta \cG_R\delta \cG_R|_{\bcG_R}$, see \Eqn{eq:Vmnrs} below. In the exact theory, the latter agrees with the (2PI-resummed) four-photon function $\Gamma^{(0,4)}_{R\,\mu\nu,\rho\sigma}(p,q,k)$ which is transverse and thus finite \cite{Reinosa:2006cm}. In contrast, at finite approximation order, the function $\bar V_{R\,\mu\nu,\rho\sigma}(p,q,k)$ is not exactly transverse -- although its longitudinal part is systematically of higher order as compared to the order of approximation -- and requires renormalization. This is the role of the four-photon counterterms $\delta \bar g_1$ and $\delta \bar g_2$ in \Eqn{eq:dGamma2app}, see \Refer{Reinosa:2006cm} for details. Using the fact that the 2PI-resummed four-photon function is transverse, the latter can be fixed by means of the consistency conditions at a given renormalization point $k=k^*$:
\beq\label{eq:renG3}
k_*^\mu\,k_*^\nu\,\bar V_{R\,\mu\nu,\rho\sigma}(k^*,k^*,k^*)=k_*^\mu\,k_*^\nu\,\Gamma^{(0,4)}_{R\,\mu\nu\rho\sigma}(k^*,k^*,k^*)=0\,.
\eeq
As shown in \cite{Reinosa:2006cm}, conditions \eqn{eq:renG1}-\eqn{eq:renG3} ensure that the difference $\delta\bar Z_3-\delta Z_3$ and the counterterms $\delta\bar M^2$, $\delta\bar\lambda$, $\delta\bar g_1$ and $\delta\bar g_2$ are systematically of higher order as compared to the order of approximation. 

The counterterms $\delta\bar Z_0=\delta\bar m/ m_R$ and $\delta\bar Z_2$ are fixed by the consistency conditions
\beq
\label{eq:renfermionmass}
{\rm tr}\Big[\bar\Sigma_R(p)\Big]_{p=p_1^*}={\rm tr}\Big[\Sigma_R(p)\Big]_{p=p_1^*}
\eeq
and
\beq
\label{eq:rendiff}
\Big\{\bar\Sigma(p+k)-\bar\Sigma(p)\Big\}_{(p^*_2,k^*_2)}=\Big\{\Sigma_R(p+k)-\Sigma_R(p)\Big\}_{(p^*_2,k^*_2)}.
\eeq
Finally, the coupling counterterms $\delta\bar Z_1$ and $\delta\tilde Z_1$ are fixed by demanding that the vector parts of their respective Lorentz decomposition, see Eq.~\eqn{eq:lorstructvertex}, agree at the renormalization point ($p'=p+k$): 
\beq\label{eq:renvertices}
\left\{k^\mu V^{(2,1)}_{R\,\mu}(p',p)\right\}_{(p^*_3,k^*_3)}=\left\{k^\mu\tilde V^{(2,1)}_{R\,\mu}(p',p)\right\}_{(p^*_3,k^*_3)}=\left\{k^\mu\Gamma_{R\,\mu}^{(2,1)}(p',p)\right\}_{(p^*_3,k^*_3)}\!.
\eeq
Again, these conditions guarantee that the differences $\delta\bar Z_i-\delta Z_i$, for $i=0\ldots 2$ and $\delta\tilde Z_1-\delta Z_1$ are systematically of higher order as compared to the order of approximation, just as $\delta\bar Z_3-\delta Z_3$. It is clear that this is true for arbitrary sets of renormalization conditions, provided one imposes the corresponding consistency conditions.

It follows from the previous considerations that for an arbitrary choice of renormalization conditions and corresponding consistency conditions, the combination of counterterms $\bar Z_1\Delta_2-\bar Z_2\Delta_1$ appearing in \Eqn{eq:2PIWIFourier2} is systematically of higher order than the approximation order and, therefore, that the standard Ward-Takahashi identity relating the renormalized 2PI-resummed two- and three-point functions $\Gamma_R^{(2,0)}$ and $\Gamma_R^{(2,1)}$ holds up to higher order corrections.
Moreover, as discussed previously, it is always possible to adjust the counterterms (i.e. choose renormalization conditions) in such a way that  the combination $\bar Z_1\Delta_2-\bar Z_2\Delta_1$ vanishes exactly at any approximation order. We now show that our choice of renormalization conditions \eqn{eq:rendiff2} and \eqn{eq:ren_coup} and corresponding consistency conditions \eqn{eq:rendiff} and \eqn{eq:renvertices} has this property. Indeed, combining the 2PI Ward-Takahashi identities \eqn{eq:2PIWIFourier1} and \eqn{eq:2PIWIFourier2} with \Eqn{eq:rendiff}, one gets
\beq
 \Delta_2-\Delta_1\frac{\bar Z_2}{\bar Z_1}= \frac{\bar Z_2}{\bar Z_1}\frac{1}{e_R}\left\{{k}^\mu \Big(\Gamma_{R\,\mu}^{(2,1)}(p+k,p)-V_{R\,\mu}^{(2,1)}(p+k,p)\Big)\right\}_{(p^*_2,k^*_2)}\!.
\eeq
The RHS of this equation vanishes provided one chooses $\smash{p^*_2=p^*_3}$ and $\smash{k^*_2=k^*_3}$. Together with Eq.~(\ref{eq:2PIWIFourier1}), this implies that
\begin{equation}\label{eq:Z1Z2}
\frac{Z_2}{Z_1}=\frac{\bar Z_2}{\bar Z_1}=1\,,
\end{equation}
as announced earlier. This guarantees that the gauge-symmetry \eqn{eq:GammaBareinv} of the bare 2PI effective action is exactly preserved after renormalization. Of course, this can be achieved by other choices of renormalization conditions as well. For instance, standard on-shell renormalization conditions also have this property, see \App{appsec:Physicalon-mass-shellrenormalizationconditions}.

\subsection{The renormalization procedure: executive summary}\label{sec:summary}

We end this section by describing how to implement 2PI renormalization in practice. The complete renormalization procedure involves three independent steps, each one involving the determination of a subclass of counterterms independent of the others. Depending on the particular application, one may need to go only through the first and/or second steps. Most existing applications of 2PI techniques actually only involve the first step. 
The three steps are:
\begin{itemize}
\item[1.] Renormalization of $\bcG_R[\varphi_R=0]$ i.e., in the present context, of the photon and fermion two-point functions $\bar G_R$ and $\bar D_R$ or, equivalently of the corresponding self energies $\bSigma_R^{AA}$ and $\bSigma_R^{\bar\psi\psi}$. This has been described in detail in \Refer{Reinosa:2006cm}. For a given 2PI approximation, this requires one to solve the coupled equations of motion for $\bar G_R$ and $\bar D_R$, see \Eqn{eq:stat}, together with a Bethe-Salpeter--type equation for the four-photon function $\bar V_R$ discussed in the previous section, whose kernel involves $\bar G_R$ and $\bar D_R$, simultaneously imposing the conditions \eqn{eq:projectors}-\eqn{eq:rendiff} with \eqn{eq:lolorentz}-\eqn{eq:crochet}. This fixes the counterterms\footnote{Beyond the two-loop approximation of the 2PI effective action, one also has to determine charge and, beyond three-loop, four-photon counterterms in $\delta\Gamma_2^{\rm BPHZ}$ in \Eqn{eq:dGamma2app}, see \Refer{Reinosa:2006cm} for details.} $\delta\bar Z_3$, $\delta\bar M^2$, $\delta\bar\lambda$, $\delta\bar g_1$, $\delta\bar g_2$, $\delta\bar Z_2$ and $\delta\bar m$ independently of all the other counterterms in the 2PI effective action.

This first step is enough for all practical applications where one is essentially interested in the propagator $\bcG_R[\varphi_R=0]$. This is actually the case of most existing applications of 2PI techniques, for instance the calculation of far-from-equilibrium dynamics in scalar \cite{Berges:2000ur} and/or fermionic \cite{Berges:2002wr} field theories, or the calculation of thermodynamic properties of scalar, fermionic, or gauge field theories \cite{Blaizot:1999ip,Berges:2004hn,Blaizot:2005wr}. An explicit application in QED has been performed recently in \Refer{Borsanyi:2007bf}. 

\item[2.] Renormalization of $\bcG_R[\varphi_R]$ or, equivalently, of 2PI $n$-point vertices, see \Eqn{eq:V}.  In the present case, only the 2PI three-point vertices $\delta\bSigma_R^{\bar\psi\psi}/\delta A_R$ and $\delta\bSigma_R^{\bar\psi A}/\delta \psi_R$ need renormalization.\footnote{Charge-conjugation invariance implies that the three-photon vertex $\delta\bSigma_R^{AA}/\delta A_R=0$. Moreover, we show later that the four-photon function $\delta^{2}\bSigma_R^{AA}/\delta A_R^2$, whose superficial degree of divergence is zero, is actually finite.} The latter requires one to solve the linear integral equation \eqn{eq:2PI3_bpap} together with the condition \eqn{eq:renvertices} for $\tilde V_R$. This fixes the counterterm $\delta\tilde Z_1$. No extra work is needed to renormalize $\delta\bSigma_R^{\bar\psi\psi}/\delta A_R$ since the counterterm $\delta\bar Z_1$ is fixed by the symmetry identity \eqn{eq:equalWard}: $\delta\bar Z_1=\delta\bar Z_2$. We show in the next sections that these counterterms are actually sufficient to remove all divergences from all 2PI vertex functions. 

We stress that this level of description is actually enough for a complete determination of all (2PI) vertex functions of the theory. A typical application concerns the calculation of transport coefficients in gauge theories using 2PI techniques \cite{Aarts:2003bk}.

\item[3.] In some specific cases, one might be interested in computing the 2PI-resummed vertex functions, which provide an alternative description of the theory, see e.g. \cite{Berges:2004hn}. In particular, the latter are interesting for they exactly satisfy Ward Takahashi identities at any approximation order \cite{Reinosa:2007vi}, see e.g. \eqn{eq:transmom}. 

Renormalization of 2PI-resummed vertices is discussed in \Sec{sec:renormalization_2PI_resummed_vertices}. There, we show that once 2PI vertices have been renormalized, there only remain global divergences in the 2PI-resummed two- and three-point vertex functions which can be eliminated by mere subtractions. Imposing the renormalization conditions \eqn{eq:2PIresrenG}-\eqn{eq:ren_coup} leads to a direct determination of the remaining counterterms $\delta Z_3$, $\delta m$, $\delta Z_2$ and $\delta Z_1$ in \Eqn{eq:Sintapp}. In the renormalization scheme proposed here, the latter is actually fixed by the symmetry identity \eqn{eq:Z1Z2}: $\delta Z_1=\delta Z_2$.

\end{itemize}

\section{Renormalization of 2PI vertex functions}
\label{sec:renormalization_2PI_vertices}

This section and the following present a detailed proof of renormalization: We show that the counterterms introduced in Eqs.~\eqn{eq:Sintapp}-\eqn{eq:dGamma2app}, and in turn the corresponding renormalization and consistency conditions, see Sec.~\ref{sec:rencond}, are enough to eliminate the (sub)divergences of all vertex functions of the theory, up to those involving composite operators. The symmetry identities discussed in Sec.~\ref{sec:WTI} play a crucial role for they constrain the structure of UV divergences in precisely such a way that only those counterterms allowed by the underlying gauge symmetry, that is those which do not spoil the 2PI Ward-Takahashi identities, are actually needed. This generalizes the results of standard (1PI) renormalization theory. 

A simple example is given by subdiagrams with four photon-legs which, because they have a vanishing superficial degree of divergence, are potentially divergent. As already mentioned, the treatment of four-photon subdivergences is actually a crucial issue in renormalizing the two-point functions $\bD_R$ and $\bG_R$ \cite{Reinosa:2006cm}. They are absorbed by four-photon counterterms such as $\delta \bar g_1$ and $\delta \bar g_2$, which are allowed by gauge symmetry. In contrast, we shall see that 2PI and 2PI-resummed vertex functions involve other classes of four-photon subdiagrams which, if divergent, would call for non-invariant counterterms in $\delta\Gamma_{\rm int}[\varphi_R,\cG_R]$. For instance global divergences in the 2PI and 2PI-resummed four-photon functions $\delta^{2}\bSigma_R^{AA}/\delta A_R^2$ and $\delta^{4}\Gamma_R/\delta A_R^4$ would necessitate counterterms $\propto A_R^2G_R$ and $\propto A_R^4$ respectively. Fortunately, thanks to the symmetry identities they satisfy, both functions turn out to be free of UV divergences and there is no need for such symmetry-breaking counterterms. 

\subsection{2PI two-point vertex functions}
\label{subsec:renormalization_2PI_vertices_2}

As already mentioned, the renormalization of the 2PI two-point vertex functions $\bSigma_R^{AA}$ and $\bSigma_R^{\bar\psi\psi}$ is discussed in detail in \Refer{Reinosa:2006cm}. The main results of this analysis have been summarized in Sec.~\ref{sec:rencond} above. For the purpose of the discussion to follow, we recall that, after the counterterms $\delta\bar M^2$, $\delta\bar\lambda$, $\delta\bar g_1$, $\delta\bar g_2$, $\delta\bar Z_2$ and $\delta\bar m$, as well as possible charge and four-photon counterterms in $\delta\Gamma_2^{\rm BPHZ}$, have been fixed through appropriate renormalization conditions, the functions $\bar G_R$ and $\bar D_R$ as well as all 2PI $2n$-point kernels $\delta^{n}\Gammaint^R/\delta\cG_R^n|_{\bcG[0]}$ with negative superficial degree of divergence are finite.\footnote{The renormalization of 2PI kernels is actually the role of the contribution $\delta\Gamma^{\rm BPHZ}_2[\cG_R]$ in \Eqn{eq:dGamma2app}. We emphasize that the presence of mixed components of the superpropagator $\cG$ does not modify the analysis of \Refer{Reinosa:2006cm}.} This includes all 2PI kernels with $n\ge2$ but the four-photon kernel $\delta^{2}\Gammaint^R/\delta G_R^2|_{\bcG[0]}$. The latter is actually finite up to constant contributions from the counterterms $\delta\bar g_1$ and $\delta\bar g_2$, whose role is to make the four-photon vertex function $\bar V_R$ finite, as discussed previously~\cite{Reinosa:2006cm}. This plays an important role in what follows for these kernels appear as building blocks of (2PI and 2PI-resummed) vertex functions (see for instance Eqs.~\eqn{eq:a1} and \eqn{eq:a2}).

\subsection{2PI three-point vertex functions}
\label{subsec:renormalization_2PI_vertices_3}

Here and in the following sections, it is implicitly understood that we work in momentum space. However, for simplicity, we shall never write explicitly neither momentum variables nor momentum loop integrals, which are pretty straightforward. We associate diagrammatic representations to the relevant equations from which momentum variables and loop integrals can be recovered using standard diagrammatic rules.

\begin{figure}[htbp]
\begin{center}
\includegraphics[width=8.2cm]{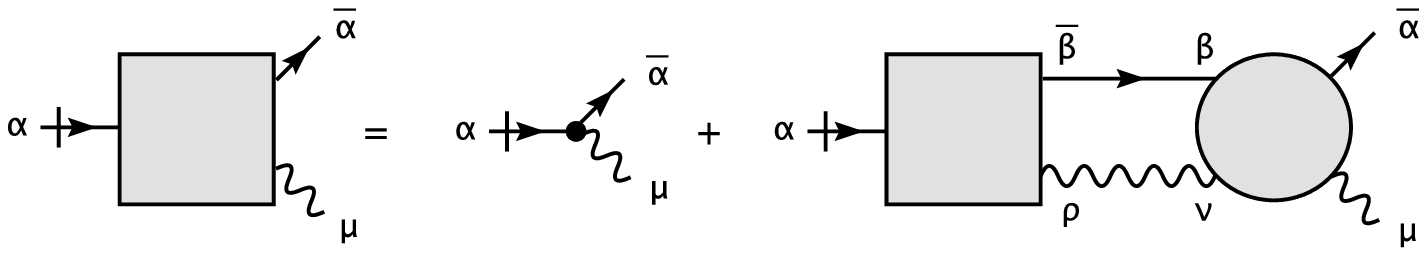}
\caption{\small Diagrammatic representation of Eq.~(\ref{eq:2PI3_bpap_bef}) for the 2PI three-point vertex $\delta\bSigma_R^{\bar\psi A}/\delta\psi_R$, represented by a grey box. The 2PI-kernels $i\delta^2\Gammaint^R/\delta\psi_R\delta\bar K_R$ and $i\delta^2\Gammaint^R/\delta K_R\delta\bar K_R|_{\bcG_R}$ are respectively represented by a dot (tree-level vertex, see \Eqn{eq:kernel_pbK}) and a grey disk.\label{fig:2PI3_bpap}}
\end{center}
\end{figure}

The 2PI three-point vertex $\delta\bSigma^{\bar\psi A}_R/\delta\psi_R$ is given in terms of the self-consistent equation (\ref{eq:2PI3_bpap}), assuming that the relevant replacements have been made to obtain the renormalized version of the equation, as explained in Sec.~\ref{sec:definition}. This equation can be rewritten as\footnote{To simplify notations in the following, we use different greek letters to distinguish photon and fermion legs: the first letters of the alphabet $\alpha,\cdots,\eta$ denote fermion $\psi$-like legs; $\bar\alpha,\cdots,\bar\eta$ denote fermion $\bar\psi$-like legs; any other (greek) letter denotes a photon leg.}
\beq\label{eq:2PI3_bpap_bef}
\frac{\delta\bSigma_{R\,\bar\alpha\mu}^{\,\bar\psi
A}}{\delta\psi^{\alpha}_{\!R}}=\frac{i\delta^2\Gammaint^R}{\delta\psi^{\alpha}_{\!R}\delta\bar K_R^{\mu\bar\alpha}}+
\frac{\delta\bSigma_{R\,\bar\beta\rho}^{\,\bar\psi
A}}{\delta\psi^{\alpha}_{\!R}}\tilde M^{\rho\bar\beta,\beta\nu}\,\tilde\Lambda_{\beta\nu,\mu\bar\alpha}\,,
\eeq
where we have introduced the notations
\beq\label{eq:notassion}
\tilde M^{\rho\bar\beta,\beta\nu}\equiv\bG_R^{\rho\nu}\bD_R^{\beta\bar\beta} \;\;{\rm and}\;\;
\tilde\Lambda_{\beta\nu,\mu\bar\alpha}\equiv\left.\frac{i\delta^2\Gammaint^R}{\delta K_R^{\beta\nu}\delta
\bK_R^{\mu\bar\alpha}}\right|_{\bcG_R}\,.
\eeq
It is to be noticed that, independently of the approximation, the 2PI kernel $i\delta^2\Gammaint^R/\delta\psi_R\delta\bar K_R$  appearing in Eq.~(\ref{eq:2PI3_bpap_bef}) reads, in momentum space:
\beq\label{eq:kernel_pbK}
\frac{i\delta^2\Gammaint^R}{\delta\psi^{\alpha}_{\!R}\delta\bar K_R^{\mu\bar\alpha}}=-ie_R(1+\delta\tilde Z_1)\,\gamma_{\mu,\bar\alpha\alpha}\,.
\eeq
Equation (\ref{eq:2PI3_bpap_bef}) is represented diagrammatically in Fig.~\ref{fig:2PI3_bpap}. It resums an infinite series of ladder diagrams with rungs given by the function $\tilde\Lambda$.

Renormalization of $\delta\bSigma_R^{\bar\psi A}/\delta\psi_R$ means that one is able to choose the counterterms, in particular $\delta\tilde Z_1$, in such a way that $\delta\bSigma_R^{\bar\psi A}/\delta\psi_R$ becomes independent of the regulator as the latter is taken away (here $\epsilon=4-d\to0$). The aim of this section is to prove the existence of $\delta \tilde Z_1$ at any order of approximation of the 2PI loop expansion. To this aim, it is convenient to organize the ladder diagrams contributing to $\delta\bSigma_R^{\bar\psi A}/\delta\psi_R$ in terms of their number of rungs. Formally, we introduce an expansion parameter $\rho$ which operates a rescaling $\smash{\tilde\Lambda \rightarrow\rho\,\tilde\Lambda}$ of the rung $\tilde\Lambda$ in Eq.~(\ref{eq:2PI3_bpap_bef}) and which is eventually taken to one. The three-point vertex $\delta\bSigma_R^{\bar\psi A}/\delta\psi_R$ thus becomes a function of $\rho$ which can be formally expanded as
\begin{equation}\label{eq:dec_tilde}
\frac{\delta\bSigma_{R\,\bar\alpha\mu}^{\,\bar\psi A}}{\delta\psi^{\alpha}_{\!R}}\equiv-ie_R\sum_{n=0}^\infty \tilde{\cal S}^{(n)}_{\mu,\bar\alpha\alpha}\,\rho^n\,,
\end{equation}
and a similar expansion can be considered for the counterterm $\delta\tilde Z_1$:
\begin{equation}\label{eq:dec_tilde_Z}
\delta\tilde Z_1\equiv\sum_{n=1}^\infty \delta\tilde Z_1^{(n)}\,\rho^n\,.
\end{equation}
From Eq.~(\ref{eq:2PI3_bpap_bef}), it is easy to obtain a recursive set of equations for $\tilde{\cal S}^{(n)}$:
\begin{eqnarray}\label{eq:recursive}
\tilde{\cal S}^{(0)}_{\mu,\bar\alpha\alpha} & = & \gamma_{\mu,\bar\alpha\alpha}\,,\nonumber\\
\tilde{\cal S}^{(n)}_{\mu,\bar\alpha\alpha} & = & \delta \tilde Z_1^{(n)}\gamma_{\mu,\bar\alpha\alpha}+\tilde{\cal S}^{(n-1)}_{\rho,\bar\beta\alpha}\tilde M^{\rho\bar\beta,\beta\nu}\,\tilde\Lambda_{\beta\nu,\mu\bar\alpha}\,,\;\;\; n\geq 1\,,
\end{eqnarray}
which can be solved and yields an explicit expression for $\smash{n \geq 1}$:
\beq\label{eq:sol_recursive}
\tilde{\cal S}^{(n)}_{\mu,\bar\alpha\alpha}= \delta\tilde Z^{(n)}_1\gamma_{\mu,\bar\alpha\alpha}+\gamma_{\rho,\bar\beta\alpha}\tilde M^{\rho\bar\beta,\beta\nu}\,\tilde V^{(n)}_{\beta\nu,\mu\bar\alpha}+\sum_{p=1}^{n-1}\delta\tilde
Z^{(p)}_1\gamma_{\rho,\bar\beta\alpha}\tilde M^{\rho\bar\beta,\beta\nu}\,\tilde V^{(n-p)}_{\beta\nu,\mu\bar\alpha},
\eeq
where $\tilde V^{(n)}$ denotes a ladder diagram made of $n$ rungs $\tilde\Lambda$ connected to each other by $\tilde M$, that is, using short-hand notations, $\smash{\tilde V^{(n)}\equiv\tilde\Lambda(\tilde M\tilde\Lambda)^{n-1}}$. As an aside remark, useful for later purposes, we notice that, plugging back this explicit expression in Eq.~(\ref{eq:dec_tilde}) and taking the limit $\smash{\rho\rightarrow 1}$, one obtains the solution to Eq.~(\ref{eq:2PI3_bpap_bef}) in the form:\footnote{This solution can alternatively be obtained by solving Eq.~(\ref{eq:2PI3_bpap_bef}) iteratively. Then, plugging this solution in Eq.~(\ref{eq:dec_tilde}), one obtains the explicit expression (\ref{eq:sol_recursive}).}
\beq\label{eq:2PI3_bpap_sol}
\frac{\delta\bSigma_{R\,\bar\alpha\mu}^{\,\bar\psi
A}}{\delta\psi^{\alpha}_{\!R}}=\frac{i\delta^2\Gammaint^R}{\delta\psi^{\alpha}_{\!R}\delta\bar K_R^{\mu\bar\alpha}}
+\frac{i\delta^2\Gammaint^R}{\delta\psi^{\alpha}_{\!R}\delta\bar K_R^{\rho\bar\beta}}
\tilde M^{\rho\bar\beta,\beta\nu}\,\tilde V_{\beta\nu,\mu\bar\alpha}\,,
\eeq
where $\smash{\tilde V=\sum_{n \geq 1}\tilde V^{(n)}}$ fulfills the Bethe-Salpeter--like equation
\beq\label{eq:BS_pa}
\tilde V_{\beta\nu,\mu\bar\alpha}=\tilde\Lambda_{\beta\nu,\mu\bar\alpha}+\tilde\Lambda_{\beta\nu,\sigma\bar\eta}\,
\tilde M^{\sigma\bar\eta,\eta\rho}\,\tilde V_{\eta\rho,\mu\bar\alpha}\,.
\eeq

\begin{figure}[htbp]
\begin{center}
\includegraphics[width=3.4cm]{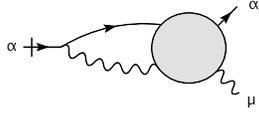}
\caption{Loop contribution to $\tilde{\cal  S}^{(1)}$. It only has a global divergence, which can be absorbed in the contribution $\delta\tilde Z_1^{(1)}$ to the counterterm $\delta\tilde Z_1$, see Eq.~(\ref{eq:tildeS1}).\label{fig:2PI3_bpap_sub3}}
\end{center}
\end{figure}

\begin{figure}[htbp]
\begin{center}
\includegraphics[width=3.5cm]{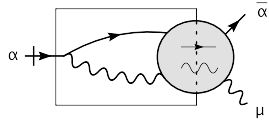}
\caption{The box represents the structure of potentially divergent subgraphs in the diagram of \Fig{fig:2PI3_bpap_sub3}, which involves a two-particle cut through the rung $\tilde\Lambda$. No such divergent subgraphs exist since the ladder rung originates from a closed two-particle-irreducible diagram.\label{fig:2PI3_bpap_sub2}}
\end{center}
\end{figure}

The utility of introducing the formal expansions (\ref{eq:dec_tilde}) and (\ref{eq:dec_tilde_Z}) is that it is now relatively easy to prove that the counterterms $\delta \tilde Z_1^{(n)}$ can be chosen such that each $\tilde{\cal  S}^{(n)}$, and in turn $\delta\bSigma_R^{\bar\psi A}/\delta\psi_R$, is finite. To do so, let us proceed recursively. Because $\smash{\tilde S^{(0)}_{\mu,\bar\alpha\alpha}=\gamma_{\mu,\bar\alpha\alpha}}$ is trivially finite, we first consider
\begin{equation}\label{eq:tildeS1}
\tilde{\cal  S}^{(1)}_{\mu,\bar\alpha\alpha}=\delta \tilde Z_1^{(1)}\gamma_{\mu,\bar\alpha\alpha}+\gamma_{\rho,\bar\beta\alpha}\tilde M^{\rho\bar\beta,\beta\nu}\,\tilde\Lambda_{\beta\nu,\mu\bar\alpha}\,.
\end{equation}
The second contribution in the RHS of this equation is effectively a one-loop contribution involving the building blocks $\tilde M$ and $\tilde \Lambda$, which we represent in Fig.~\ref{fig:2PI3_bpap_sub3}. 
Since $\tilde M$ and $\tilde \Lambda$ are already finite, see \Sec{subsec:renormalization_2PI_vertices_2}, potential subdivergences can only arise from subgraphs obtained by combining lines of these building blocks. Exploiting the 1PI character of proper vertex functions, one finds that the only potentially divergent such subgraphs are those depicted in Fig.~\ref{fig:2PI3_bpap_sub2}. They involve one fermion and one photon internal line -- i.e. a two-particle cut -- of the ladder rung $\tilde\Lambda$. 
But since the latter originates from a closed 2PI diagram, no such subgraph exists. We conclude that the second contribution in the RHS of Eq.~(\ref{eq:tildeS1}) is void of subdivergences. The remaining global divergence\footnote{The tensor structure of the RHS of Eq.~(\ref{eq:dtildeZ1}) is fixed by usual power counting arguments together with global symmetries, see \App{appsec:lorstruct}. Here and in the following, it is implicitly assumed that Lorentz symmetry, parity and charge-conjugation invariance are left unbroken by both the regularization and gauge-fixing procedures. Under these assumptions, the overall divergence of any globally divergent fermion photon three-point vertex function is of the form $a_{\rm div}\gamma_\mu$ with $a_{\rm div}$ a constant.} can thus be absorbed in the divergent part of $\delta\tilde Z_1^{(1)}$ (we denote by $[{\cal D}]_{\bar\infty}$ the overall divergence of a globally divergent sum of diagrams ${\cal D}$):
\begin{equation}\label{eq:dtildeZ1}
 -\Big[\delta \tilde Z_1^{(1)}\gamma_{\mu,\bar\alpha\alpha}\Big]_{\bar\infty}=\left[\gamma_{\rho,\bar\beta\alpha}\tilde M^{\rho\bar\beta,\beta\nu}\,\tilde\Lambda_{\beta\nu,\mu\bar\alpha}\right]_{\bar\infty},
\end{equation}
so that $\tilde{\cal  S}^{(1)}$ is now finite.

\begin{figure}[htbp]
\begin{center}
\includegraphics[width=4.2cm]{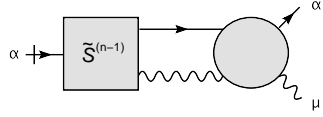}
\caption{Loop contribution to $\tilde{\cal  S}^{(n)}$. It only has a global divergence, which can be absorbed in the contribution $\delta\tilde Z_1^{(n)}$ to the counterterm $\delta\tilde Z_1$, see Eq.~(\ref{eq:tildeSn}).\label{fig:2PI3_bpap_2}}
\end{center}
\end{figure}

\begin{figure}[htbp]
\begin{center}
\includegraphics[width=6.5cm]{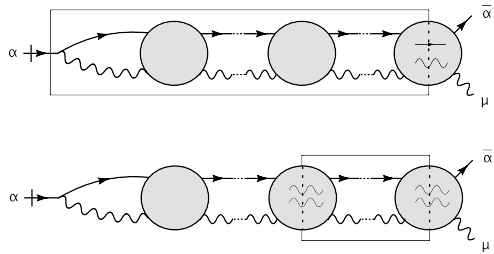}
\caption{The boxes represent the potentially divergent subgraphs in the 2PI three-point vertex $\delta\bSigma_R^{\bar\psi A}/\delta\psi_{R}$. Because the ladder rung $\tilde\Lambda$ (grey bubble) originates from a closed 2PI diagram, it follows that no such subgraph actually exist.\label{fig:2PI3_bpap_n_no}}
\end{center}
\end{figure}

Let us now assume that $\tilde{\cal  S}^{(n-1)}$ has been made finite by a choice of the counterterms $\delta \tilde Z_1^{(p)}$ for $\smash{p\leq n-1}$, and let us consider
\begin{equation}\label{eq:tildeSn}
\tilde{\cal  S}^{(n)}_{\mu,\bar\alpha\alpha}=\delta \tilde Z_1^{(n)}\gamma_{\mu,\bar\alpha\alpha}+\tilde{\cal  S}^{(n-1)}_{\rho,\bar\beta\alpha}\tilde M^{\rho\bar\beta,\beta\nu}\,\tilde\Lambda_{\beta\nu,\mu\bar\alpha}\,.
\end{equation}
Again, the second term of this equation, represented in \Fig{fig:2PI3_bpap_2}, is effectively a one-loop contribution involving the three building blocks $\tilde M$, $\tilde \Lambda$ and $\tilde{\cal  S}^{(n-1)}$. 
Since $\tilde M$ and $\tilde \Lambda$ have already been renormalized, and $\tilde{\cal  S}^{(n-1)}$ is assumed finite by hypothesis, subdivergences in the RHS of Eq.~(\ref{eq:tildeSn}) can only originate from combinations of lines belonging to different building blocks. Again, exploiting the 1PI character of proper vertex functions, one finds that the only potentially divergent such subgraphs are those depicted in \Fig{fig:2PI3_bpap_n_no}, where the ladder structure of $\tilde S^{(n-1)}$ is made explicit, see \Eqn{eq:sol_recursive}.
No such subgraph exists since they all involve a two-particle cut through the kernel $\tilde\Lambda$. The second contribution in the RHS of \Eqn{eq:tildeSn} is thus void of subdivergences and the remaining global divergence can be absorbed in the divergent part of $\delta\tilde Z_1^{(n)}$:
\begin{eqnarray}
-\Big[\delta \tilde Z_1^{(n)}\gamma_{\mu,\bar\alpha\alpha}\Big]_{\bar\infty} & = & \left[\tilde{\cal  S}^{(n-1)}_{\rho,\bar\beta\alpha}\tilde M^{\rho\bar\beta,\beta\nu}\,\tilde\Lambda_{\beta\nu,\mu\bar\alpha}\right]_{\bar\infty}\nonumber\\
\label{eq:deltatildeZ1}
& = & \left[\gamma_{\rho,\bar\beta\alpha}\tilde M^{\rho\bar\beta,\beta\nu}\,\tilde V^{(n)}_{\beta\nu,\mu\bar\alpha}+\sum_{p=1}^{n-1}\delta\tilde
Z^{(p)}_1\gamma_{\rho,\bar\beta\alpha}\tilde M^{\rho\bar\beta,\beta\nu}\,\tilde V^{(n-p)}_{\beta\nu,\mu\bar\alpha}\right]_{\bar\infty},\nn
\end{eqnarray}
so that $\tilde{\cal  S}^{(n)}$ is now finite. 
This ends the proof that the counterterms $\delta\tilde Z_1^{(n)}$ and, in turn, $\delta\tilde Z_1$ can be adjusted such that all $\tilde{\cal  S}^{(n)}$ and, in turn, $\delta\bSigma_R^{\bar\psi A}/\delta\psi_R$ are finite. 

\begin{figure}[htbp]
\begin{center}
\includegraphics[width=8.2cm]{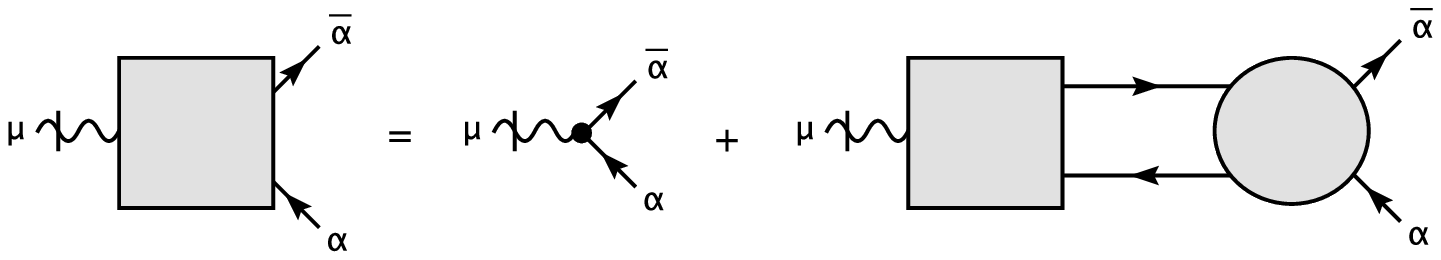}
\caption{\small Diagrammatic representation of Eq.~(\ref{eq:2PI3_bppa_bef}) for the 2PI three-point vertex $\delta\bSigma_R^{\bar\psi\psi}/\delta A_R$ represented by a grey box. The 2PI-kernels $-i\delta^2\Gammaint^R/\delta A_R\delta D_R$ and $i\delta^2\Gammaint^R/\delta D_R\delta D_R|_{\bcG_R}$ are respectively represented by a dot (tree-level vertex, see \Eqn{eq:kernel_AD}) and a grey disk.\label{fig:2PI3_bppa}}
\end{center}
\end{figure}

We next consider the 2PI three-point vertex $\delta\bSigma_R^{\bar\psi\psi}/\delta A_R$. Introducing the notations: 
\beq\label{eq:notascion2}
\bM^{\gamma\bar\gamma,\beta\bar\beta}\equiv\bD_R^{\gamma\bar\beta}\,\bD_R^{\beta\bar\gamma} \;\;{\rm and}\;\;
\bLambda_{\beta\bar\beta,\alpha\bar\alpha}\equiv\left.\frac{-i\delta^2\Gammaint^R}{\delta D_R^{\beta\bar\beta}\delta D_R^{\alpha\bar\alpha}}\right|_{\bcG_R}\,,
\eeq
we obtain, from \Eqn{eq:2PI3_bppa},
\beq\label{eq:2PI3_bppa_bef}
\frac{\delta\bSigma_{R\,\bar\alpha\alpha}^{\,\bar\psi\psi}}{\delta A_R^\mu}=\frac{-i\delta^2\Gammaint^R}{\delta A_R^\mu\delta D_R^{\alpha\bar\alpha}}+\frac{\delta\bSigma_{R\,\bar\gamma\gamma}^{\,\bar\psi\psi}}{\delta A_R^\mu}\bM^{\gamma\bar\gamma,\beta\bar\beta}\,\bLambda_{\beta\bar\beta,\alpha\bar\alpha}\,,
\eeq
where the first term on the RHS reads, explicitly,
\beq\label{eq:kernel_AD}
\frac{-i\delta^2\Gammaint^R}{\delta A_R^\mu\delta D_R^{\alpha\bar\alpha}}=-ie_R(1+\delta\bar
Z_1)\,\gamma_{\mu,\bar\alpha\alpha}\,.
\eeq
Eq.~(\ref{eq:2PI3_bppa_bef}) is represented diagrammatically in Fig.~\ref{fig:2PI3_bppa}.

\begin{figure}[htbp]
\begin{center}
\includegraphics[width=6.5cm]{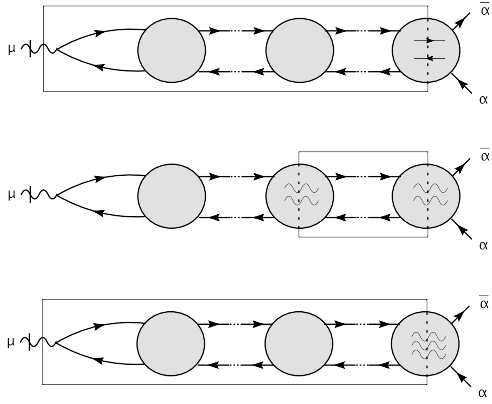}
\caption{The boxes represent the potentially divergent subgraphs in the 2PI three-point vertex $\delta\bSigma_R^{\bar\psi \psi}/\delta A_{R}$. Because the ladder rung $\bar\Lambda$ (grey bubble) originates from a closed 2PI diagram, it follows that all such subgraphs but the last one are ruled out.\label{fig:2PI3_bppa_n_all_2}}
\end{center}
\end{figure}

For simple enough approximations, that is up to and including three-loop order in the loop-expansion of $\Gammaint^R[\varphi_R,\cG_R]$, the analysis of $\delta\bSigma_R^{\bar\psi\psi}/\delta A_R$ can proceed along the same lines as that of $\delta\bSigma_R^{\bar\psi A}/\delta\psi_R$ above. Since this is a rather simple proof, which still concerns already quite nontrivial 2PI approximations, we first consider this case. We give a complete proof, valid at four-loop order and beyond, later on. Following the previous discussion, we write:
\beq\label{eq:dec_bar}
\frac{\delta\bSigma_{R\,\bar\alpha\alpha}^{\,\bar\psi\psi}}{\delta A_R^\mu}\equiv-ie_R\sum_{n=0}^\infty
\bar{\cal  S}^{(n)}_{\mu,\bar\alpha\alpha}\,\rho^n
\eeq
and 
\begin{equation}
\delta\bar Z_1\equiv\sum_{n=1}^\infty \delta\bar Z_1^{(n)}\,\rho^n\,,
\end{equation}
with $\bar{\cal  S}^{(n)}_{\mu,\bar\alpha}$ given either in terms of a recursive set of equations
\begin{eqnarray}\label{eq:recursive_2}
\bar {\cal  S}^{(0)}_{\mu,\bar\alpha\alpha} & = & \gamma_{\mu,\bar\alpha\alpha}\,,\nonumber\\
\label{eq:srec}
\bar {\cal  S}^{(n)}_{\mu,\bar\alpha\alpha} & = & \delta \bar Z_1^{(n)}\gamma_{\mu,\bar\alpha\alpha}+\bar {\cal  S}^{(n-1)}_{\mu,\bar\gamma\gamma}\bar M^{\bar\gamma\gamma,\bar\beta\beta}\,\bar\Lambda_{\bar\beta\beta,\bar\alpha\alpha}\,,\;\;\; n\geq 1\,,
\end{eqnarray}
or explicitly as ($\smash{n\geq 1}$)
\beq\label{eq:sol_recursive_2}
\bar{\cal  S}^{(n)}_{\mu,\bar\alpha\alpha}=\delta\bar Z^{(n)}_1\gamma_{\mu,\bar\alpha\alpha}+\gamma_{\mu,\bar\gamma\gamma}\bM^{\gamma\bar\gamma,\beta\bar\beta}\,\bV^{(n)}_{\beta\bar\beta,\alpha\bar\alpha}+\sum_{p=1}^{n-1}\delta\bar
Z^{(p)}_1\gamma_{\mu,\bar\gamma\gamma}\bM^{\gamma\bar\gamma,\beta\bar\beta}\,\bV^{(n-p)}_{\beta\bar
\beta,\alpha\bar\alpha},
\eeq
where $\smash{\bar V^{(n)}\equiv\bar\Lambda(\bar M\bar\Lambda)^{n-1}}$ denotes the ladder contribution involving $n$ ladder rungs $\bar\Lambda$ connected to each other by $\bar M$. 
Again, it is useful to introduce the function $\smash{\bar V=\sum_{n \geq 1}\bar V^{(n)}}$, which satisfies the following Bethe-Salpeter equation
\beq\label{eq:oubli2}
\bar V_{\beta\bar\beta,\alpha\bar\alpha}=\bar\Lambda_{\beta\bar\beta,\alpha\bar\alpha}+\bar\Lambda_{\beta\bar\beta,\gamma\bar\gamma}\,
\bar M^{\gamma\bar\gamma,\eta\bar\eta}\,\bar V_{\eta\bar\eta,\alpha\bar\alpha}\,,
\eeq
and in terms of which \Eqn{eq:2PI3_bppa_bef} can be solved as
\beq\label{eq:oubli}
\frac{\delta\bSigma_{R\,\bar\alpha\alpha}^{\,\bar\psi\psi}}{\delta A_R^\mu}=\frac{-i\delta^2\Gammaint^R}{\delta A_R^\mu\delta D_R^{\alpha\bar\alpha}}+\frac{-i\delta^2\Gammaint^R}{\delta A_R^\mu\delta D_R^{\gamma\bar\gamma}}\bM^{\gamma\bar\gamma,\beta\bar\beta}\,\bar V_{\beta\bar\beta,\alpha\bar\alpha}\,.
\eeq
Following the same steps as before, one recursively shows that it is possible to adjust the infinite parts of the contributions $\delta \bar Z_1^{(n)}$ to the counterterms $\delta \bar Z_1$ as
\begin{equation}\label{eq:dZbn}
-\Big[\delta \bar Z_1^{(n)}\gamma_{\mu,\bar\alpha\alpha}\Big]_{\bar\infty}=\left[\bar{\cal  S}^{(n-1)}_{\mu,\bar\gamma\gamma}\bar M^{\bar\gamma\gamma,\bar\beta\beta}\,\bar\Lambda_{\bar\beta\beta,\bar\alpha\alpha}\right]_{\bar\infty}
\end{equation}
so that all $\bar{\cal  S}^{(n)}$ and, in turn, $\delta\bSigma_R^{\bar\psi\psi}/\delta A_R$ are finite. A key point in this recursive proof is that the expression in brackets in Eq.~(\ref{eq:dZbn}) be void of subdivergences. The potentially divergent subgraphs in $\bar{\cal  S}^{(n-1)}\bar M\bar\Lambda$ which are not subgraphs of neither $\bar{\cal  S}^{(n-1)}$ nor $\bar M$ nor $\Lambda$ are depicted in Fig.~\ref{fig:2PI3_bppa_n_all_2}. Using the fact that the ladder rung $\bar\Lambda$ originates from a closed 2PI diagram, they can all be ruled out except the last one which involves a three-photon cut through $\bar\Lambda$. Such cuts occur whenever $\Gammaint^R[\varphi_R,\cG_R]$ contains three-photon-reducible diagrams, that is at and beyond four-loop order. The above proof is thus valid up to the 2PI three-loop approximation. The analysis of higher orders requires a careful treatment of diagrams arising from three-photon cuts, which we now present.

Assume that we isolate one particular three-photon-reducible piece of the ladder rung $\bar\Lambda$
\beq\label{eq:cut}
\bar\Lambda_{\alpha\bar\alpha,\beta\bar\beta}=L_{\alpha\bar\alpha}^{\nu\rho\sigma}\,\bar G_{R\,\nu\bar\nu}\,\bar
G_{R\,\rho\bar\rho}\,\bar G_{R\,\sigma\bar\sigma}\,R_{\beta\bar\beta}^{\bar\nu\bar\rho\bar\sigma}+\dots
\eeq
It generates in \Eqn{eq:srec} (logarithmically) divergent four-photon subgraphs of the form ($\smash{n\geq 1}$), see \Fig{fig:2PI3_bppa_n_all_2}:
\begin{equation}\label{eq:div4}
\bar{\cal  S}^{(n-1)}_{\mu,\alpha\bar\alpha}\,\bM^{\alpha\bar\alpha,\beta\bar\beta}\,L_{\beta\bar\beta}^{\nu\rho\sigma}\,.
\end{equation}
Such divergences are problematic since there is no counterterm allowed by the gauge symmetry which could absorb them. Fortunately, these potentially dangerous terms actually sum up to a finite contribution in the limit $\smash{\rho\rightarrow 1}$. Indeed, as shown in \App{appsec:four_photon}, the following four-photon function
\beq\label{eq:sumsubstruct}
 -ie_R\sum_{n\geq 1}\,\bar{\cal  S}^{(n-1)}_{\mu,\alpha\bar\alpha}\bM^{\alpha\bar\alpha,\beta\bar\beta} L^{\nu\rho\sigma}_{\beta\bar\beta}=\frac{\delta\bSigma_{R,\bar\alpha\alpha}^{\,\bar\psi\psi}}{\delta A_R^\mu}\bM^{\alpha\bar\alpha,\beta\bar\beta} L^{\nu\rho\sigma}_{\beta\bar\beta}
\eeq
is transverse with respect to the external momentum carrying the Lorentz index $\mu$:
\beq\label{eq:symsum}
 k^\mu\frac{\delta\bSigma_{R,\bar\alpha\alpha}^{\,\bar\psi\psi}}{\delta A_R^\mu}\bM^{\alpha\bar\alpha,\beta\bar\beta} L^{\nu\rho\sigma}_{\beta\bar\beta}=0\,,
\eeq
for any possible function $L_{\beta\bar\beta}^{\nu\rho\sigma}$ in Eq.~\eqn{eq:cut}. This key property is a non-trivial consequence of the underlying gauge symmetry, valid at any loop order. It implies that the divergent four-photon subgraphs (\ref{eq:div4}) sum up to a contribution (\ref{eq:sumsubstruct}) whose actual superficial degree of divergence is lowered to $-1$. It is thus finite.

As it stands the above discussion is not completely rigorous. The point is that, in order to deduce the finiteness of (\ref{eq:sumsubstruct}) from its transversality, one has first to show that (\ref{eq:sumsubstruct}) is void of divergent subgraphs, see App.~\ref{appsec:four_photon}. However, among the subgraphs of (\ref{eq:sumsubstruct}) appears $\delta\bSigma_R^{\bar\psi\psi}/\delta A_R$ itself, which we want to prove finite in the first place. To avoid such a circular reasoning, we consider a formal coupling-expansion:
\beq
\frac{\delta\bSigma_{R\,\bar\alpha\alpha}^{\,\bar\psi\psi}}{\delta A_R^\mu}\equiv-ie_R\sum_{n=0}^\infty
\bar S^{(2n)}_{\mu,\bar\alpha\alpha}\,e_R^{2n}\,,
\eeq
and
\beq
\delta \bar Z_1\equiv\sum_{n=1}^\infty \Delta \bar Z_1^{(2n)}\,e_R^{2n}\,.
\eeq

Let us again recursively show that it is possible to tune the $\Delta\bar Z_1^{(n)}$'s in such a way that each $\bar S^{(2n)}$ and, in turn, $\delta\bSigma_R^{\bar\psi\psi}/\delta A_R$ are finite. We first consider
\begin{equation}
\bar S^{(2)}_{\mu,\bar\alpha\alpha}=\Delta \bar Z_1^{(2)}\gamma_{\mu,\bar\alpha\alpha}+\bar S^{(0)}_{\mu,\bar\gamma\gamma}\left\{\bar M^{\bar\gamma\gamma,\bar\beta\beta}\,\bar\Lambda_{\bar\beta\beta,\bar\alpha\alpha}\right\}_{2}\,,
\end{equation}
where $\left\{{\cal  D}\right\}_n$ denotes the order $e_R^{n}$ coefficient in the expansion of ${\cal  D}$. To order $e^2_R$, $\bM \bLambda$ is given by its tree level contribution. Thus $\bar S^{(0)}\bM\bLambda$ is a one-loop perturbative diagram. It only has a global divergence which can be absorbed in $\Delta Z_1^{(2)}$:
\begin{equation}
-\Big[\Delta \bar Z_1^{(2)}\gamma_{\mu,\bar\alpha\alpha}\Big]_{\bar\infty}=\left[\bar S^{(0)}_{\mu,\bar\gamma\gamma}\left\{\bar M^{\bar\gamma\gamma,\bar\beta\beta}\,\bar\Lambda_{\bar\beta\beta,\bar\alpha\alpha}\right\}_{2}\right]_{\bar\infty}\,.
\end{equation}
We then assume that the $\bar S^{(2p)}$ have been made finite for $\smash{p\leq n-1}$ by an appropriate choice of the counterterms $\Delta Z_1^{(2p)}$ and consider $\bar S^{(2n)}$. For the sake of the argument, we concentrate on the potentially dangerous contributions discussed above, namely the divergent subgraphs in $\bar S^{(2n)}$ which originate from three-photon cuts in the rung $\bar\Lambda$, see \Fig{fig:2PI3_bppa_n_all_2}. Let us thus consider a particular three-photon-reducible contribution to the ladder rung $\bar\Lambda$ as in \Eqn{eq:cut}, schematically: 
\beq
 \bar\Lambda=L\,\bar G_R^3R+\ldots
\eeq 
This cut generates four-photon divergences in $\bar S^{(2n)}$ of the form, see \Eqn{eq:2PI3_bppa_bef}:
\beq\label{eq:trivially}
\bar S^{(2n)}_{\mu} = \Delta \bar Z_1^{(2n)}\gamma_{\mu}+\sum_{p+q=N}\, \left\{\frac{i}{e_R}\frac{\delta\bSigma_R^{\bar\psi\psi}}{\delta A_R^\mu}\bar M L\right\}_{l+2p}\Big\{\bar G_R^3R\Big\}_{r+2q}+\ldots
\eeq
where $\smash{2N\equiv 2n-r-l}$ and, $l$ and $r$ denote the order of the lowest-order contributions to $\bar M L$ and $G_R^3 R$ respectively (note that $l,r\ge3$ and thus $p+q\le n-3$).

As before, we see that all possible four-photon subdiagrams corresponding to a given three-photon cut in $\bar\Lambda$ sum up to transverse contributions, see \Eqn{eq:symsum}:
\beq
\label{eq:original}
 k^\mu\left\{\frac{i}{e_R}\frac{\delta\bSigma_R^{\bar\psi\psi}}{\delta A_R^\mu}\bar M L\right\}_{l+2p}=0
\eeq
for $0\le p\le N$.
According to the general result of App.~\ref{appsec:result}, to prove that each such contribution is finite, it is therefore enough to show that they are void of subdivergences. We check that this is the case by inspection. Notice first that, here, we do not need to worry about subdivergences originating from neither $\bar M$ nor the function $L$, since they all have already been taken into account at this stage, without modifying the transversality property (\ref{eq:symsum}). Neither can the subdivergences originate from $\delta\bSigma_R^{\bar\psi\psi}/\delta A_R$ for it involves the $\bar S^{(2p)}$'s up to order $\smash{2p=2N<2n}$, which are finite by assumption. The only possible divergent subgraphs are four-photon subgraphs belonging to the same class as the original structure in Eq.~(\ref{eq:original}) but strictly included in it, which can occur if $L$ is itself three-photon-reducible. In that case, $L$ can be further split as $\smash{L=L'\bar G^3 C}$ and one is lead to analyze a similar structure as in \eqn{eq:original}, with $L\to L'$. Applying this argument iteratively, one is ultimately lead to discuss the case where the function $L$ in Eq.~\eqn{eq:original} is no longer three-photon-reducible. It is thus void of subdivergences and, consequently, finite since transverse. One concludes that, as a consequence of the underlying gauge symmetry, the four-photon function appearing in \Eqn{eq:original} is finite for any choice of $L$.
 
It follows that the only divergence in $\bar S^{(2n)}$ is a global one, which can be absorbed in the ${\cal O}(e_R^{2n})$ contribution $\Delta\bar Z_1^{(2n)}$ to the counterterm $\delta \bar Z_1$, see \Eqn{eq:trivially}.

\subsection{Higher 2PI vertex functions}
\label{subsec:renormalization_higher_2PI_vertices}

The next step is to show that once 2PI two- and three-point vertex functions have been renormalized, higher 2PI vertex functions are void of subdivergences. We present this technical proof in App.~\ref{appsec:renormalization_higher_2PI_vertices}. It follows that all 2PI vertex functions with negative superficial degree of divergence are UV convergent. Only the 2PI four-photon vertex $\delta^{2}\bSigma_R^{AA}/\delta A_R^2$, whose superficial degree of divergence is zero, might contain an overall divergence. As already emphasized, if present, such a divergence could only be removed by including a symmetry-breaking counterterm $\propto A_R^2 G_R$ in \Eqn{eq:dGammaint1loop2app}. Fortunately, as shown in the previous section, the renormalization procedure which removes the subdivergences of $\delta^{2}\bSigma_R^{AA}/\delta A_R^2$ does not spoil gauge-invariance, we can thus use the symmetry identity (\ref{eq:tata}) and the result in App.~\ref{appsec:result} to conclude that because $\delta^{2}\bSigma_R^{AA}/\delta A_R^2$ is free of subdivergences, it can neither have overall divergences and it is thus finite. This completes the proof that all 2PI vertex functions can be made finite by a suitable choice of the (gauge-invariant) counterterms of Eqs.~\eqn{eq:dGammaint1loop2app} and \eqn{eq:dGamma2app}.

\section{Renormalization of 2PI-resummed vertex functions}
\label{sec:renormalization_2PI_resummed_vertices}

We finally come to the third step of the renormalization procedure described in Sec.~\ref{sec:summary}, i.e. the renormalization of 2PI-resummed vertex functions. As we saw in \Sec{sec:barevertex}, see also \cite{Berges:2005hc,Reinosa:2007vi}, the latter can be expressed in a closed form, involving a finite number of contributions, in terms of the propagators $\bD_R$ and $\bG_R$ and higher 2PI vertex functions.
Using this fact, we show in this section that all possible subdivergences of 2PI-resummed vertex functions are eliminated by the renormalization of 2PI vertices. Once the latter have been made finite, there remain at most global divergences in the former, which can be eliminated by mere subtractions -- i.e. local counterterms. As discussed in Sec.~\ref{sec:WTI}, see also \Refer{Reinosa:2007vi}, 2PI-resummed vertex functions exactly satisfy standard Ward-Takahashi identities at any approximation order and, therefore, so do their global divergences. As a consequence, only those counterterms permitted by the underlying gauge symmetry, see \Eqn{eq:Sintapp}, are needed. 
(N.B.: The remark concerning notations at the beginning of Sec.~\ref{subsec:renormalization_2PI_vertices_3} applies to the present section as well.)

\subsection{2PI-resummed two-point functions}
\label{subsec:renormalization_2PI_resummed_vertices_2}

\begin{figure}[htbp]
\begin{center}
\includegraphics[width=12.0cm]{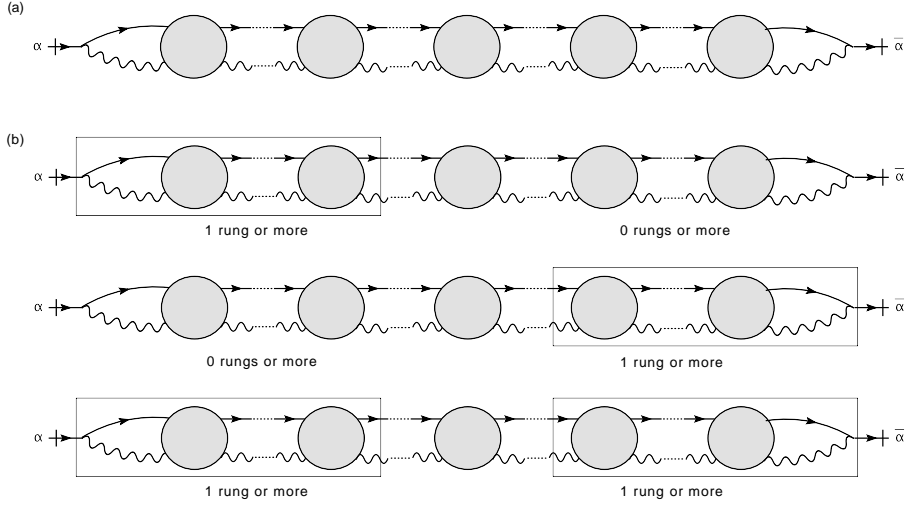}
\caption{Loop contributions to $\Gamma^{(2,0)}_R$. Fig.a represents a ladder diagrams originating from the second term of Eq.~(\ref{eq:2PIr2_pbp_ren}). The boxes in Fig.b show  possible coupling subdivergences of the latter. These are absorbed by the last three terms of Eq.~(\ref{eq:2PIr2_pbp_ren}).\label{fig:2PIr2_pbp_sub1}}
\end{center}
\end{figure}

We first consider the fermion two-point function $\smash{\Gamma^{(2,0)}_R\equiv\delta^{2}\Gamma_R/\delta \psi_R\delta \bar\psi_R}$, see Eq.~(\ref{eq:2PIr2_pbp}). Using Eq.~(\ref{eq:2PI3_bpap_sol}), it can be written as\footnote{We use the same convention as in the previous section to distinguish photon from fermion legs: the first letters of the alphabet $\alpha,\ldots,\eta$ denote fermion $\psi$-like legs; $\bar\alpha,\cdots,\bar\eta$ denote fermion $\bar\psi$-like legs; any other (greek) letter denotes a photon leg.}
\beq\label{eq:2PIr2_pbp_sol}
\Gamma^{(2,0)}_{R\,\bar\alpha\alpha}=iD_{0,R\,\bar\alpha\alpha}^{-1}\!\!+\frac{\delta^2\Gammaint^R}{\delta\psi_R^\alpha\delta\bar\psi_R^{\bar\alpha}\!}-i\frac{i\delta^2\Gammaint^R}{\delta \psi_R^\alpha\delta\bK_R^{\nu\bar\beta}}\,\tilde M^{\nu\bar\beta,\gamma\sigma}\tilde W_{\gamma\sigma,\rho\bar\gamma}\tilde M^{\rho\bar\gamma,\beta\mu}\,\frac{i\delta^2\Gammaint^R}{\delta K_R^{\beta\mu}\delta \bar\psi_R^{\bar\alpha}}\,, 
\eeq
where we have defined\footnote{Alternatively, the function $\tilde W$ can be defined through $\smash{\tilde V=\tilde\Lambda\tilde M\tilde W}$.} $\smash{\tilde W_{\gamma\sigma,\rho\bar\gamma}\equiv\tilde M^{-1}_{\gamma\sigma,\rho\bar\gamma}+\tilde V_{\gamma\sigma,\rho\bar\gamma}}$ with $\smash{\tilde M^{-1}_{\gamma\sigma,\rho\bar\gamma}\equiv\bar G_{R,\sigma\rho}^{-1}\bar D_{R,\bar\gamma\gamma}^{-1}}$, see Eqs.~\eqn{eq:notassion} \begin{figure}[tbp]
\begin{center}
\includegraphics[width=10.0cm]{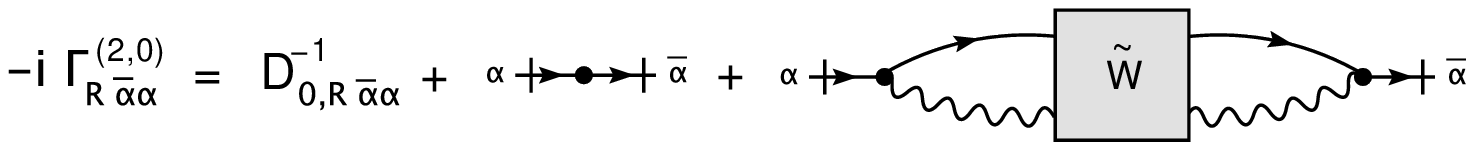}
\caption{Diagrammatic representation of Eq.~(\ref{eq:2PIr2_pbp_sol}) for the 2PI-resummed two-point function $-i\Gamma_R^{(2,0)}\equiv-i\delta^2\Gamma_R/\delta \psi_R\delta \bar\psi_R$ in terms of $\smash{\tilde W}$, defined below Eq.~(\ref{eq:2PIr2_pbp_sol}).\label{fig:2PIr2_pbp_sol}}
\end{center}
\end{figure}
and \eqn{eq:BS_pa}. The diagrammatic representation of \Eqn{eq:2PIr2_pbp_sol} is given in \Fig{fig:2PIr2_pbp_sol},
where we have made explicit the fact that the 2PI kernels $\delta^2\Gammaint^R/\delta\psi_R\delta\bar\psi_R$ and $i\delta^2\Gammaint^R/\delta\psi_R\delta\bK_R=i\delta^2\Gammaint^R/\delta K_R\delta\bar\psi_R$ only contribute at tree level, see \Eqn{eq:kernel_pbK} and:
\beq\label{eq:plugg}
\frac{\delta^2\Gammaint^R}{\delta \psi_R^\alpha\delta\bar\psi_R^{\bar\alpha}}(p)=\left(\delta Z_2\slashchar{p}-\delta m\right)_{\bar\alpha\alpha}\,.
\eeq
Plugging Eqs.~\eqn{eq:kernel_pbK} and \eqn{eq:plugg} into Eq.~\eqn{eq:2PIr2_pbp_sol}, we obtain
\bea
\Gamma^{(2,0)}_{R\,\bar\alpha\alpha}(p)&=&iD_{0,R\,\bar\alpha\alpha}^{-1}+(\delta Z_2\slashchar{p}-\delta m)_{\bar\alpha\alpha}\nn
\label{eq:2PIr2_pbp_ren}
&+&\tilde T_{\bar\alpha\alpha}(p)+\delta \tilde Z_1\tilde T_{\bar\alpha\alpha}(p)+\tilde T_{\bar\alpha\alpha}(p)\delta\tilde Z_1+\delta\tilde Z_1\tilde T_{\bar\alpha\alpha}(p)\delta\tilde Z_1\,,
\eea
where we introduced the function
\beq
\tilde T_{\bar\alpha\alpha}\equiv ie_R^2\,\gamma_{\nu,\bar\beta\alpha}\tilde M^{\nu\bar\beta,\gamma\sigma}\tilde W_{\gamma\sigma,\rho\bar\gamma}\tilde M^{\rho\bar\gamma,\beta\mu}\gamma_{\mu,\bar\alpha\beta}\,.
\eeq
Using Eq.~\eqn{eq:BS_pa}, one sees that it is made of ladder diagrams with rungs $\tilde\Lambda$, see Fig.~\ref{fig:2PIr2_pbp_sub1}.a, which a priori contain (sub)divergences. Using the diagrammatic interpretation of the counterterm $\delta\tilde Z_1$ obtained previously, see \Eqn{eq:deltatildeZ1}, one checks that that the terms involving $\delta\tilde Z_1$ in \Eqn{eq:2PIr2_pbp_ren} absorb the class of subdivergences depicted in Fig.~\ref{fig:2PIr2_pbp_sub1}.b. 
One can show by inspection that there are, in fact, no other subdivergences in Eq.~(\ref{eq:2PIr2_pbp_ren}). The superficial degree of divergence of the two-point function $\delta^{2}\Gamma_R/\delta \psi_R\delta \bar\psi_R$ being $+1$, the remaining global
divergence is a polynomial of order $1$ in momentum space, which Lorentz structure must be $A\slashchar{p}+B$, with $A$ and $B$ divergent numbers. The latter can, therefore, be absorbed in the counterterms $\delta Z_2$ and $\delta m$:
\beq
-\Big[(\delta Z_2\slashchar{p}-\delta m)_{\bar\alpha\alpha}\Big]_{\bar\infty}=\Big[\tilde T_{\bar\alpha\alpha}(p)+\delta \tilde Z_1\tilde T_{\bar\alpha\alpha}(p)+\tilde T_{\bar\alpha\alpha}(p)\delta\tilde Z_1+\delta\tilde Z_1\tilde T_{\bar\alpha\alpha}(p)\delta\tilde Z_1\Big]_{\bar\infty}\,.
\eeq

\begin{figure}[htbp]
\begin{center}
\includegraphics[width=10.0cm]{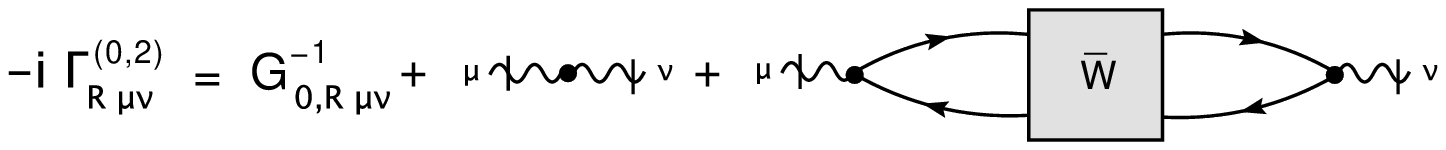}
\caption{Diagrammatic representation of Eq.~(\ref{eq:2PIr2_aa_sol}) for the 2PI-resummed two-point function $-i\Gamma_R^{(0,2)}\equiv-i\delta^2\Gamma_R/\delta A_R\delta A_R$ in terms of $\bW$.\label{fig:2PIr2_aa_sol}}
\end{center}
\end{figure}

\begin{figure}[htbp]
\begin{center}
\includegraphics[width=12.0cm]{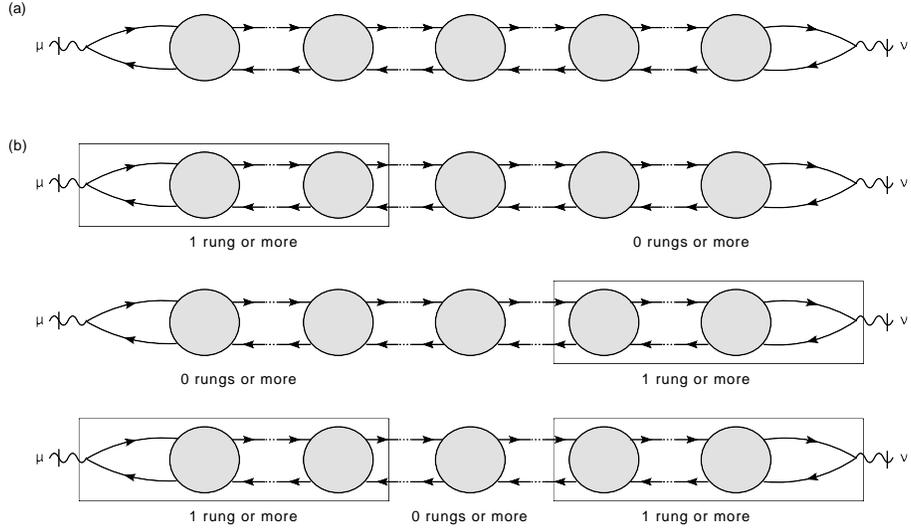}
\caption{Loop contributions to $\Gamma_R^{(0,2)}$. Fig.a represents a ladder diagram originating from the second term of Eq.~(\ref{eq:2PIr2_aa_ren}). Boxes in Fig.b show possible coupling subdivergences of the latter. These are to be absorbed by the last three terms of Eq.~(\ref{eq:2PIr2_aa_ren}).\label{fig:2PIr2_aa_sub1}}
\end{center}
\end{figure}

We next consider the photon two-point function $\delta^2\Gamma_R/\delta A_R\delta A_R$. Following the same steps as before, we get, from Eqs.~\eqn{eq:2PIr2} and \eqn{eq:oubli},
\beq\label{eq:2PIr2_aa_sol}
\Gamma^{(0,2)}_{R\,\mu\nu}=iG_{0,R\,\mu\nu}^{-1}+\frac{\delta^2\Gammaint^R}{\delta A_R^\mu\delta
A_R^\nu}+i\frac{-i\delta^2\Gammaint^R}{\delta A_R^\mu\delta D_R^{\beta\bar\beta}}\,\bM^{\beta\bar\beta,\zeta\bar\zeta}\bW_{\zeta\bar\zeta,\eta\bar\eta}\bM^{\eta\bar\eta,\alpha\bar\alpha}\,\frac{-i\delta^2\Gammaint^R}{\delta D_R^{\alpha\bar\alpha}\delta A_R^\nu}\,, 
\eeq
where we have defined $\smash{\bW_{\zeta\bar\zeta,\eta\bar\eta}\equiv\bM^{-1}_{\zeta\bar\zeta,\eta\bar\eta}+\bV_{\zeta\bar\zeta,\eta\bar\eta}}$ with $\smash{\bM^{-1}_{\zeta\bar\zeta,\eta\bar\eta}=\bar D^{-1}_{R,\bar\zeta\eta}\bar D^{-1}_{R,\bar\eta\zeta}}$, see Eqs.~\eqn{eq:notascion2} and \eqn{eq:oubli2}. The diagrammatic representation of Eq.~\eqn{eq:2PIr2_aa_sol} is given in Fig.~\ref{fig:2PIr2_aa_sol} 
where we have made explicit that the 2PI kernels $\delta^2\Gammaint^R/\delta A_R\delta A_R$ and $-i\delta^2\Gammaint^R/\delta A_R\delta D_R$ only receive tree-level contributions, see Eq.~\eqn{eq:kernel_AD}, and
\beq\label{eq:plug2}
\frac{\delta^2\Gammaint^R}{\delta A_R^\mu\delta A_R^\nu}(k)=-\delta
Z_3\left(g_{\mu\nu}k^2-k_\mu k_\nu\right)\equiv-\delta Z_3 k^2 P^T_{\mu\nu}(k)\,,
\eeq
where $P^T_{\mu\nu}(k)$ is the transverse projector introduced in \Eqn{eq:lolorentz}. Plugging Eqs.~\eqn{eq:kernel_AD} and \eqn{eq:plug2} into Eq.~\eqn{eq:2PIr2_aa_sol}, we get
\bea
\Gamma^{(0,2)}_{R\,\mu\nu}(k)&=&iG_{0,R\,\mu\nu}^{-1}(k)-\delta
Z_3 k^2P^T_{\mu\nu}(k)\nn
\label{eq:2PIr2_aa_ren}
&+&\bar T_{\mu\nu}(k)+\delta\bar Z_1\bar T_{\mu\nu}(k)+\bar T_{\mu\nu}(k)\delta\bar Z_1+\delta\bar Z_1\bar T_{\mu\nu}(k)\delta\bar Z_1\,,
\eea
where we defined the function
\beq\label{eq:thefunction}
\bar T_{\mu\nu}\equiv -ie_R^2\,\gamma_{\mu,\bar\beta\beta}\bM^{\beta\bar\beta,\zeta\bar\zeta}\bW_{\zeta\bar\zeta,\eta\bar\eta}\bar
M^{\eta\bar\eta,\bar\alpha\alpha}\gamma_{\nu,\bar\alpha\alpha}\,.
\eeq
Using \Eqn{eq:oubli2}, one sees that it is made of ladder diagrams with rungs $\bar\Lambda$, see Fig.~\ref{fig:2PIr2_aa_sub1}.a, which contain (sub)divergences. Similar to the case of $\Gamma_R^{(2,0)}$ discussed previously, one easily checks that the terms involving $\delta\bar Z_1$ in \Eqn{eq:2PIr2_aa_ren} absorb three-point subdivergences of the type depicted in Fig.~\ref{fig:2PIr2_aa_sub1}.b. 

\begin{figure}[htbp]
\begin{center}
\includegraphics[width=11.5cm]{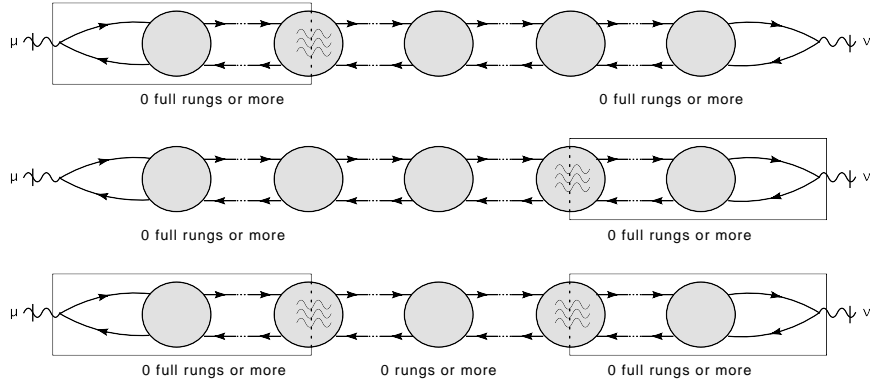}
\caption{Divergent four-photon subgraphs in the 2PI-resummed photon two-point function $\Gamma_R^{(0,2)}$. All such subgraphs sum up to a UV convergent function.\label{fig:2PIr2_aa_sub4}}
\end{center}
\end{figure}

But in the present case, these are not the only potential subdivergences. Indeed, there can be -- starting at the four-loop approximation of the 2PI effective action -- four-photon subdiagrams such as those depicted in Fig.~\ref{fig:2PIr2_aa_sub4}. These are the very same four-photon subdiagrams we encountered in discussing the renormalization of $\delta\bSigma_R^{\bar\psi\psi}/\delta A_R$ in the previous section. We already showed that they in fact sum up to finite contributions, thanks to the underlying gauge symmetry.

\begin{figure}[htbp]
\begin{center}
\includegraphics[width=10.5cm]{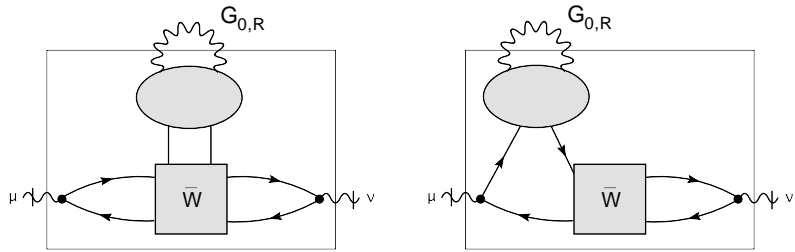}
\caption{Divergent four-photon subgraphs in $\Gamma_R^{(0,2)}$, obtained by combining the two external photon legs of the original diagram with perturbative photon lines $G_{0,R}$ from the internal structure of the resummed propagators $\bar D_R$ and $\bar G_R$ (grey blobs)
.\label{fig:2PIr2_aa_sub2}}
\end{center}
\end{figure}

Finally, there are potentially divergent four-photon subgraphs which arise from the internal structure of the resummed propagator-lines $\bD_R$ and $\bG_R$. Recalling that the latter resum infinite subclasses of perturbative diagrams, one can combine some of the photon lines of these perturbative diagrams with the external legs of the original diagram to form (potentially divergent) four-photon subgraphs, as depicted in \Fig{fig:2PIr2_aa_sub2}.
These subgraphs are such that two of their external legs are those of the original 2PI-resummed photon two-point function $\delta^2\Gamma_R/\delta A_R\delta A_R$ and the other two are obtained by opening a perturbative photon line $G_{0,R}$ somewhere in the diagram. Writing
\beq
 \frac{\delta^2\Gamma_R}{\delta A_R\delta A_R}=iG_{0,R}^{-1}-i\Pi_R\,,
\eeq
we deduce that the (infinite) sum of these subgraphs can be written as a functional derivative of the self-energy $\Pi_R$ with respect to the tree-level propagator $G_{0,R}$: $\delta\Pi_R/\delta G_{0,R}$.\footnote{More precisely, it corresponds to a derivative at fixed $D_{0,R}$ and fixed counterterms.} We show in App.~\ref{appsubsec:four_photon_2} that this quantity is actually related to the 2PI four-photon vertex $\delta^2\bSigma_R^{AA}/\delta A_R\delta A_R$:
\beq\label{eq:fourphotondeux}
G_{0,R}^{\,\rho\lambda}\,\frac{\delta\Pi_{R\,\mu\nu}}{\delta G_{0,R}^{\,\lambda\xi}}\,G_{0,R}^{\,\xi\sigma}=\bG_R^{\rho\lambda}\,\frac{\delta^2\bSigma^{AA}_{R\,\lambda\xi}}{\delta A_{R}^{\mu}\delta A_{R}^{\nu}}\,\bG_R^{\xi\sigma}\,.
\eeq
Since $\bar G_R$ and $\delta^2\bSigma_R^{AA}/\delta A_R\delta A_R$ have already been made finite, \Eqn{eq:fourphotondeux} implies that $\delta\Pi_R/\delta G_{0,R}$ is finite as well. Moreover, using the 2PI Ward identity \eqn{eq:tata}, one concludes that it is transverse with respect to the momenta carrying the labels $\mu$ and $\nu$ (those originating from field derivatives), schematically:
\beq
k^\mu\frac{\delta\Pi_{R\,\mu\nu}}{\delta G_{0,R}^{\,\lambda\eta}}=k^\nu\frac{\delta\Pi_{R\,\mu\nu}}{\delta G_{0,R}^{\,\lambda\eta}}=0\,.
\eeq

We conclude that once 2PI resummed vertices have been properly renormalized, there are no subdivergence left in $\Gamma_R^{(0,2)}$. The remaining global divergence is thus a polynomial of degree $2$ in momentum space, constrained by Lorentz and gauge symmetry, see Eq.~\eqn{eq:transmom}, to be $\propto A(g^{\mu\nu}k^2-k^\mu k^\nu)$, with $A$ a constant. It can, therefore, be absorbed in the infinite part of $\delta Z_3$, see Eqs.~\eqn{eq:2PIr2_aa_ren}-\eqn{eq:thefunction}:
\beq
-\Big[\delta Z_3\,P^T_{\mu\nu}(k)\Big]_{\bar\infty}=\left[\frac{\delta\bSigma_{R,\bar\beta\beta}^{\bar\psi\psi}}{\delta A_R^\mu}\,\bM^{\beta\bar\beta,\alpha\bar\alpha}\,\frac{-i\delta^2\Gammaint^R}{\delta D_R^{\alpha\bar\alpha}\delta A_R^\nu}\right]_{\bar\infty}\,.
\eeq

\subsection{2PI-resummed three-point function}
\label{subsec:renormalization_2PI_resummed_vertices_3}

\begin{figure}[htbp]
\begin{center}
\includegraphics[width=12.0cm]{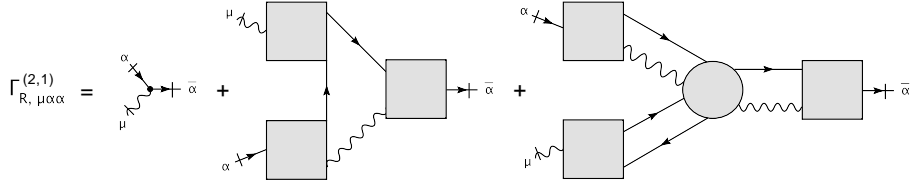}
\caption{Diagrammatic representation of \Eqn{eq:une_equation}. The grey blob represents the 2PI kernel $i\delta^3\Gammaint^R/\delta D_R\delta K_R\delta\bar K_R|_{\bcG_R}$. The grey boxes represent the 2PI three-point vertices $\delta\bSigma^{\bar\psi A}_R/\delta\psi_R$, $\delta\bSigma^{A\psi}_R/\delta \bar\psi_R$ and $\delta\bSigma^{\bar\psi\psi}_R/\delta A_R$.}
\label{fig:2PIr3_bis}
\end{center}
\end{figure}

\begin{figure}
\begin{center}
\includegraphics[width=6.0cm]{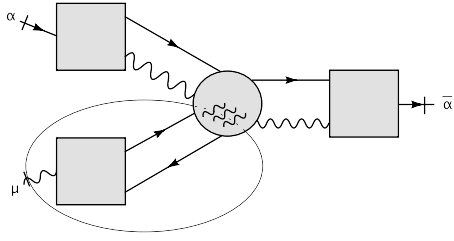}
\caption{These subgraphs are potentially divergent. Fortunately, the 2PI Ward-Takahashi identities ensure that these subgraphs are all finite.\label{fig:2PIr3_sub1}}
\end{center}
\end{figure}

Our starting point for the analysis of the 2PI-resummed three-point function is \Eqn{eq:2PIr34}, which involves, in terms of renormalized quantities, the various 2PI two- and three-point functions as well as the 2PI four-point vertex function $\delta^2\bSigma_R^{\bar\psi A}/\delta A_R\delta\psi_R$. The latter fulfills a linear integral equation, see Eq.~\eqn{eq:explicitely}, which can be conveniently solved in terms of the function $\tilde V$ introduced previously, see Eq.~\eqn{eq:BS_pa}:
\bea
\frac{\delta^2\bar\Sigma^{\bar\psi A}_{R\,\bar\alpha\nu}}{\delta A_R^\mu\delta\psi_R^\alpha} \!\!\!\! & = & \!\!\!\!\left(\bD_R \frac{\delta\bSigma_R^{\bar\psi\psi}}{\delta A_R^\mu}\bD_R\frac{\delta\bSigma_R^{\bar\psi
A}}{\delta\psi_R^\alpha}\bG_R\right)_{\!\!\!\beta\rho}\!\!\tilde V_{\beta\rho,\nu\bar\alpha}\nonumber\\
&&\hspace*{-.9cm}+\left(\bD_R\frac{\delta\bSigma_R^{\bar\psi
A}}{\delta\psi_R^\alpha}\bG_R\right)_{\!\!\!\beta\rho}\!\!\left(\bD_R\frac{\delta\bSigma_R^{\bar\psi\psi}}{\delta
A_R^\mu}\bD_R\right)_{\!\!\!\gamma\bar\gamma}\!\left.\frac{i\delta^3\Gammaint^R}{\delta D_R^{\gamma\bar\gamma}\delta K_R^{\beta\rho}\delta
\bK_R^{\nu\bar\alpha}}\right|_{\bcG_R}\nonumber\\
&&\hspace*{-.9cm}+\left(\bD_R\frac{\delta\bSigma_R^{\bar\psi
A}}{\delta\psi_R^\alpha}\bG_R\right)_{\!\!\!\beta\rho}\!\!\left(\bD_R\frac{\delta\bSigma_R^{\bar\psi\psi}}{\delta
A_R^\mu}\bD_R\right)_{\!\!\!\gamma\bar\gamma}\!\left.\frac{i\delta^3\Gammaint^R}{\delta D_R^{\gamma\bar\gamma}\delta K_R^{\beta\rho}\delta
\bK_R^{\sigma\bar\delta}}\right|_{\bcG_R}\!\!\tilde M^{\sigma\bar\delta,\delta\lambda}\,\tilde V_{\delta\lambda,\nu\bar\alpha}.\nn
\eea
Plugging this equation into Eq.~(\ref{eq:2PIr34}) and using Eq.~(\ref{eq:2PI3_bpap_sol}), one finally obtains the following expression for the 2PI-resummed three-point function:
\bea\label{eq:une_equation}
&&\hspace*{-0.5cm}i\Gamma_{R\,\mu\bar\alpha\alpha}^{(2,1)} =
-ie_R(1+\delta Z_1)\gamma_{\mu,\bar\alpha\alpha}+\tr\left(\bD_R\frac{\delta\bSigma_R^{\bar\psi \psi}}{\delta
A_R^\mu}\bD_R\frac{\delta\bSigma_R^{\bar\psi
A}}{\delta\psi_R^\alpha}\bG_R\frac{\delta\bSigma_R^{A\psi}}{\delta\bar\psi_R^{\bar\alpha}}\right)\nonumber\\
&& \hspace*{-0.2cm}+ \left(\bG_R\frac{\delta\bSigma_R^{A\psi}}{\delta
\bar\psi_R^{\bar\alpha}}\bD_R\right)_{\!\!\!\rho\bar\beta}\!\!\left(\bD_R\frac{\delta\bSigma_R^{\bar\psi
A}}{\delta\psi_R^\alpha}\bG_R\right)_{\!\!\!\beta\nu}\!\!\left(\bD_R\frac{\delta\bSigma_R^{\bar\psi\psi}}{\delta
A_R^\mu}\bD_R\right)_{\!\!\!\gamma\bar\gamma}\!\left.\frac{i\delta^3\Gammaint^R}{\delta D_R^{\gamma\bar\gamma}\delta K_R^{\beta\nu}\delta \bar K_R^{\rho\bar\beta}}\right|_{\bcG_R}\!\!,\nn
\eea
which is particularly suited for hunting potential subdivergences. Here, we used 
\beq\label{eq:2PIres3vertex}
\frac{i\delta^3\Gammaint^R}{\delta A_R^\mu\delta\psi_R^\alpha\delta\bar\psi_R^{\bar\alpha}}=-ie_R(1+\delta Z_1)\gamma_{\mu,\bar\alpha\alpha}\,.
\eeq 
Equation \eqn{eq:une_equation} is represented diagrammatically in Fig.~\ref{fig:2PIr3_bis}, where the exact 2PI kernel \eqn{eq:2PIres3vertex} is represented by a tree-level three-point vertex. 
It is a rather easy task to see that the diagrams of Fig.~\ref{fig:2PIr3_bis} are, in fact, void of subdivergences. First, their building blocks -- namely the resummed propagators $\bar D_R$ and $\bar G_R$, the 2PI three-point vertices $\delta\bSigma_R^{\bar\psi A}/\delta\psi_R$, $\delta\bSigma_R^{A\psi}/\delta\bar\psi_R$ and $\delta\bSigma_R^{\bar\psi\psi}/\delta A_R$, and the 2PI six-point kernel $i\delta\Gammaint^R/\delta D_R\delta K_R\delta \bar K_R|_{\bcG_R}$ -- are all UV convergent. As a consequence, possible subdivergences -- i.e. subgraphs with non-negative superficial degree of divergence -- can only arise by combining some of these, or part of these together. A careful inspection demonstrates that there are no such subgraphs in the second diagram of Fig.~\ref{fig:2PIr3_bis}. The third diagram contains possible four-photon subgraphs, as represented in Fig.~\ref{fig:2PIr3_sub1}. These have a similar structure as those encountered previously in the renormalization of $\delta\bSigma^{\bar\psi\psi}_R/\delta A_R$, see Eq.~\eqn{eq:sumsubstruct}. A similar analysis shows that they are finite.

There only remains a global divergence in $\Gamma_{R\,\mu}^{(2,1)}$, whose Lorentz structure is $\propto A\gamma_\mu$ with $A$ a constant, and which can, therefore, be absorbed in the counterterm $\delta Z_1$:
\bea
&&\hspace*{-0.8cm}\Big[ie_R\delta Z_1\gamma_{\mu,\bar\alpha\alpha}\Big]_{\bar\infty}= \left[\tr\left(\bD_R\frac{\delta\bSigma_R^{\bar\psi \psi}}{\delta A_R^\mu}\bD_R\frac{\delta\bSigma_R^{\bar\psi A}}{\delta\psi_R^\alpha}\bG_R\frac{\delta\bSigma_R^{A\psi}}{\delta\bar\psi_R^{\bar\alpha}}\right)\right]_{\bar\infty}\nonumber\\
&&\hspace*{-0.6cm}+ \!\!\left[\left(\bG_R\frac{\delta\bSigma_R^{A\psi}}{\delta
\bar\psi_R^{\bar\alpha}}\bD_R\right)_{\!\!\!\rho\bar\beta}\!\!\!\left(\bD_R\frac{\delta\bSigma_R^{\bar\psi
A}}{\delta\psi_R^\alpha}\bG_R\right)_{\!\!\!\beta\nu}\!\!\!\left(\bD_R\frac{\delta\bSigma_R^{\bar\psi\psi}}{\delta
A_R^\mu}\bD_R\right)_{\!\!\!\gamma\bar\gamma}\!\!\left.\frac{i\delta^3\Gammaint^R}{\delta D_R^{\gamma\bar\gamma}\delta K_R^{\beta\nu}\delta \bar K_R^{\rho\bar\beta}}\right|_{\bcG_R}\right]_{\bar\infty}\!\!.\nn
\eea

\subsection{Higher 2PI-resummed vertex functions}
\label{subsec:renormalization_higher_2PI_resummed_vertices}

A similar analysis as that performed above for 2PI-resummed two- and three-point vertex functions demonstrates that higher $n$-point functions are void of subdivergences as well. This is detailed in \App{appsec:renormalization_higher_2PI_vertices}. Thus 2PI-resummed vertex functions whose superficial degree of divergence is negative are automatically UV convergent. This includes all $n$-point vertices with $n\ge4$ but the four-photon function $\Gamma_R^{(0,4)}=\delta^4\Gamma_R/\delta A_R^4$. As explained previously (see also \cite{Reinosa:2007vi}) the latter exactly satisfies the usual Ward-Takahashi identities at any approximation order in the 2PI expansion, which states that it is transverse with respect to its four external momenta. Its actual degree of divergence is, therefore, $-4$ and it is consequently finite. This completes the proof that all 2PI and 2PI-resummed vertex functions of the theory can be renormalized\footnote{As already emphasized, we do not discuss the renormalization of composite operators.} by means of the gauge-symmetric counterterms in Eqs.~\eqn{eq:Sintapp}-\eqn{eq:dGamma2app}.

\section*{Acknowledgments}
We would like to thank Sz.~Bors\'anyi as well as J.-P.~Blaizot and E.~Iancu for fruitful collaboration on related topics. UR acknowledges support from the Alexander von Humboldt foundation during the early stages of this work.

\appendix

\section{Global symmetries}
\label{appsec:charge_conjugation_invariance}
We work out some consequences of the global (Lorentz, parity, charge-conjugation and global $U(1)$) symmetries of the gauge-fixed QED classical action for 2PI and 2PI-resummed vertex functions. The results of this section apply indifferently to either bare or renormalized vertex functions. For notational convenience, we present results for the former. 

Let us first recall some general results concerning linearly realized symmetries, see \Refer{Reinosa:2007vi}. It is an easy exercise to show that any linearly realized (global or local) symmetry of the classical action, i.e. 
\beq
 S[\varphi]=S[\varphi']
\eeq
under a transformation 
\beq\label{appeq:trsf}
\varphi_m(x)\to\varphi'_m(x)=\int_y{\cal A}_{mn}(x,y)\varphi_n(y)+{\cal B}_m(x)
\eeq
with $\varphi$-independent (invertible) supermatrix and supervector fields\footnote{We restrict our attention to the case where
bosonic and fermionic components of the superfield are not mixed by the symmetry transformation. This means that ${\cal
A}_{mn}\neq0$ iff $(-1)^{q_m}=(-1)^{q_n}$ or, equivalently, $|q_m|=|q_n|$. In particular, the only nonvanishing
components of the matrix ${\cal A}$ are $c$-numbers.} ${\cal A}(x,y)$ and ${\cal B}(x)$, translates at the level of the 2PI effective action as:
\beq\label{appeq:symac}
 \Gammatpi[\varphi,\cG]=\Gammatpi[\varphi',\cG']\,,
\eeq
where 
\beq
 \cG_{mn}'(x,y)=\int_{u,v}{\cal A}_{mp}(x,u){\cal A}_{nq}(y,v)\cG_{pq}(u,v)\,.
\eeq
From \Eqn{appeq:symac} one can deduce a number of symmetry identities relating the various $n$-point functions of the theory, see \Refer{Reinosa:2007vi}. For instance, assuming that $\bcG[\varphi]$ is unique, one concludes that it transforms covariantly:
\beq\label{appeq:cov}
 \bcG[\varphi']=\bcG'[\varphi]\,.
\eeq
or, equivalently (space-time variables implicit),  
\beq\label{appeq:covself}
 \bSigma_{mn}[\varphi']=\bSigma_{pq}[\varphi]\,{\cal A}_{qn}^{-1}\,{\cal A}_{pm}^{-1}\,.
\eeq
It follows that:
\beq\label{appeq:2PIsym}
\left.\frac{\delta^p\bSigma_{nm}}{\delta\varphi_p\cdots\delta\varphi_1}\right|_{\varphi}=\left.\frac{\delta^p{\bSigma}_{u_nu_m}}{\delta\varphi_{u_p}\cdots\delta\varphi_{u_1}}\right|_{\varphi'}{\cal A}_{u_1 1}\cdots{\cal A}_{u_p p}{\cal A}_{u_n n}\,{\cal A}_{u_m m}\,,
\eeq
from which one deduces symmetry identities for the 2PI vertex functions \eqn{eq:V}.

Finally, using \Eqn{appeq:cov}, one sees that the 2PI-resummed effective action \eqn{eq:1PI} shares the same symmetry property as the classical action:
\beq
 \Gamma[\varphi]=\Gamma[\varphi']\,.
\eeq
It follows that 
\beq\label{appeq:2PIressym}
\left.\frac{\delta^p\Gamma}{\delta\varphi_p\cdots\delta\varphi_1}\right|_{\varphi}=\left.\frac{\delta^p\Gamma}{\delta\varphi_{u_p}\cdots\delta\varphi_{u_1}}\right|_{\varphi'}{\cal A}_{u_1 1}\cdots{\cal A}_{u_p p}\,,
\eeq 
which leads to symmetry identities for 2PI-resummed vertex functions \eqn{eq:npoint}. 

In the following, we analyze the consequences of the global $U(1)$, parity, charge-conjugation and Lorentz symmetries of the QED classical action. These all correspond to global, purely linear transformations of the form \eqn{appeq:trsf}, possibly involving a space-time transformation, i.e.:
\beq\label{appeq:global}
 {\cal A}_{mn}(x,y)={\cal M}_{mn}\,\delta^{(4)}\!\left(\Lambda^{-1}x-y\right)\quad{\rm and}\quad {\cal B}_m(x)=0\,,
\eeq
with $\Lambda$ a constant invertible $4\times4$ matrix and ${\cal M}$ a constant invertible ($12\times12$) supermatrix. For the symmetry identities derived in this section to be valid, the 2PI approximation at hand must satisfy \Eqn{appeq:symac} for each symmetry under consideration. All global symmetries of QED listed above are respected by the 2PI loop-expansion at any approximation order.

\subsection{Global $U(1)$}

The global $U(1)$ symmetry of QED corresponds to the transformation \eqn{appeq:trsf}, \eqn{appeq:global} with $\Lambda=\mathds{1}_{4\times4}$ the $4\times4$ identity and ${\cal M}={\rm diag}\,(0_{4\times4},e^{i\alpha}\mathds{1}_{4\times4},e^{-i\alpha}\mathds{1}_{4\times4})$ where $\alpha$ is an arbitrary real number. In this case, \Eqn{appeq:2PIsym} implies that, for $\varphi=0$,
\beq
\frac{\delta^p\bSigma_{nm}}{\delta\varphi_p\cdots\delta\varphi_1}=0 \quad {\rm unless} \quad q_n+q_m+q_1+\cdots+q_p=0\,.
\eeq
Similarly, it follows from \Eqn{appeq:2PIressym} that 2PI-resummed vertices satisfy ($\varphi=0$)
\beq
\frac{\delta^n\Gamma}{\delta\varphi_n\cdots\delta\varphi_1}=0 \quad {\rm unless} \quad q_1+\cdots+q_n=0\,.
\eeq
More generally, since $\Gammaint[\varphi,\cG]$ is invariant, one concludes that any quantity having a nonvanishing fermion number vanishes for $\varphi=0$.\footnote{This result is extensively used in the previous section. Notice that the fermion number of a propagator is the opposite as that of a self-energy. For instance, the fermion number of $\delta\cG_{mn}/\delta\varphi_1$ is $q_m+q_n-q_1$: $\delta\bcG_{\psi A}/\delta\psi\neq 0$, whereas $\delta\bcG_{\psi A}/\delta\bar\psi= 0$ for $\varphi=0$.} 
For instance, $\delta^2\Gammaint/\delta\cG_{mn} \delta\varphi_1|_{\bcG}=0$ unless $q_1+q_m+q_n=0$, or $\delta^2\Gammaint/\delta\cG_{mn}\delta\cG_{rs}|_{\bcG}=0$ unless $q_m+q_n+q_r+q_s=0$ etc.

\subsection{Lorentz symmetry and parity}

Lorentz transformations correspond to \eqn{appeq:trsf}, \eqn{appeq:global} with $\Lambda$ the transformation matrix of space-time variables and ${\cal M}$ the bloc-diagonal supermatrix ${\cal M}={\rm diag}\,(\Lambda,S(\Lambda),[S^{-1}(\Lambda)]^t)$, where $S(\Lambda)$ is the Lorentz transformation matrix for Dirac spinors. In that case, Eqs.~\eqn{appeq:cov}-\eqn{appeq:2PIressym} lead to the following relations for the fermion and photon inverse propagators, in momentum space:
\beq\label{appeq:lor1}
 \bar D^{-1}(\Lambda p)=S^{-1}(\Lambda)\bar D^{-1}(p)S(\Lambda)\quad,\quad
 \bar G^{-1}_{\mu\nu}(\Lambda p)={\Lambda_\mu}^\rho{\Lambda_\nu}^\sigma\,\bar G^{-1}_{\rho\sigma}(p)
\eeq
and similarly for the 2PI-resummed fermion and photon two-point functions $\Gamma^{(2,0)}(p)$ and $\Gamma^{(0,2)}(p)$. Similarly, the three-point functions (see \Eqn{eq:mom3}) satisfy 
\beq\label{appeq:lor2}
 V^{(2,1)}_\mu(\Lambda p',\Lambda p)={\Lambda_\mu}^\nu \,S^{-1}(\Lambda)V^{(2,1)}_\nu(p',p)S(\Lambda)
\eeq
and similarly for the other 2PI three-point vertices $\tilde V^{(2,1)}=i\delta\bSigma_{\bar\psi A}/\delta\psi$ and $\delta\bSigma_{A\psi}/\delta\bar\psi$ as well as the 2PI-resummed one $\Gamma^{(2,1)}$.

Parity transformations can be cast in a similar form as Lorentz transformations with $\Lambda\to P\equiv{\rm diag}\,(1,-1,-1,-1)$ the space-time inversion matrix and $S(P)=\gamma_0$. Parity constraints on vertex functions can be obtained from the above equations with this simple replacement.

\subsection{Charge-conjugation}
Charge-conjugation takes the form \eqn{appeq:trsf}, \eqn{appeq:global} with $\Lambda=\mathds{1}_{4\times4}$ and 
\bea
 {\cal M}=\left(
 \begin{tabular}{ccc}
 $-\mathds{1}_{4\times4}$&$0$&$0$\\
 $0$&$0$&$C$\\
 $0$&$C^{-1}$&$0$
 \end{tabular}
 \right)\,,
\eea
where $C$ is the usual matrix for charge-conjugation of Dirac spinors, defined by the equation
$C\,\gamma_\mu^t\,C^{-1}=-\gamma_\mu$.\footnote{For simplicity, we use a basis where $C^t=C^\dagger=C^{-1}=-C$.} 

For $n$-photon vertex functions $V^{(0,n)}=\delta^{n-2}\bSigma_{AA}/\delta A^{n-2}|_{\varphi=0}$, the constraints \eqn{appeq:2PIsym} and \eqn{appeq:2PIressym} translate to ($\varphi=0$ implicit)
\beq
\frac{\delta^{n-2}\bSigma_{AA}}{\delta A_{n-2}\cdots\delta A_1}=(-1)^n\frac{\delta^{n-2}\bSigma_{AA}}{\delta A_{n-2}\cdots\delta A_1}\,,
\eeq
and similarly for 2PI-resummed photon vertices: $\Gamma^{(0,n)}=(-1)^n\Gamma^{(0,n)}$. This implies that 2PI and 2PI-resummed photon vertices with odd number of legs are identically zero as expected from Furry theorem. 

For vertices with fermionic legs, charge-conjugation symmetry relates functions with $\psi$- and $\bar\psi$-components exchanged. For instance, one obtains, for the fermion inverse propagator and the three-point vertex function in momentum space, see \eqn{eq:mom1}-\eqn{eq:mom3}:
\beq\label{appeq:chconj}
 \bar D^{-1}(p)=C\Big[\bar D^{-1}(-p)\Big]^tC^{-1} \quad {\rm and}\quad
 V^{(2,1)}_\mu(p',p)=-C\Big[V^{(2,1)}_\mu(-p,-p')\Big]^tC^{-1}
\eeq
and similarly for the 2PI-resummed two- and three-points vertex functions $\Gamma^{(2,0)}$ and $\Gamma^{(2,1)}$.Here, the transposition refers to Dirac indices only.  As for the other 2PI three-point functions, $\delta\bSigma_{\bar\psi A}/\delta\psi$ and $\delta\bSigma_{A\psi}/\delta\bar\psi$, one obtains \Eqn{eq:3p}. Equivalently, the latter reads, in momentum space, with obvious notations:
\beq\label{appeq:Crel}
\frac{\delta\bSigma_{A\psi}^{\mu\alpha}}{\delta\bar\psi_{\bar\alpha}}(p',p)=-C_{\alpha\bar\beta}\,\frac{\delta\bSigma_{\bar\psi A}^{\bar\beta\mu}}{\delta\psi_\beta}(-p,-p')\,C^{-1}_{\beta\bar\alpha}\,.
\eeq

\subsection{Lorentz structure of two- and three-point vertex functions and their divergences}
\label{appsec:lorstruct}
By construction, the momentum space fermion and photon inverse propagators have the following properties:\footnote{These follow from the fact that the 2PI effective action is real and from the (anti)commutating properties of the various field variables.}
\beq\label{appeq:rel}
 \Big[i\bar D^{-1}(p)\Big]^\dagger=\gamma_0\,i\bar D^{-1}(p)\,\gamma_0\quad{\rm and}\quad\Big[i\bar G^{-1}_{\mu\nu}(p)\Big]^*=i\bar G^{-1}_{\nu\mu}(p)=i\bar G^{-1}_{\mu\nu}(p)\,,
\eeq
and similarly for the 2PI-resummed fermion and photon two-point functions $\Gamma^{(2,0)}(p)$ and $\Gamma^{(0,2)}(p)$.
Here, the hermitic conjugation refers to Dirac indices only. Similarly, the 2PI and 2PI-resummed three-point vertex $V^{(2,1)}=\delta\bSigma_{\bar\psi\psi}/\delta A$ and  $\Gamma^{(2,1)}(p',p)$ behave, under hermitic conjugation, as
\beq\label{appeq:rel2}
 \Big[V^{(2,1)}_\mu(p',p)\Big]^\dagger=\gamma_0 \,V^{(2,1)}_\mu(p,p')\,\gamma_0\,,
\eeq
and similarly for  $\Gamma^{(2,1)}(p',p)$. 

Using the relations \eqn{appeq:rel} and the symmetry constraints \eqn{appeq:lor1} and \eqn{appeq:chconj}, it is a simple exercise to show that the fermion and photon two-point functions admit the following Dirac and Lorentz decompositions respectively:
\beq\label{appeq:structstruct}
 i\bar D^{-1}(p)=\bar d_0(p^2)\mathds{1}+\bar d_1(p^2)\slashchar{p}\quad{\rm and}\quad i\bar G^{-1}_{\mu\nu}(p)=\bar g_0(p^2)g_{\mu\nu}+\bar g_2(p^2)p_\mu p_\nu
\eeq
where $\bar d_0$, $\bar d_1$, $\bar g_0$ and $\bar g_2$ are real functions. Similarly, using \Eqn{appeq:lor2} for Lorentz and parity symmetries as well as charge-conjugation symmetry \Eqn{appeq:chconj} and the relation \eqn{appeq:rel2}, it is straightforward to check that the three point function $V^{(2,1)}_\mu$ admits the following Dirac decomposition:
\beq\label{appeq:lorstructvertex}
 V^{(2,1)}_\mu(p',p)=\bar V^S_\mu(p',p)\mathds{1}+\bar V^V_{\mu,\nu}(p',p)\gamma^\nu+i\bar V^A_{\mu,\nu}(p',p)\gamma_5\gamma^\nu+\frac{i}{2}\bar V^T_{\mu,\nu\rho}(p',p)\sigma^{\nu\rho}
\eeq
with $\sigma^{\nu\rho}=\frac{i}{2}[\gamma^\nu,\gamma^\rho]$, where ($q=p'+p$, $k=p'-p$)
\bea
 \bar V^S_\mu(p',p) &=& \bar \alpha_S q_\mu+\bar \delta_S k_\mu\,,\\
 \bar V^V_{\mu,\nu}(p',p) &=& \bar \alpha_V g_{\mu\nu}+\bar \beta_V q_\mu q_\nu+\bar \gamma_V k_\mu k_\nu
 +\bar \delta_V q_\mu k_\nu+\bar \epsilon_V k_\mu q_\nu\,,\\
 \bar V^A_{\mu,\nu}(p',p) &=& \bar \alpha_A\epsilon_{\mu\nu\rho\sigma}q^\rho k^\sigma\,,\\
\label{appeq:vertexstruct}
 \bar V^T_{\mu,\nu\rho}(p',p) &=& \bar \alpha_T(g_{\mu\nu}k_\rho-g_{\mu\rho}k_\nu)+\bar \beta_T q_\mu(q_\nu k_\rho-k_\nu q_\rho)\nn
&+&\bar \delta_T(g_{\mu\nu}q_\rho-g_{\mu\rho}q_\nu)+\bar \epsilon_T k_\mu(q_\nu k_\rho-k_\nu q_\rho)\,,
 \eea
where the form factors $\bar \alpha_S\equiv\bar \alpha_S(x,y,z)$, $\bar \delta_S\equiv\bar \delta_S(x,y,z)$ etc. are all real functions of $x=q^2$, $y=k^2$ and $z=q\cdot k$. The $\bar \alpha$'s, $\bar \beta$'s and $\bar \gamma$'s are even functions of $z$, whereas the $\bar \delta$'s and $\bar \epsilon$'s are odd in $z$.\footnote{Projecting the QED vertex on the fermion mass shell by means of the Dirac spinor $u(p)$ and $\bar u(p)$, right and left eigenvectors of $\slashchar{p}$ with eigenvalue $m$, and using standard Gordon identities, one recovers the usual result:
$$
 \bar u(p')V^{(2,1)}_\mu(p',p)u(p)= \bar u(p')\left[\bar F_1(k^2)\gamma_\mu+i\bar F_2(k^2)\sigma_{\mu\nu}k^\nu\right]u(p)
$$
where the electric and magnetic form factors read
\bea
\bar  F_1(k^2)&=&\bar \alpha_V(k^2) +2m\bar \alpha_S(k^2)  +4m^2\bar \beta_V (k^2) +k^2\bar \alpha_A(k^2)+mk^2\bar \beta_T(k^2)\nn
\bar  F_2(k^2)&=&-\bar \alpha_S(k^2) -2m\bar \beta_V(k^2)-2m\bar \alpha_A(k^2) +\bar \alpha_T(k^2)-2k^2\bar \beta_T(k^2)\nonumber
\eea
where $\bar \alpha_V(y)\equiv\bar\alpha_V(4m^2-y,y,0)$ etc.}
The 2PI-resummed two- and three-point vertex functions $\Gamma^{(2,0)}(p)$, $\Gamma^{(0,2)}_{\mu\nu}(p)$ and $\Gamma^{(2,1)}_\mu(p',p)$ admit similar decompositions with corresponding form factors $d_0,\dots,\epsilon_T$. 

The 2PI Ward-Takahashi identity \eqn{eq:2PIWIFourier1} implies (for $p$ and $p'$ noncollinear)
\bea
z\bar \alpha_S(x,y,z)+y\bar \delta_S(x,y,z)&=&- e\left[\bar d_0(p'^2)-\bar d_0(p^2)\right]\,,\\
 z\bar \beta_V(x,y,z) + y\bar \epsilon_V(x,y,z) &=& -\frac{e}{2}\left[\bar d_1(p'^2)-\bar d_1(p^2)\right]\,,\\
\label{appeq:symidentityimply}
 \bar \alpha_V(x,y,z) + y\bar \gamma_V(x,y,z) + z\bar \delta_V(x,y,z) &=& -\frac{e}{2}\left[\bar d_1(p'^2)+\bar d_1(p^2)\right]
\eea
and
\beq
 z\bar \beta_T(x,y,z)+y\bar \epsilon_T(x,y,z)-\bar \delta_T(x,y,z)=0\,.
\eeq
Note that, for $p'^2=p^2$ and $k^2 =0$ (i.e. $y=z=0$), in particular for on-shell momenta, these relations boil down to:
\beq\label{appeq:simpler}
 \bar \alpha_V(4p^2,0,0)=-e\bar d_1(p^2)\,.
\eeq
Similar relations hold for the form factors of 2PI-resummed two- and three-point functions $\Gamma^{(2,0)}(p)$ and $\Gamma^{(2,1)}(p',p)$, which satisfy a similar Ward-Takahashi identity, see Eqs. \eqn{eq:2PIWIFourier2} and \eqn{eq:cafaitzero}.
The 2PI-resummed two-photon function $\Gamma^{(0,2)}(p)$ is further constrained by gauge symmetry, see \Eqn{eq:transmom}, which implies that $g_0(p^2)=-p^2 g_2(p^2)$.

It follows from Eqs. \eqn{appeq:structstruct}-\eqn{appeq:vertexstruct} and the fact that, once all subdivergences have been eliminated, the global divergence of a given graph is a polynomial (in momentum space) of degree equal to the superficial degree of divergence of the graph \cite{Weinberg}, that the global divergences -- denoted by $[\dots]_{\bar\infty}$ -- of the two- and three-point functions $i\bar D^{-1}$, $i\bar G^{-1}$ and $V^{(2,1)}$ have the following structure:
\beq
 \left[i\bar D^{-1}(p)\right]_{\bar\infty}=\bar a_m\mathds{1}+\bar a_z\slashchar{p}\quad{\rm and}\quad\left[i\bar G^{-1}_{\mu\nu}(p)\right]_{\bar\infty}=(\bar a_M+\bar a_Zp^2)g_{\mu\nu}+a_\lambda p_\mu p_\nu
\eeq
and
\beq
 \left[V^{(2,1)}_\mu(p',p)\right]_{\bar\infty}=\bar a_v\gamma_\mu
\eeq
where $\bar a_m$, $\bar a_z$, $\bar a_M$, $\bar a_Z$, $\bar a_\lambda$ and $\bar a_v$ are real constants. The symmetry identity \eqn{appeq:symidentityimply}, or \eqn{appeq:simpler}, implies that $\bar a_v=-e\bar a_z$, from which one deduces, in particular, that $\bar Z_1/\bar Z_2$ is finite, see \Eqn{eq:equalWard}. The global divergences of 2PI-resummed two- and three-point functions $\Gamma^{(2,0)}(p)$, $\Gamma^{(0,2)}(p)$ and $\Gamma^{(2,1)}(p',p)$ are constrained in a similar way with corresponding constants $a_m$, $a_z$, $a_M$, $a_Z$, $a_\lambda$ and $a_v$. The symmetry identity \eqn{eq:transmom} further implies that, for the 2PI-resummed two-photon function, $a_M= a_Z+a_\lambda=0$.

Finally, similarly to \Eqn{appeq:rel2}, the two remaining 2PI three-point vertex $\delta\bSigma_{\bar\psi A}^\mu/\delta\psi$ and $\delta\bSigma_{A\psi}^\mu/\delta\bar\psi$ are related by
\beq\label{appeq:rel3}
\left(\frac{\delta\bSigma_{A\psi}^{\mu\alpha}}{\delta\bar\psi_{\bar\alpha}}(p',p)\right)^{\!\!*}=\gamma^0_{\alpha\bar\beta}\,\frac{\delta\bSigma_{\bar\psi A}^{\bar\beta\mu}}{\delta\psi_{\beta}}(p,p')\,\gamma^0_{\beta\bar\alpha}\,.
\eeq
It follows from this relation as well as Lorentz, parity, and charge-conjugation symmetries, see Eqs. \eqn{appeq:lor2} and \eqn{appeq:Crel}, that they admit a similar decomposition as \eqn{appeq:lorstructvertex}-\eqn{appeq:vertexstruct} with corresponding real form factors $\tilde\alpha^{(1)},\dots,\tilde\epsilon^{(1)}$ and $\tilde\alpha^{(2)},\dots,\tilde\epsilon^{(2)}$ respectively. However, the latter are neither even nor odd functions of $z$ but are, instead, related by Eqs. \eqn{appeq:Crel} and \eqn{appeq:rel3} as e.g. $\tilde\alpha_V^{(1)}(x,y,z)=\tilde\alpha_V^{(2)}(x,y,-z)$, and similarly for all $\tilde\alpha$'s, $\tilde\beta$'s and $\tilde\gamma$'s, whereas $\tilde\delta$'s and $\tilde\epsilon$'s are such that e.g. $\tilde\delta_V^{(1)}(x,y,z)=-\tilde\delta_V^{(2)}(x,y,-z)$. It follows that their global divergence is given by 
\beq
 \left[\frac{\delta\bSigma_{A\psi}^{\mu}}{\delta\bar\psi}(p',p)\right]_{\bar\infty}=\left[\frac{\delta\bSigma_{\bar\psi A}^{\mu}}{\delta\psi}(p',p)\right]_{\bar\infty}=\tilde a_v\gamma^\mu\,,
\eeq
with $\tilde a_v$ a real constant.

\section{On-mass-shell renormalization conditions}
\label{appsec:Physicalon-mass-shellrenormalizationconditions}

Here, we consider a more conventional set of renormalization conditions than the one considered in \Sec{sec:rencond}, where $m_R$ and $e_R$ are identified with the physical fermion mass and electromagnetic charge. We show, in particular, that this is also a gauge-symmetric set of renormalization conditions in the sense that it leads to \Eqn{eq:cafaitzero} and, therefore, all 2PI Ward-Takahashi identities are exactly preserved after renormalization at any approximation order.

The renormalization points are chosen such that photon ($k_*$) and fermion ($p_*$) momenta are on their respective mass-shell: $k_*^2=0$ and $p_*^2=m_R^2$. It is convenient to consider the 2PI-resummed and 2PI fermion self-energies $\Sigma_R(p)$ and $\bSigma_R(p)$ as functions of the Dirac matrix $\slashchar{p}$:
\beq
 \Sigma_{R}(p)\equiv\Sigma_R(\slashchar{p})
\eeq
and similarly for $\bSigma_R(p)$, where we keep the same letter on both sides of the equation for simplicity and make sure that no ambiguity is possible in the following. Moreover, the 2PI-resummed three-point vertex projected on physical fermion states reads, see \Sec{appsec:lorstruct},
\beq
 \bar u(p')\Gamma^{(2,1)}_{R\mu}(p',p)u(p)= \bar u(p')\left[F_1(k^2)\gamma_\mu+iF_2(k^2)\sigma_{\mu\nu}k^\nu\right]u(p)\,,
\eeq
where the Dirac spinor $u(p)$ is a solution of the Dirac equation $(\slashchar{p}-m_R)u(p)=0$ normalized as $\bar u(p) u(p)=1$. The functions $F_1$ and $F_2$ are the standard electric and --  up to a factor $2m_R$ -- magnetic form factors of the fermion. In particular, $F_1(0)$ is the total electric charge of the latter. It can be obtained as:
\beq
 F_1(0) = \frac{p^\mu}{m_R}\,\bar u(p)\Gamma^{(2,1)}_{R\mu}(p,p)u(p)\,.
\eeq

Standard  on-mass-shell renormalization conditions are formulated as follows:
\beq
 \left.\frac{d\Pi_{R}^T(k^2)}{dk^2}\right|_{k^2=0}\!\!\!=0\,,\,\,\,\Sigma_R(\slashchar{p}=m_R)=0\,,\,\,\,\left.\frac{\partial\Sigma_R(\slashchar{p})}{\partial\slashchar{p}}\right|_{\slashchar{p}=m_R}\!\!\!=0\,\,\,{\rm and}\,\,\,F_1(0)=-e_R\,.
\eeq
The corresponding consistency conditions read:
\beq\label{appeq:renG1}
 \left.\frac{d\bar\Pi_{R}^T(k^2)}{dk^2}\right|_{k^2=0}=\left.\frac{d\Pi_R^T(k^2)}{dk^2}\right|_{k^2=0}\,,
\eeq
\beq
 \bSigma_R(\slashchar{p}=m_R)=\Sigma_R(\slashchar{p}=m_R)\quad,\quad
 \left.\frac{\partial\bSigma_R(\slashchar{p})}{\partial\slashchar{p}}\right|_{\slashchar{p}=m_R}=\left.\frac{\partial\Sigma_R(\slashchar{p})}{\partial\slashchar{p}}\right|_{\slashchar{p}=m_R}
\eeq
and 
\beq
  \bar F_1(0)=\tilde F_1(0)=F_1(0)
\eeq
where $\bar F_1(k^2)$ and $\tilde F_1(k^2)$ are the electric form factors corresponding to the renormalized 2PI vertices $V^{(2,1)}_{R}$ and $\tilde V^{(2,1)}_{R}$ respectively.

Writing the 2PI Ward-Takahashi identity \eqn{eq:2PIWIFourier1} in the limit $k\to 0$, one gets the 2PI Ward identity
\beq
 -\frac{\bar Z_2}{\bar Z_1}\frac{1}{e_R}\bar F_1(0)=1-i\left.\frac{\partial\bSigma_R(\slashchar{p})}{\partial\slashchar{p}}\right|_{\slashchar{p}=m_R}\,.
\eeq
from which it follows, using the above renormalization and consistency conditions, that $\bar Z_1=\bar Z_2$. Similarly, we obtain, from \Eqn{eq:2PIWIFourier2},
\beq
 -\frac{\bar Z_2}{\bar Z_1}\frac{1}{e_R}F_1(0)=1-\Delta_2+\Delta_1\frac{\bar Z_2}{\bar Z_1}-i\left.\frac{\partial\Sigma_R(\slashchar{p})}{\partial\slashchar{p}}\right|_{\slashchar{p}=m_R}
\eeq
It follows that $\bar Z_1(1-\Delta_2)=\bar Z_2(1-\Delta Z_1)$, and then, because $\bar Z_1=\bar Z_2$, that $Z_1=Z_2$, as announced.

\section{Vertex functions in the 2PI superfield formalism}
\label{appsec:2PI_and_2PI_resummed_vertices}

We provide general tools to deal with 2PI techniques in the superfield formalism and use them to derive general expressions for various vertex functions. As usual, 2PI-resummed vertices can be expressed explicitly in terms of 2PI vertices, which fulfill self-consistent equations \cite{Berges:2005hc}. Our analysis is greatly simplified by the use of a generalized Leibniz rule for computing multiple derivatives of products of supermatrix.

\subsection{Generalized Leibniz rule}
\label{appsubsec:Leibniz_formula}

The (right) superfield-derivative of the product of two supermatrices, say ${\cal A}$ and ${\cal B}$ reads, explicitly,
\beq\label{eq:AB1}
\frac{\delta}{\delta\varphi_1}({\cal A}{\cal B})_{mn}=\frac{\delta{\cal A}_{mp}}{\delta\varphi_1}{\cal
B}_{pn}+(-1)^{q_1(q_{\cal A}+q_m+q_p)}{\cal A}_{mp}\frac{\delta{\cal B}_{pn}}{\delta\varphi_1}\,,
\eeq
where $q_j$ denotes the fermion numbers associated with the superindex $j=1,m,p$ and $q_{\cal A}$ accounts for possible field derivatives in the matrix ${\cal A}$. For instance, if $\smash{{\cal A}=\bar\cG}$, $\smash{q_{\cal A}\equiv 0}$ whereas if $\smash{{\cal A}=\delta\bar\cG/\delta \varphi_r}$, $\smash{q_{\cal A}=-q_r}$ etc. It is convenient to rewrite Eq.~(\ref{eq:AB1}) in a component-independent form. To this purpose, we notice that
\beq
(-1)^{q_1(q_m+q_p)}{\cal A}_{mp}=\left(\sigma^{q_1}{\cal A}\,\sigma^{q_1}\right)_{mp}\,,
\eeq
where $\sigma$ is the bloc-diagonal supermatrix $\smash{\sigma\equiv{\rm diag}\,(\mathds{1}_{4\times 4},-\mathds{1}_{4\times 4},-\mathds{1}_{4\times 4})}$. Defining now the
superderivative of a supermatrix as
\beq
\frac{\delta_s}{\delta \varphi_1}{\cal A}\equiv\frac{\delta}{\delta\varphi_1}(\sigma^{q_1}{\cal A})\,,
\eeq
it is possible to rewrite Eq.~(\ref{eq:AB1}) as
\beq\label{eq:AB2}
\frac{\delta_s}{\delta\varphi_1}({\cal A}{\cal B})=\frac{\delta_s{\cal A}}{\delta\varphi_1}{\cal
B}+(-1)^{q_1q_{\cal A}}{\cal A}\frac{\delta_s{\cal B}}{\delta\varphi_1}\,,
\eeq
where the sign factor $(-1)^{q_1q_{\cal A}}$ appears because the superderivative $\delta_s/\delta\varphi_1$ goes through the supermatrix ${\cal A}$ before to act on ${\cal B}$. Repeated use of Eq.~(\ref{eq:AB2})
leads to the generalized Leibniz rule for the $k$-th superderivative:
\beq\label{eq:Leibniz}
\frac{\delta^k_s}{\delta\varphi_k\cdots\delta\varphi_1}({\cal A}{\cal
B})=\sum_{J}(-1)^{\delta'_{J}}\frac{\delta^p_s{\cal
A}}{\delta\varphi_{i_p}\cdots \delta\varphi_{i_1}}\frac{\delta_s^q{\cal
B}}{\delta\varphi_{j_q}\cdots\delta\varphi_{j_1}}\,,
\eeq
where the sum runs over all possible subsets $J\equiv\{j_1,\ldots,j_q\}$ of $\{1,\ldots, k\}$ such that $\smash{j_1<\cdots<j_q}$. We denote by $I\equiv\{i_1,\ldots,i_p\}$ with $i_1<\ldots<i_p$ the complementary subset ($k=p+q$)\footnote{Either $I$ or $J$ can be the empty set, in which case the corresponding derivative is the identity operator.}: $I\cup J=\{1,\ldots,k\}$. A sign factor arises each time a superderivative $\delta/\delta\varphi_j$ goes through the supermatrix $\delta^p_s{\cal A}/\delta\varphi_{i_p}\cdots \delta\varphi_{i_1}$. This produces a factor $(-1)^{q_jq_{\cal A}}$ as well as a factor $(-1)^{q_jq_i}$ for each $i<j$:
\beq\label{eq:rule}
\delta'_{J}=q_{\cal A}q_J-\sum_{\substack{j\in J, i\in I\\i<j}}q_jq_i\,,
\eeq
where we introduced the notation $q_J=\sum_{j\in J}q_j$.
The relative sign in \eqn{eq:rule} makes no difference and is chosen negative for later convenience. In order to express formula (\ref{eq:Leibniz}) in terms of normal derivatives we use the obvious relation
\beq
\frac{\delta^k}{\delta\varphi_k\cdots\delta\varphi_1}=\sigma^{q_1+\cdots+q_k}\frac{\delta^k_s}{\delta\varphi_k\cdots\delta\varphi_1}
\eeq
to obtain
\bea\label{eq:Leibniz2}
\frac{\delta^k}{\delta\varphi_k\cdots\delta\varphi_1}({\cal A}{\cal
B})=\sum_{J}(-1)^{\delta'_{J}}\sigma^{q_J}\frac{\delta^p{\cal
A}}{\delta\varphi_{i_p}\cdots \delta\varphi_{i_1}}\sigma^{q_J}\frac{\delta^q{\cal
B}}{\delta\varphi_{j_q}\cdots\delta\varphi_{j_1}}\,.
\eea
The presence of the matrices $\sigma$ is
cumbersome. In practice however, we shall eventually be interested in evaluating derivatives for physical values of the fields, in which case the $U(1)$ symmetry of the theory implies that, see \App{appsec:charge_conjugation_invariance},
\beq\label{eq:U1}
\frac{\delta^p{\cal A}_{mn}}{\delta\varphi_{i_p}\cdots\varphi_{i_1}}=0\quad{\rm if}\quad q_{\cal A}+q_m+q_n+q_I\quad{\rm is\,\,odd}.
\eeq
It follows that
\beq
(-1)^{q_m+q_n}\frac{\delta^p{\cal
A}_{mn}}{\delta\varphi_{i_p}\cdots\varphi_{i_1}}=(-1)^{q_{\cal A}+q_I}\frac{\delta^p{\cal
A}_{mn}}{\delta\varphi_{i_p}\cdots\varphi_{i_1}}\,,
\eeq
or, in matrix notation,
\beq
\sigma\,\frac{\delta^p{\cal
A}}{\delta\varphi_{i_p}\cdots\varphi_{i_1}}\,\sigma=(-1)^{q_{\cal A}+q_I}\frac{\delta^p{\cal A}}{\delta\varphi_{i_p}\cdots\varphi_{i_1}}\,.
\eeq
Thus, provided that one keeps track of the sign factors, one can get rid of the matrices $\sigma$. We obtain, explicitly,
\beq\label{eq:Leibniz3}
\frac{\delta^k}{\delta\varphi_k\cdots\delta\varphi_1}({\cal A}{\cal
B})=\sum_{J}(-1)^{\delta_{J}}\frac{\delta^p{\cal
A}}{\delta\varphi_{i_p}\cdots \delta\varphi_{i_1}}\frac{\delta^q{\cal
B}}{\delta\varphi_{j_q}\cdots\delta\varphi_{j_1}}\,,
\eeq
with $\delta_{J}=\delta'_{J}+q_J(q_I+q_{\cal A})$, that is
\beq\label{eq:rule2}
\delta_{J}=\sum_{\substack{j\in J, i\in I\\i>j}}q_jq_i\,.
\eeq
Equation \eqn{eq:rule2} has a simple interpretation: One adds a contribution $q_i q_j$ to
$\delta_{J}$ for each pair of derivatives $\delta/\delta\varphi_i$ and $\delta/\delta\varphi_j$, not acting on the same factor (${\cal A}$ or ${\cal B}$) and appearing in the same order as on the left-hand-side of Eq.~(\ref{eq:Leibniz3}). For instance, for $k=2$, we get
\beq
\frac{\delta^2}{\delta\varphi_2\delta\varphi_1}({\cal A}{\cal B})=\frac{\delta^2{\cal
A}}{\delta\varphi_2\delta\varphi_1}{\cal B}+(-1)^{q_1q_2}\frac{\delta{\cal
A}}{\delta\varphi_2}\frac{\delta{\cal B}}{\delta\varphi_1}+\frac{\delta{\cal
A}}{\delta\varphi_1}\frac{\delta{\cal B}}{\delta\varphi_2}+{\cal A}\frac{\delta^2{\cal
B}}{\delta\varphi_2\delta\varphi_1}\,,
\eeq
which is easily checked to be correct by a direct calculation.

\subsection{2PI-resummed vertex functions}
\label{appsubsec:application_2PI_resummed_vertices}

The 2PI-resummed vertex functions are defined as field derivatives of the 2PI-resummed effective action evaluated at
vanishing fields. Taking a field derivative on Eq.~(\ref{eq:first}), we obtain the 2PI-resummed two-point vertex function as
\beq\label{eq:2PIr2app}
\frac{\delta^2\Gamma}{\delta\varphi_2\delta\varphi_1}=(-1)^{q_1}i\cG^{-1}_{0,21}+\left.\frac{\delta^2\Gammaint}{\delta\varphi_2\delta\varphi_1}\right|_{\bcG}+\frac{\delta\bar\cG_{mn}}{\delta\varphi_2}\left.\frac{\delta^2\Gammaint}{\delta\cG_{mn}\delta\varphi_1}\right|_{\bcG}\,,
\eeq
where it is understood that derivatives have to be evaluated at $\varphi=\bar\varphi=0$. Similarly, one obtains the 2PI-resummed three-point vertex function
\beq\label{eq:2PIr3}
\frac{\delta^3\Gamma}{\delta\varphi_3\delta\varphi_2\delta\varphi_1}=\left.\frac{\delta^3\Gammaint}{\delta\varphi_3\delta\varphi_2\delta\varphi_1}\right|_{\bcG}+\frac{\delta^2\bar\cG_{mn}}{\delta\varphi_3\delta\varphi_2}\left.\frac{\delta^2\Gammaint}{\delta\cG_{mn}\delta\varphi_1}\right|_{\bcG}\,,
\eeq
as well as higher ($k\geq 4$) 2PI-resummed vertex functions
\beq\label{eq:2PIrn}
\frac{\delta^k\Gamma}{\delta\varphi_k\cdots\delta\varphi_1}=\frac{\delta^{k-1}\bar\cG_{mn}}{\delta\varphi_k\cdots\delta\varphi_2}\left.\frac{\delta^2\Gammaint}{\delta\cG_{mn}\delta\varphi_1}\right|_{\bcG}\,.
\eeq
Notice that, in deriving these expressions, we have assumed that the interaction term of the classical action is cubic in the superfield. It follows in particular that $\delta^3\Gammaint/\delta\varphi^3$ and $\delta^2\Gammaint/\delta\cG\delta\varphi$ are $\varphi$- and $\cG$-independent.

From the above formulae, it follows that the determination of 2PI-resummed vertex functions essentially amounts to the evaluation of field derivatives of $\bar\cG$ at vanishing fields. The latter can be directly related to 2PI vertex functions, that is derivatives of $\bar\Sigma$ evaluated at vanishing fields, see \eqn{eq:V}. Using Eq.~(\ref{eq:Leibniz3}) with ${\cal A}=\bar\cG^{-1}$ and ${\cal B}={\bar\cG}$ and extracting explicitly the term corresponding to $\smash{J=\{1,\cdots,k\}}$ (i.e. $\smash{I=\emptyset}$), we get
\beq\label{eq:recursion}
\frac{\delta^k\bar\cG}{\delta\varphi_k\cdots\delta\varphi_1}=\sum_{I\neq\emptyset,J}(-1)^{\delta_{J}}\bar\cG\frac{\delta^p\bSigma}{\delta\varphi_{i_p}\cdots
\delta\varphi_{i_1}}\frac{\delta^q\bar\cG}{\delta\varphi_{j_q}\cdots\delta\varphi_{j_1}}\,,
\eeq
where we used \Eqn{eq:stat} on the RHS. Here again derivatives are understood to be taken at  $\varphi=\bar\varphi=0$.

\Eqn{eq:recursion} is a recursion formula which expresses the $k^{\rm th}$ derivative of $\bar\cG$
in terms of its $q^{\rm th}$ derivatives with $\smash{q\le k-1}$. Successive iterations yield
\beq\label{appeq:Gn}
\frac{\delta^k\bar\cG}{\delta\varphi_k\cdots\delta\varphi_1}=\sum_{\left\{I_r\right\}}(-1)^{\delta_{\left\{I_r\right\}}}\,\bar\cG\,\prod_{r=1}^R\left(\frac{\delta^{p_r}\bar\Sigma}{\delta\varphi_{i^{(r)}_{p_r}}\cdots\delta\varphi_{i^{(r)}_1}}\,\bar\cG\right)\,,
\eeq
where the sum runs over all possible families $\smash{\left\{I_r\right\}\equiv\{I_1,\ldots,I_R\}}$, $\smash{1 \le R \le k}$, of nonempty disjoint subsets of $\left\{1,\ldots,k\right\}$ such that $\smash{\cup_{r=1}^R
I_r=\left\{1,\ldots,k\right\}}$. The elements of a given subset $I_r$ are denoted by $I_r\equiv\{i^{(r)}_1,\ldots,i^{(r)}_{p_r}\}$ with the convention that $\smash{i^{(r)}_1<\cdots<i^{(r)}_{p_r}}$. One has the relation $k=\sum_{r=1}^R p_r$. Finally, the exponent $\delta_{\{I_r\}}$ is computed as follows: One adds a contribution
$q_sq_t$ to $\delta_{\{I_r\}}$ for each pair of derivatives $\delta/\delta\varphi_s$ and $\delta/\delta\varphi_t$ acting on different $\bar\Sigma$'s and appearing in the same order as on the LHS of Eq.~(\ref{appeq:Gn}).

\Eqn{appeq:Gn} has a simple interpretation. In order to express the $k^{\rm th}$-derivative of $\bar\cG$ solely in terms of derivatives of $\bSigma$, one needs to sum up, with appropriate sign factors, all the possible ways to alternate $\bar \cG$'s and derivatives of $\bar\Sigma$ such that: 1) each product starts and ends by a factor $\bar\cG$ and contains $k$ derivatives; 2) the derivatives acting on
a given factor $\bar\Sigma$ keep the same order as in the left-hand-side of
Eq.~(\ref{appeq:Gn}); 3) each pair of derivatives $\delta/\delta\varphi_s$ and $\delta/\delta\varphi_t$ acting on different factors $\bSigma$ and appearing in the same order as in the left-hand-side of Eq.~(\ref{appeq:Gn}) contributes a sign $(-1)^{q_sq_t}$ to the corresponding contribution. For $\smash{k=1}$, one get the trivial result:
\beq\label{appeq:trivial}
\frac{\delta\bar\cG}{\delta\varphi_1}=\bar\cG\,\frac{\delta\bSigma}{\delta\varphi_1}\,\bar\cG\,,
\eeq
which leads to
\beq\label{appeq:A21}
\frac{\delta^2\Gamma}{\delta\varphi_2\delta\varphi_1}=(-1)^{q_1}i\cG^{-1}_{0,21}+\left.\frac{\delta^2\Gammaint}{\delta\varphi_2\delta\varphi_1}\right|_{\bcG}+\left(\bar\cG\,\frac{\delta\bSigma}{\delta\varphi_2}\,\bar\cG\right)_{\!\!mn}\left.\frac{\delta^2\Gammaint}{\delta\cG_{mn}\delta\varphi_1}\right|_{\bcG}\,.
\eeq
Similarly, for $\smash{k=2}$, we obtain
\beq
\frac{\delta^2\bar\cG}{\delta\varphi_2\delta\varphi_1}=\bar\cG\,\frac{\delta^2\bar\Sigma}{\delta\varphi_2\delta\varphi_1}\,\bar\cG+(-1)^{q_1q_2}\bar\cG\,\frac{\delta\bar\Sigma}{\delta\varphi_2}\,\bar\cG\frac{\delta\bar\Sigma}{\delta\varphi_1}\,\bar\cG+\bar\cG\,\frac{\delta\bar\Sigma}{\delta\varphi_1}\,\bar\cG\frac{\delta\bar\Sigma}{\delta\varphi_2}\,\bar\cG\,,
\eeq
which leads to
\beq
\frac{\delta^3\Gamma}{\delta\varphi_3\delta\varphi_2\delta\varphi_1}=
\left.\frac{\delta^3\Gammaint}{\delta\varphi_3\delta\varphi_2\delta\varphi_1}\right|_{\bcG}+\left(\bar\cG\,\frac{\delta^2\bar\Sigma}{\delta\varphi_3\delta\varphi_2}\,\bar\cG+2\,\bar\cG\,\frac{\delta\bar\Sigma}{\delta\varphi_2}\,\bar\cG\frac{\delta\bar\Sigma}{\delta\varphi_3}\,\bar\cG\right)_{\!\!mn}\left.\frac{\delta^2\Gammaint}{\delta\cG_{mn}\delta\varphi_1}\right|_{\bcG}.
\eeq
Here, we have used the property
\beq
(-1)^{q_2q_3}\left(\bar\cG\,\frac{\delta\bar\Sigma}{\delta\varphi_3}\,\bar\cG\frac{\delta\bar\Sigma}{\delta\varphi_2}\,\bar\cG\right)_{mn}\frac{\delta^2\Gammaint}{\delta\cG_{mn}\delta\varphi_1}=\left(\bar\cG\,\frac{\delta\bar\Sigma}{\delta\varphi_2}\,\bar\cG\frac{\delta\bar\Sigma}{\delta\varphi_3}\bar\cG\right)_{mn}\frac{\delta^2\Gammaint}{\delta\cG_{mn}\delta\varphi_1}\,,
\eeq
which follows from $U(1)$ global symmetry and the symmetry relations
\beq
\bcG_{mn}=(-1)^{q_mq_n}\bcG_{nm} \quad {\rm and} \quad \bSigma_{mn}=(-1)^{q_mq_n+q_m+q_n}\bSigma_{nm}\,.
\eeq
Higher 2PI-resummed vertex functions are obtained in a similar way by plugging Eq.~(\ref{appeq:Gn}) into Eq.~(\ref{eq:2PIrn}), that is
\beq\label{appeq:gamma_n}
\frac{\delta^k\Gamma}{\delta\varphi_k\cdots\delta\varphi_1}=\sum_{\left\{I_r\right\}}(-1)^{\delta_{\left\{I_r\right\}}}\,\left(\bar\cG\,\prod_{r=1}^R\left(\frac{\delta^{p_r}\bar\Sigma}{\delta\varphi_{i^{(r)}_{p_r}}\cdots\delta\varphi_{i^{(r)}_1}}\,\bar\cG\right)\right)_{\!\!mn}\left.\frac{\delta^2\Gammaint}{\delta\cG_{mn}\delta\varphi_1}\right|_{\bar\cG}\,.
\eeq
As announced, all 2PI-resummed vertex function are now explicitly expressed in terms of 2PI vertex functions.

\subsection{2PI vertex functions}
\label{appsubsec:application_2PI_vertices}

2PI vertex functions are defined as field derivatives of $\bSigma[\varphi]$ evaluated at vanishing fields, see \eqn{eq:V}. To obtain self-consistent equations for the latter, we start from the definition
\beq
(-1)^{q_m}\bSigma_{nm}[\varphi]=2i\left.\frac{\delta\Gammaint}{\delta\cG_{mn}}\right|_{\bar\cG[\varphi]}\,.
\eeq
When evaluating field derivatives of the RHS, one has to take into account both the explicit and implicit -- through $\bcG[\varphi]$ -- $\varphi$-dependences of $\delta\Gammaint/\delta\cG|_{\bcG}$. One gets, for the first derivative, 
\beq\label{eq:lisa}
\frac{\delta}{\delta\varphi_1}\left[\frac{\delta\Gammaint}{\delta\cG_{mn}}\right]_{\bcG}=\left.\frac{\delta^2\Gammaint}{\delta\varphi_1\delta\cG_{mn}}\right|_{\bar\cG}+\frac{\delta\bar\cG_{ab}}{\delta\varphi_1}\!\left.\frac{\delta^2\Gammaint}{\delta\cG_{ab}\delta\cG_{mn}}\right|_{\bar\cG}\,,
\eeq
where the notation on the LHS stresses the fact that one takes a total $\varphi$-derivative.
Using \Eqn{appeq:trivial}, one obtains the following self-consistent equation for the 2PI three-point function $\delta\bSigma/\delta\varphi$ at $\varphi=0$:
\beq\label{eq:abc}
\frac{\delta\bSigma_{nm}}{\delta\varphi_1}(-1)^{q_m}=\left.\frac{2i\,\delta^2\Gammaint}{\delta\varphi_1\delta\cG_{mn}}\right|_{\bar\cG}+\left(\bcG\,\frac{\delta\bSigma}{\delta\varphi_1}\,\bcG\right)_{\!\!ab}\left.\frac{2i\,\delta^2\Gammaint}{\delta\cG_{ab}\delta\cG_{mn}}\right|_{\bar\cG}\,.
\eeq
To obtain the corresponding equations for higher 2PI vertex functions, we consider total
derivatives of Eq.~(\ref{eq:lisa}) and use the previously derived Leibniz rule, \Eqn{eq:Leibniz3}. Using similar notations as in Eqs.~\eqn{eq:Leibniz}-\eqn{eq:Leibniz3}, we get (derivatives are eventually evaluated at $\smash{\varphi=0}$)
\beq\label{eq:relkernels}
\frac{\delta^k}{\delta\varphi_k\cdots\delta\varphi_1}\left[\frac{\delta\Gammaint}{\delta\cG}\right]_{\bcG}={\sum_{J}}^*(-1)^{\delta_{J}}\,\frac{\delta^{p}\bar\cG_{ab}}{\delta\varphi_{i_p}\cdots\delta\varphi_{i_1}}\,\frac{\delta^q}{\delta\varphi_{j_q}\cdots\delta\varphi_{j_1}}\left[\frac{\delta^2\Gammaint}{\delta\cG_{ab}\delta\cG_{mn}}\right]_{\bar\cG}\,,
\eeq
where $\sum_J^*$ means that the sum is restricted on subsets $J\equiv\{j_1,\ldots,j_q\}$ of $\{1,\ldots,k\}$ such that the complementary subset $I\equiv\{i_1,\ldots,i_p\}$ always contains the element $i_1=1$.\footnote{In other words, $J$ is a subset of $\{2,\ldots,k\}$ such that $I\cup J=\{1,\ldots,k\}$.} This is due to the fact that the first derivative $\delta/\delta\varphi_1$ always acts on the first factor under the sum on the RHS. This formula relates the $k^{th}$ derivative of $\delta\Gammaint/\delta\cG|_{\bcG}$ to lower order ($q\le k-1$) derivatives of the 2PI kernel $\delta^2\Gammaint/\delta\cG^2|_{\bcG}$. Following a similar procedure, the latter can be expressed in terms of total $\varphi$-derivatives of the higher 2PI kernel $\delta^3\Gammaint/\delta\cG^3|_{\bcG}$ etc. In fact, \Eqn{eq:relkernels} still holds if one replaces 
$\delta\Gammaint/\delta\cG|_{\bcG}\to\delta^p\Gammaint/\delta\cG^p|_{\bcG}$ on the LHS and $\delta^2\Gammaint/\delta\cG^2|_{\bcG}\to\delta^{p+1}\Gammaint/\delta\cG^{p+1}|_{\bcG}$ on the RHS.

This leads to a set of recursion formulae from which one can derive a general expression for the $k^{ th}$ derivative of $\delta\Gammaint/\delta \cG|_{\bcG}$ (and thus of $\bSigma[\varphi]$) with $k\ge2$ in terms of 2PI kernels $\delta^l\Gammaint/\delta\cG^l|_{\bcG}$ with $2\le l\le k+1$ and of $q^{th}$ derivatives of $\bar\cG[\varphi]$ (and thus of $\bSigma[\varphi]$, see \eqn{appeq:Gn}) with $1\le q\le k$. Using similar notations as in \Eqn{appeq:Gn}, we get, for $\varphi=0$:
\bea\label{appeq:totalder}
&&\hspace{-1.5cm}\frac{\delta^k}{\delta\varphi_k\cdots\delta\varphi_1}
\left[\frac{\delta\Gammaint}{\delta\cG_{mn}}\right]_{\bar\cG}=\nn
&&\hspace{-0.5cm}={\sum_{\{I_r\}}}^{\star}(-1)^{\delta_{\{I_r\}}}\,\left(\prod_{r=1}^R\frac{\delta^p\bar\cG_{a_rb_r}}{\delta\varphi_{i^{(r)}_{p_r}}\cdots\delta\varphi_{i^{(r)}_1}}\right)\left.\frac{\delta^{R+1}\Gammaint}{\delta\cG_{a_Rb_R}\cdots\delta\cG_{a_1b_1}\delta\cG_{mn}}\right|_{\bar\cG},
\eea
where the sum is restricted to families $\smash{\left\{I_r\right\}=\{I_1,\ldots,I_R\}}$ of nonempty disjoint subsets of $\left\{1,\ldots,k\right\}$ such that $I_1$ always contains the element $i_1=1$ and the smallest element of the subsets $I_{r\ge2}$ are $\smash{i_1^{(r)}\geq r}$. One adds a contribution $q_rq_s$ to $\delta_{\{I_r\}}$ for each pair of derivatives $\delta/\delta\varphi_r$ and $\delta/\delta\varphi_s$ acting on different $\bar\cG$'s and appearing in the same order as the one defined by the left-hand-side of Eq.~(\ref{appeq:totalder}). 

As an illustration, we get, for $k=2$,
\beq\label{eq:lisa2}
\frac{\delta^2}{\delta\varphi_2\delta\varphi_1}\left[\frac{\delta\Gammaint}{\delta\cG_{mn}}\right]_{\bar\cG}=\frac{\delta\bar\cG_{ab}}{\delta\varphi_1}\frac{\delta\bar\cG_{cd}}{\delta\varphi_2}\left.\frac{\delta^3\Gammaint}{\delta\cG_{cd}\delta\cG_{ab}\delta\cG_{mn}}\right|_{\bar\cG}+\frac{\delta^2\bar\cG_{ab}}{\delta\varphi_2\delta\varphi_1}\left.\frac{\delta^2\Gammaint}{\delta\cG_{ab}\delta\cG_{mn}}\right|_{\bcG}\,,
\eeq
from which it follows, using the results of the previous subsection, that
\bea\label{eq:eqeq}
\frac{\delta^2\bar\Sigma_{nm}}{\delta\varphi_2\delta\varphi_1} (-1)^{q_m}\!\!\!\! & = & \!\!\!\!\left(\bcG\frac{\delta\bSigma}{\delta\varphi_1}\bcG\right)_{\!\!ab}\!\!\left(\bcG\frac{\delta\bSigma}{\delta\varphi_2}\bcG\right)_{\!\!cd}\left.\frac{2i\,\delta^3\Gammaint}{\delta\cG_{cd}\delta\cG_{ab}\delta\cG_{mn}}\right|_{\bcG}\nonumber\\
& + &\!\!\!\!
\left(2\,\bcG\,\frac{\delta\bSigma}{\delta\varphi_1}\,\bcG\,\frac{\delta\bSigma}{\delta\varphi_2}\,\bcG+\bcG\,\frac{\delta^{2}\bSigma}{\delta\varphi_2\delta\varphi_1}\,\bcG\right)_{\!\!\!ab}\left.\frac{2i\,\delta^2\Gammaint}{\delta\cG_{ab}\delta\cG_{mn}}\right|_{\bcG}.
\eea
Notice that this equation defines $\delta^2\bSigma/\delta\varphi^2$ self-consistently. For a given 2PI approximation, once the three-point function $\delta\bSigma/\delta\varphi$ is known from \Eqn{eq:abc}, the linear self-consistent equation \eqn{eq:eqeq} can be solved by iterations. This feature appears at any order: the function $\delta^k\bSigma/\delta\varphi^k$, obtained from Eq.~(\ref{appeq:totalder}), satisfies a linear integral equation since the RHS of Eq.~(\ref{appeq:totalder}) only contains derivatives $\delta^q\bSigma/\delta\varphi^q$ with $q\le k$.

\section{Four-photon graphs}
\label{appsec:four_photon}  

We gather some results concerning diagrams with four external photon legs and their UV structure. We first recall a useful result which states that whenever a four-photon structure is, in momentum space, transverse with respect to at least one of its external momenta, it is finite provided it is free of subdivergences. We then provide two important examples of such classes of diagrams, which appear in the analysis of Secs.~\ref{sec:renormalization_2PI_vertices}, \ref{sec:renormalization_2PI_resummed_vertices} and \App{appsec:renormalization_higher_2PI_vertices}.

\subsection{Transversality and UV structure}
\label{appsec:result}

Consider a sum $I^{\mu\nu\rho\sigma}$ of bare four-photon diagrams, transverse, in momentum space, with respect to one of its external momenta, say the one corresponding to the index $\mu$:
\beq\label{eq:transC}
k_\mu\,I^{\mu\nu\rho\sigma}=0\,.
\eeq
Assume that all subdivergences of $I$ can be removed by means of a renormalization procedure which does not violate the transversality relation (\ref{eq:transC}). Therefore, it can only contain global divergences which are polynomials in its external momenta, of order equal to the superficial degree of divergence of $I$, i.e. zero. The most general Lorentz structure for the latter is therefore, a linear combination of $g^{\mu\nu}g^{\rho\sigma}$, $g^{\mu\rho}g^{\nu\sigma}$ and $g^{\mu\sigma}g^{\nu\rho}$: we have
\beq
I^{\mu\nu\rho\sigma}=\alpha\,g^{\mu\nu}g^{\rho\sigma}+\beta\,g^{\mu\rho}g^{\nu\sigma}+\gamma\,g^{\mu\sigma}g^{\nu\rho}+I_{\rm CV}^{\mu\nu\rho\sigma}\,,
\eeq
 where $I_{\rm CV}$ is UV convergent and $\alpha$, $\beta$ and $\gamma$ are momentum independent functions of $\epsilon$ which could diverge as $\epsilon\rightarrow 0^+$. The transversality condition \eqn{eq:transC} reads
\beq\label{eq:transC2}
\alpha\,k^\nu g^{\rho\sigma}+\beta\,k^\rho g^{\nu\sigma}+\gamma\,k^\sigma g^{\nu\sigma}+k_\mu\,I_{\rm CV}^{\mu\nu\rho\sigma}=0\,.
\eeq
from which it follows that $\alpha$, $\beta$, $\gamma$ and thus $I$ are in fact UV convergent (i.e. finite).

\subsection{First class of transverse graphs}
\label{appsubsec:four_photon_1}
In Sec.~\ref{sec:renormalization_2PI_vertices}, we deal with a particular class of four-photon subgraphs arising from three-photon-reducible contributions in $\Gammaint[\varphi,\cG]$. A given such contribution, denoted here by $3\gamma R$, can be written as
\beq\label{appeq:3gr}
i\Gammaint^{3\gamma R}[\varphi,\cG]=\cC_1^{\nu\rho\sigma}G_{\nu\bar\nu}G_{\rho\bar\rho}G_{\sigma\bar\sigma}\,\cC_2^{\bar\nu\bar\rho\bar\sigma}
\eeq
where $\cC_{1,2}$ are 1PI three-photon subgraphs. This leads to three-photon-reducible contributions of the form \eqn{eq:cut} to the 2PI kernel $\bar\Lambda$:
\beq
\bLambda^{3\gamma R}_{\alpha\bar\alpha,\beta\bar\beta}= L^{\nu\rho\sigma}_{\alpha\bar\alpha}\bar G_{\nu\bar\nu}\bar G_{\rho\bar\rho}\bar G_{\sigma\bar\sigma} R^{\bar\nu\bar\rho\bar\sigma}_{\beta\bar\beta}\,,
\eeq
where $L^{\nu\rho\sigma}_{\alpha\bar\alpha}=\delta \cC_i^{\nu\rho\sigma}/\delta D_{\alpha\bar\alpha}|_\bcG$ with $i=1$ or $2$ and similarly for $R$. As argued in \Sec{sec:renormalization_2PI_vertices} such three-photon-reducible contributions lead to potentially divergent four-photon subgraphs of the form, see \Eqn{eq:sumsubstruct},
\beq\label{eq:hello}
\frac{\delta\bSigma^{\bar\psi\psi}_{\bar\alpha\alpha}}{\delta A^\mu}\,\bM^{\alpha\bar\alpha,\beta\bar\beta}L^{\nu\rho\sigma}_{\beta\bar\beta}\,,\eeq
which we want to prove to be transverse, see Eq.~\eqn{eq:symsum}. To this purpose notice that, since fermion propagator lines $D$ are only involved in closed fermion loops in the functions $\cC_{1,2}$ in Eq.~\eqn{appeq:3gr}, we can always write
\beq\label{appeq:trulu}
L^{\nu\rho\sigma}_{\alpha\bar\alpha} = \sum_{F}\left.(-ie)^n\frac{\delta F^{\mu_1\cdots\mu_n}}{\delta D^{\alpha\bar\alpha}}\right|_\bcG\bar G_{\mu_1\bar\mu_1}\cdots \bar G_{\mu_n\bar\mu_n}C_F^{\bar\mu_1\cdots\bar\mu_n;\nu\rho\sigma}\,,
\eeq
where the sum runs over all fermion loops of $\cC_{1,2}$ with $n\ge3$ QED vertex insertions:
\beq
F^{\mu_1\cdots\mu_n}={\rm tr}\,\left(D\gamma^{\mu_1}D\cdots D\gamma^{\mu_n}\right)\,.
\eeq
Noticing that
\beq\label{appeq:trala}
\left.\frac{\delta F^{\mu_1\cdots\mu_n}}{\delta D^{\alpha\bar\alpha}}\right|_{\bcG}=\left(\gamma^{\mu_1}\bar D\cdots \bar D\gamma^{\mu_n}\right)_{\bar\alpha\alpha}+{\rm cyclic\; permutations\; of\;} (1,\dots,n)\,,
\eeq
we are lead to evaluate the $(n+1)$-photon diagram 
\begin{equation}
{I_\mu}^{\mu_1\cdots\mu_n}=\frac{\delta\Sigma^{\bar\psi\psi}_{\bar\alpha\alpha}}{\delta A^\mu}\bM^{\alpha\bar\alpha,\beta\bar\beta}\left(\gamma^{\mu_1}\bD\cdots \bD\gamma^{\mu_n}\right)_{\bar\beta\beta}\,.
\end{equation}
In momentum space, denoting by $k$ and $p_1,\dots ,p_n$ the external incoming momenta, we have, see Eq.~\eqn{eq:mom3},
\begin{equation}
{I_\mu}^{\mu_1\cdots\mu_n}(k,p_1,\dots,p_n)=\int_q{\rm tr}\Big[V_\mu^{(2,1)}(q+k,q)\bD(q)\gamma^{\mu_1}\bD(q-p_1)\cdots\gamma^{\mu_n}\bD(q+k)\Big]\,.
\end{equation}
Using the Ward-Takahashi identity \eqn{eq:2PIWIFourier1}, the cyclicity of the trace and performing an appropriate change of variable under the momentum integral\footnote{Here, it is crucial to use a gauge-invariant regulator.}, we get
\begin{equation}\label{appeq:trolo}
k^\mu {I_\mu}^{\mu_1\cdots\mu_n}(k,p_1,\dots,p_n)=I^{\mu_1\mu_2\cdots\mu_n}(p_1,p_2,\dots, p_n)-I^{\mu_2\cdots\mu_n\mu_1}(p_2,\dots,p_n,p_1)
\end{equation}
where
\beq
I^{\mu_1\cdots\mu_n}(p_1,\dots,p_n) = -ie\int_q{\rm tr}\Big[\bD(q)\gamma^{\mu_1}\bD(q-p_{1})\cdots\gamma^{\mu_{n-1}}\bD(q+k+p_n)\gamma^{\mu_n}\Big]\,.
\eeq
Putting together Eqs.~\eqn{eq:hello}, \eqn{appeq:trulu}, \eqn{appeq:trala} and \eqn{appeq:trolo}, we finally obtain the desired result, \Eqn{eq:symsum}:
\begin{equation}
k^\mu\frac{\delta\bSigma^{\bar\psi\psi}_{\bar\alpha\alpha}}{\delta A^\mu}\,\bM^{\alpha\bar\alpha,\beta\bar\beta}L^{\nu\rho\sigma}_{\beta\bar\beta}=0\,,
\end{equation}
where $L=\delta \cC/\delta D|_\bcG$.

\subsection{Second class of transverse graphs}
\label{appsubsec:four_photon_2}
The analysis of the 2PI-resummed two-photon function $\delta^2\Gamma/\delta A\delta A$ in \Sec{sec:renormalization_2PI_resummed_vertices} reveals the presence of potentially divergent four-photon subgraphs which involve the two external photon legs of the original function and the two ends of an internal perturbative photon propagator $G_0$ taken anywhere in the full diagram, see \Fig{fig:2PIr2_aa_sub2}. As argued in \Sec{sec:renormalization_2PI_resummed_vertices}, the sum of such subdiagrams is given by the derivative $\delta\Pi/\delta G_{0}$ of the 2PI-resummed photon self-energy $\Pi=\Sigma_{AA}$ with respect to the free photon propagator. We show here that the latter is directly related to the 2PI four-photon function $\delta^2\bSigma_{AA}/\delta A\delta A$ through \Eqn{eq:fourphotondeux}, which implies that it is finite and (doubly) transverse.\footnote{For simplicity of notations, we work here with bare quantities. The calculation is identically the same in terms of renormalized quantities.} To this aim, it proves convenient to consider the more general function $\delta\Sigma/\delta \cG_0$, where $\Sigma$ is the 2PI-resummed self-energy, defined as
\beq
 (-1)^{q_1}\frac{\delta^2\Gamma}{\delta\varphi_2\delta\varphi_1}=i\cG_{0,21}^{-1}-i\Sigma_{21}\,.
\eeq 
Note that the derivative $\delta/\delta\cG_0$ has to be understood as restricted to the subspace of function $\cG_0$ of the form \eqn{eq:scor0}. Taking a derivative with respect to $\cG_0$ at fixed counterterms of \Eqn{appeq:A21} yields
\beq\label{eq:D1}
(-1)^{q_1}\frac{\delta\Sigma_{21}}{\delta \cG_0}=\frac{1}{2}\left(\frac{\delta}{\delta \cG_0}\frac{\delta\bSigma_{qp}}{\delta\varphi_2}\right)\!\cM_{pq,mn}\!\!\left.\frac{2i\delta^2\Gammaint}{\delta\cG_{mn}\delta\varphi_1}\right|_\bcG+\left(\frac{\delta\bar\cG}{\delta \cG_0}\,\frac{\delta\bSigma}{\delta\varphi_2}\,\bar\cG\right)_{\!\!mn}\!\!\left.\frac{2i\delta^2\Gammaint}{\delta\cG_{mn}\delta\varphi_1}\right|_\bcG\!.
\eeq
To evaluate $\delta/\delta\cG_0(\delta\bSigma/\delta\varphi)$, we take a derivative with respect to $\cG_0$ at fixed counterterms of \Eqn{eq:abc}. We obtain a linear integral equation which can be solved as
\bea
\frac{\delta}{\delta \cG_0}\frac{\delta\bSigma_{nm}}{\delta\varphi_1} & = & \left(\frac{\delta\bar\cG}{\delta \cG_0}\,\frac{\delta\bSigma}{\delta\varphi_1}\,\bar\cG\right)_{\!\!ab}\cV_{ab,mn}\nonumber\\
& + & \left(\bar\cG\,\frac{\delta\bSigma}{\delta\varphi_1}\,\bar\cG\right)_{\!\!ab}\frac{\delta\bar\cG_{cd}}{\delta \cG_0}\left.(-1)^{q_m}\frac{2i\delta^3\Gammaint}{\delta\cG_{cd}\delta\cG_{ab}\delta\cG_{mn}}\right|_\bcG\nonumber\\
& + & \frac{1}{2}\left(\bar\cG\,\frac{\delta\bSigma}{\delta\varphi_1}\,\bar\cG\right)_{\!\!ab}\frac{\delta\bar\cG_{cd}}{\delta \cG_0}\left.(-1)^{q_r}\frac{2i\delta^3\Gammaint}{\delta\cG_{cd}\delta\cG_{ab}\delta\cG_{rs}}\right|_\bcG\cM_{rs,pq}\,\cV_{pq,mn}\,,\nn
\eea
where the function $\cV$ fulfills the Bethe-Salpeter--type equation
\beq\label{appeq:VVVV}
\cV_{pq,mn}=\cL_{pq,mn}+\frac{1}{2}\cL_{pq,rs}\,\cM_{rs,tu}\,\cV_{tu,mn}\,.
\eeq
Plugging this back into Eq.~(\ref{eq:D1}), we find 
\bea\label{eq:def}
(-1)^{q_1}\frac{\delta\Sigma_{21}}{\delta \cG_0}&=&\left(\frac{\delta\bcG}{\delta\cG_0}\frac{\delta\bSigma}{\delta\varphi_2}\bcG\right)_{\!\!mn}{\cal T}_{mn,1}\nn
&+&\frac{1}{2}\left(\bcG\frac{\delta\bSigma}{\delta\varphi_2}\bcG\right)_{\!\!ab}\!\!\frac{\delta\bcG_{cd}}{\delta\cG_0}\left.(-1)^{q_p}\frac{2i\delta^3\Gammaint}{\delta\cG_{cd}\delta\cG_{ab}\delta\cG_{pq}}\right|_\bcG\cM_{pq,mn}{\cal T}_{mn,1}\,,\nn
\eea
where the function
\beq
{\cal T}_{mn,1}=\left.\frac{2i\delta^2\Gammaint}{\delta\cG_{mn}\delta\varphi_1}\right|_\bcG+\frac{1}{2}\,\cV_{mn,pq}\,\cM_{pq,rs}\left.\frac{2i\delta^2\Gammaint}{\delta\cG_{rs}\delta\varphi_1}\right|_\bcG\,.
\eeq
turns out to be simply related to $\delta\bSigma_{nm}/\delta\varphi_1$. Indeed, solving Eq.~(\ref{eq:abc}) in terms of $\cV$, we get
\beq\label{eq:ghi}
\frac{\delta\bSigma_{nm}}{\delta\varphi_1}=\left.(-1)^{q_m}\frac{2i\delta\Gammaint}{\delta\varphi_1\delta\cG_{mn}}\right|_\bcG+\left.\frac{1}{2}(-1)^{q_r}\frac{2i\delta^2\Gammaint}{\delta\varphi_1\delta\cG_{rs}}\right|_\bcG\cM_{rs,pq}\,\cV_{pq,mn}\,.
\eeq
Using the fact that the total fermion number of each factor in this equation is zero as well as the symmetry properties $\cM_{rs,pq}=\cM_{pq,rs}$, $\cV_{pq,mn}=(-1)^{q_p+q_n}\cV_{mn,pq}$ and
\beq
\left.\frac{2i\delta^2\Gammaint}{\delta\varphi_1\delta\cG_{mn}}\right|_\bcG=\left.(-1)^{q_1}\frac{2i\delta^2\Gammaint}{\delta\cG_{mn}\delta\varphi_1}\right|_\bcG\,,
\eeq
one easily checks that
\beq
\frac{\delta\bSigma_{nm}}{\delta\varphi_1}=(-1)^{q_1+q_m}{\cal T}_{mn,1}\,.
\eeq
It follows that Eq.~(\ref{eq:def}) can be rewritten as
\bea\label{eq:last}
\frac{\delta\Sigma_{21}}{\delta\cG_0}&=&\left(\frac{\delta\bSigma}{\delta\varphi_2}\bar\cG\frac{\delta\bSigma}{\delta\varphi_1}\right)_{\!\!nm}(-1)^{q_m}\frac{\delta\bar\cG_{mn}}{\delta \cG_0}\nn
&+&\frac{1}{2}\left(\bar\cG\frac{\delta\bSigma}{\delta\varphi_2}\bcG\right)_{\!\!ab}\!\!\left(\bcG\frac{\delta\bSigma}{\delta\varphi_1}\bcG\right)_{\!\!cd}\!\!\!\!\left.(-1)^{q_m}\frac{2i\delta^3\Gammaint}{\delta\cG_{cd}\delta\cG_{ab}\delta\cG_{mn}}\right|_\bcG(-1)^{q_m}\frac{\delta\bar\cG_{mn}}{\delta \cG_0}\,.\nn
\eea
Notice now that, from Eqs. \eqn{eq:stat}-\eqn{eq:sig}, one obtains the linear equation
\beq\label{eq:autreequation}
\frac{\delta\bSigma_{nm}}{\delta\cG_{0,tu}}=\frac{1}{2}\frac{\delta\bcG_{pq}}{\delta\cG_{0,tu}}\,\cL_{pq,mn}=\frac{1}{2}\left[-\frac{\delta\cG_{0,sr}^{-1}}{\delta\cG_{0,tu}}+\frac{\delta\bSigma_{sr}}{\delta\cG_{0,tu}}\right]\cM_{rs,pq}\,\cL_{pq,mn}\,,
\eeq
which can be solved in terms of $\cV$ as
\beq\label{eq:uneautreequation}
\frac{\delta\bSigma_{nm}}{\delta \cG_{0,tu}}=-\frac{1}{2}\frac{\delta\cG_{0,sr}^{-1}}{\delta\cG_{0,tu}}\,\cM_{rs,pq}\,\cV_{pq,mn}=\frac{1}{2}\,(\cG^{-1}_0\bcG)_{uq}\,(\bcG\cG^{-1}_0)_{pt}\,\cV_{pq,mn}\,.
\eeq
From this it is not difficult to arrive at (we use that $q_m=q_p$) 
\beq\label{eq:julien}
(-1)^{q_m}\frac{\delta\bar\cG_{mn}}{\delta \cG_0}=-\left[\cM_{mn,pq}+\frac{1}{2}\cM_{mn,rs}\,\cV_{rs,tw}\,\cM_{tw,pq}\right]\frac{\delta\cG^{-1}_{0,qp}}{\delta \cG_0}(-1)^{q_p}\,,
\eeq
which can be plugged into Eq.~(\ref{eq:last}) and allows a simple comparison with Eq.~(\ref{eq:eq3}) below. We find:
\beq\label{appeq:eqdfhfjru}
\frac{\delta\Sigma_{21}}{\delta \cG_{0,rs}} = -\left[\frac{1}{2}\frac{\delta^2\bSigma_{nm}}{\delta\varphi_1\delta\varphi_2}+\left(\frac{\delta\bSigma}{\delta\varphi_2}\,\bar\cG\,\frac{\delta\bSigma}{\delta\varphi_1}\right)_{\!\!nm}\right]\cM_{mn,pq}\,\frac{\delta\cG^{-1}_{0,qp}}{\delta \cG_{0,rs}}(-1)^{q_p}\,,
\eeq
or, equivalently,
\beq
\cM_{0;pq,rs}\frac{\delta\Sigma_{21}}{\delta \cG_{0,rs}}= (-1)^{q_p}\cM_{pq,mn}\left[\frac{1}{2}\frac{\delta^2\bSigma_{nm}}{\delta\varphi_1\delta\varphi_2}+\left(\frac{\delta\bSigma}{\delta\varphi_2}\,\bar\cG\,\frac{\delta\bSigma}{\delta\varphi_1}\right)_{\!\!nm}\right]\,,
\eeq
where $\cM_{0;pq,rs}=\cG_{0,ps}\cG_{0,rq}$. Choosing $\smash{\varphi_1=A_\mu}$, $\smash{\varphi_2=A_\nu}$ and $\smash{\cG_{0,rs}=G_{0,\rho\sigma}}$ in Eq.~\eqn{appeq:eqdfhfjru}, the second term between brackets vanishes and we obtain the desired result, see \Eqn{eq:fourphotondeux},
\beq\label{appeq:ffff}
G_{0,\rho\lambda}\frac{\delta\Sigma_{AA}^{\mu\nu}}{\delta G_{0,\lambda\xi}}G_{0,\xi\sigma}=\bG_{\rho\lambda}\frac{\delta^2\bSigma_{AA}^{\lambda\xi}}{\delta A_\mu\delta A_\nu}\bG_{\xi\sigma}\,.
\eeq

\section{Renormalization of higher vertex functions}
\label{appsec:renormalization_higher_2PI_vertices}

Here, we show that, once 2PI two- and three-point vertex functions, $\bcG^{-1}_R$ and $V^{(3)}_R\propto\delta\bSigma_R/\delta\varphi_R$ have been made finite according to the procedure described in Secs.~\ref{subsec:renormalization_2PI_vertices_2} and \ref{subsec:renormalization_2PI_vertices_3}, all (2PI and 2PI-resummed) vertex functions are void of subdivergences. As explained in Secs.~\ref{subsec:renormalization_higher_2PI_vertices} and \ref{subsec:renormalization_higher_2PI_resummed_vertices}, this completes our proof of renormalization. All quantities appearing in this section are meant as renormalized ones. However, for notational simplicity we omit the subscript `$R$'.

\subsection{Renormalization of $\mathcal{V}$}
Because it will play a role in the following, we first show that $\cV$ is finite. Notice that, due to charge conservation, $\smash{\cV_{pq,mn}=0}$ unless $\smash{q_p+q_q+q_m+q_n=0}$. Moreover, it follows from the symmetry identities, see \eqn{eq:sympropG},
\beq\label{appeq:symofV}
\cV_{pq,mn}=(-1)^{q_pq_q}\cV_{qp,mn}=(-1)^{q_mq_n+q_m+q_n}\cV_{pq,nm}=(-1)^{q_n+q_p}\cV_{mn,pq}
\eeq
that the only independent components of $\cV$ are the four-photon function $\cV_{\rho\sigma,\mu\nu}$, the two-fermion--two-photon functions  $\cV_{\alpha\nu,\mu\bar\alpha}$ and $\cV_{\alpha\bar\alpha,\mu\nu}$ and the four-fermion function $\cV_{\alpha\bar\alpha,\beta\bar\beta}$.\footnote{We use the same convention as in Secs.~\ref{sec:renormalization_2PI_vertices} and \ref{sec:renormalization_2PI_resummed_vertices} to distinguish photon from fermion legs: the first letters of the alphabet $\alpha,\ldots,\eta$ denote fermion $\psi$-like legs; $\bar\alpha,\cdots,\bar\eta$ denote fermion $\bar\psi$-like legs; any other (greek) letter denote a photon leg.} It is easy to check from \Eqn{appeq:VVVV} that the two-fermion--two-photon function $\cV_{\alpha\nu,\mu\bar\alpha}$ can be identified with the function $\tilde V_{\alpha\nu,\mu\bar\alpha}$ introduced in \Sec{sec:renormalization_2PI_vertices}, see \Eqn{eq:BS_pa}. It resums an infinite number of ladder diagrams $\tilde\Lambda(\tilde M\tilde\Lambda)^n$, which are trivially finite since the building blocks $\tilde\Lambda$ and $\tilde M$ are and there is no way one can combine lines of these different building blocks to form a subgraph with positive superficial degree of divergence. 

The case of the other components of $\cV$ is slightly more complicated because they are linearly coupled to each other. This is, for instance, the case of $\cV_{\rho\sigma,\mu\nu}$ and $\cV_{\alpha\bar\alpha,\mu\nu}$. However, one can rewrite the equation for $\cV_{\rho\sigma,\mu\nu}$ as 
\beq\label{eq:Vmnrs}
\cV_{\rho\sigma,\mu\nu}={\cal K}_{\rho\sigma,\mu\nu}+\frac{1}{2}{\cal K}_{\rho\sigma,\lambda\omega}\cM_{\lambda\omega,\xi\kappa}\cV_{\xi\kappa,\mu\nu}
\eeq
where ${\cal K}_{\rho\sigma,\mu\nu}$ depends only on the 2PI kernel $\cL$, not on $\cV$:
\beq\label{appeq:Kfunction}
{\cal K}_{\rho\sigma,\mu\nu}=\cL_{\rho\sigma,\mu\nu}+\cL_{\rho\sigma,\alpha\bar\alpha}\cM_{\alpha\bar\alpha,\beta\bar\beta}\cL_{\beta\bar\beta,\mu\nu}+\cL_{\rho\sigma,\alpha\bar\alpha}\cM_{\alpha\bar\alpha,\beta\bar\beta}{\cal J}_{\beta\bar\beta,\gamma\bar\gamma}\cM_{\gamma\bar\gamma,\delta\bar\delta}\cL_{\delta\bar\delta,\mu\nu}
\eeq
where ${\cal J}$ resums fermion ladders:
\beq\label{appeq:Jfunction}
{\cal J}_{\beta\bar\beta,\alpha\bar\alpha}=\cL_{\beta\bar\beta,\alpha\bar\alpha}+\cL_{\beta\bar\beta,\gamma\bar\gamma}\cM_{\gamma\bar\gamma,\delta\bar\delta}{\cal J}_{\delta\bar\delta,\alpha\bar\alpha}\,.
\eeq
With slightly different notations, Eq.~(\ref{eq:Vmnrs}) for $\cV_{\rho\sigma,\mu\nu}$ is nothing but the equation for the four-photon function $\bar V_{\rho\sigma,\mu\nu}$, derived in \Refer{Reinosa:2006cm}. There, it was shown that the latter resums four-photon subdivergences in $\bSigma_{AA}$ and $\bSigma_{\bar\psi\psi}$and is made finite by properly adjusting the counterterms $\delta\bar g_1$ and $\delta\bar g_2$ in \Eqn{eq:dGamma2app}. It is then straightforward to show by direct inspection that the only possibly divergent subgraphs of $\cV_{\alpha\bar\alpha,\mu\nu}$ and $\cV_{\beta\bar\beta,\alpha\bar\alpha}$ are those of $\cV_{\rho\sigma,\mu\nu}$: once the latter has been renormalized, the former are finite. 

Here, we pause a moment to give a new derivation of a general result of 2PI renormalization theory, namely the fact that the Bethe-Salpeter like equation \eqn{appeq:VVVV} resums the four-point subdivergences of the two-point function $\bSigma[\varphi=0]$ \cite{VanHees:2001pf,Reinosa:2005pj,Reinosa:2006cm}. Four-point subgraphs are obtained by expanding $\bSigma$  in perturbative diagrams and opening a perturbative line -- corresponding to the free propagator $\cG_0$ -- in all possible ways.\footnote{Strictly speaking, this is only true for $\varphi=0$. For nonvanishing field, there exists other types of four-point subgraphs} The sum of all such subgraphs is, therefore, encoded in the functional derivative $\delta\bSigma/\delta \cG_0$. According to Eq.~(\ref{eq:uneautreequation}) the latter is such that
\beq\label{eq:uneautreequationbis}
\cM_{0;rs,pq}\,\frac{\delta\bSigma_{nm}}{\delta \cG_{0,pq}}=\frac{1}{2}\cM_{rs,pq}\,\cV_{pq,mn}\,.
\eeq
where $\cM_{0;rs,pq}=\cG_{0,rq}\cG_{0,ps}$. We see that the function $\cV$ resums all four-point subgraphs of $\bSigma$, from which the announced result follows.

\subsection{2PI vertex functions}
We treat explicitly the case of the 2PI four-point vertex function. Higher 2PI vertex functions can be treated along similar lines as sketched at the end of this subsection. Our starting point is Eq.~(\ref{eq:eqeq}) which we rewrite as
\bea\label{eq:eqeq2}
\frac{\delta^2\bar\Sigma_{nm}}{\delta\varphi_2\delta\varphi_1} & = & \left(\bcG\frac{\delta\bSigma}{\delta\varphi_1}\bcG\right)_{\!\!ab}\!\!\left(\bcG\frac{\delta\bSigma}{\delta\varphi_2}\bcG\right)_{\!\!cd}(-1)^{q_m}\left.\frac{2i\,\delta^3\Gammaint}{\delta\cG_{cd}\delta\cG_{ab}\delta\cG_{mn}}\right|_{\bcG}\nonumber\\
& + &
\left(\bcG\,\frac{\delta\bSigma}{\delta\varphi_1}\,\bcG\,\frac{\delta\bSigma}{\delta\varphi_2}\bcG\right)_{\!\!\!ab}{\cal L}_{ab,mn}+\frac{1}{2}\frac{\delta^{2}\bSigma_{sr}}{\delta\varphi_2\delta\varphi_1}\,{\cal M}_{rs,pq}\,{\cal L}_{pq,mn}\,,
\eea
where we have introduced the notations
\beq\label{eq:not}
\cM_{rs,pq}\equiv \bar\cG_{rq}\,\bar\cG_{ps} \quad {\rm and} \quad \cL_{pq,mn}\equiv (-1)^{q_m}\left.\frac{4i\delta^2\Gammaint}{\delta\cG_{pq}\delta\cG_{mn}}\right|_{\bcG}\,.
\eeq
Equation (\ref{eq:eqeq2}) can be solved using ${\cal V}$
\bea\label{eq:eq3}
\frac{\delta^2\bSigma_{nm}}{\delta\varphi_2\delta\varphi_1} & = & \left(\bar\cG\,\frac{\delta\bSigma}{\delta\varphi_1}\,\bar\cG\,\frac{\delta\bSigma}{\delta\varphi_2}\,\bar\cG\right)_{\!\!ab}\cV_{ab,mn}\nonumber\\
& + & \left(\bar\cG\,\frac{\delta\bSigma}{\delta\varphi_1}\,\bar\cG\right)_{\!\!ab}\left(\bar\cG\,\frac{\delta\bSigma}{\delta\varphi_2}\,\bar\cG\right)_{\!\!cd}\!(-1)^{q_m}\left.\frac{2i\delta^3\Gammaint}{\delta\cG_{cd}\delta\cG_{ab}\delta\cG_{mn}}\right|_{\bcG}\nonumber\\
& + &  \frac{1}{2}\left(\bar\cG\,\frac{\delta\bSigma}{\delta\varphi_1}\,\bar\cG\right)_{\!\!ab}\left(\bar\cG\,\frac{\delta\bSigma}{\delta\varphi_2}\,\bar\cG\right)_{\!\!cd}\!(-1)^{q_r}\left.\frac{2i\delta^3\Gammaint}{\delta\cG_{cd}\delta\cG_{ab}\delta\cG_{rs}}\right|_{\bcG}\,\cM_{rs,pq}\,\cV_{pq,mn}\,,\nn
\eea

\begin{figure}[htbp]
\begin{center}
\includegraphics[width=12.0cm]{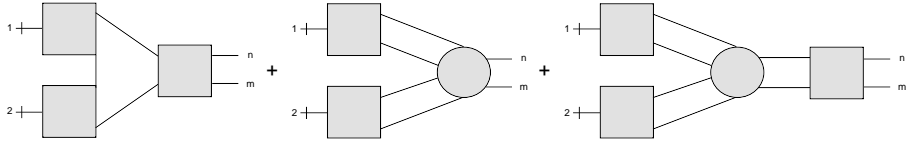}
\caption{Diagrammatic representation of the three contributions to $\delta^2\bSigma/\delta\varphi_2\delta\varphi_1$ in the r.h.s of Eq.~(\ref{eq:eq3}). The grey blob represent the 2PI kernel $2i\delta\Gammaint/\delta\cG^3|_{\bcG}$. The grey boxes with three legs represent the 2PI three-point vertex $\delta\bSigma/\delta\psi$ whereas those with four legs represent the four-point function ${\cal V}$.\label{fig:2PI4_1}}
\end{center}
\end{figure}

\begin{figure}[htbp]
\begin{center}
\includegraphics[width=10.0cm]{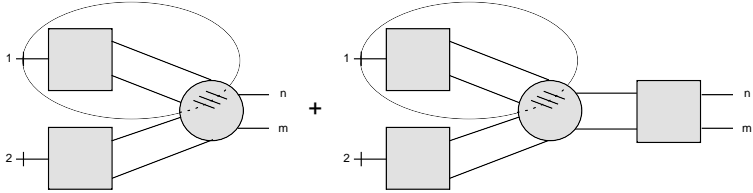}
\caption{These subraphs are potentially divergent when their external legs are all photons. 2PI Ward-Takahashi identities guarantee that these subgraphs are all finite.\label{fig:2PI4_2_div2}}
\end{center}
\end{figure}

This equation, represented diagrammatically in Fig.~\ref{fig:2PI4_1}, happens to be quite convenient to show that $\delta^2\bSigma/\delta\varphi^2$ is void of subdivergences. The various contributions on the RHS of Eq.~(\ref{eq:eq3}) involve four different building blocks: The propagator $\bcG$, the 2PI vertex $\delta\bSigma/\delta\varphi$, the function $\cV$ which sums ladder diagrams with rungs $\cL\propto\delta^2\Gammaint^R/\delta\cG^2|_\bcG$ and the six-point 2PI kernel $\delta^3\Gammaint^R/\delta\cG^3|_\bcG$. Each of these building blocks being finite, subdivergences in $\delta^2\bSigma/\delta\varphi^2$ can only originate from two-, three- or four-point 1PI subgraphs which combine lines of different building blocks. Using the fact that the kernels $\delta^n\Gammaint^R/\delta\cG^n|_\bcG$ originate from closed 2PI diagrams, it is easy to convince oneself that the only potentially divergent subgraphs are four-photon ones. 
These can be of two types only: The first type is depicted in Fig.~\ref{fig:2PI4_2_div2}, where it is understood that the subgraph is potentially divergent only when its four external legs are photons. We have already encountered such four-photon substructures in the analysis of two- and three-point functions in Secs \ref{sec:renormalization_2PI_vertices} and \ref{sec:renormalization_2PI_resummed_vertices}, see e.g. Eqs. \eqn{eq:sumsubstruct}-\eqn{eq:symsum}, where we have shown, using 2PI Ward-Takahashi identities, that they actually do not give rise to subdivergences.

\begin{figure}[htbp]
\begin{center}
\includegraphics[width=10.0cm]{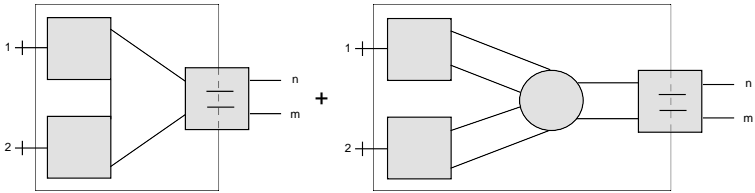}
\caption{These subraphs are potentially divergent when their external legs are all photons. 2PI Ward-Takahashi identities ensure that these subgraphs add up to a finite contribution.\label{fig:2PI4_3_div2}}
\end{center}
\end{figure}

 The second type of four-photon subgraphs is illustrated in \Fig{fig:2PI4_3_div2}: It involves two internal (photon) lines of $\cV$ as well as the two external legs of the original function $\delta^2\bSigma/\delta\varphi^2$ corresponding to (photon) field derivatives. It can, therefore, only arise in the two-photon--two-fermion function $\delta^2\bSigma_{\psi\bar\psi}/\delta A\delta A$ or in the four-photon function $\delta^2\bSigma_{AA}/\delta A\delta A$. We conclude that the two other four-point functions $\delta^2\bSigma_{\psi A}/\delta A\delta\bar\psi$ and $\delta^2\bSigma_{AA}/\delta\psi\delta\bar\psi$ are void of subdivergences.

The functions $\delta^2\bSigma_{\psi\bar\psi}/\delta A\delta A$ and $\delta^2\bSigma_{AA}/\delta A\delta A$ satisfy coupled linear integral equations, see Eq.~\eqn{eq:eqeq2}. For the sake of the argument, we introduce a simplified notation which emphasizes the important aspects of the relevant equations and hides the unimportant details. We note $V_{\gamma f}\equiv\delta^2\bSigma_{\psi\bar\psi}/\delta A\delta A$ and $V_{\gamma\gamma}\equiv\delta^2\bSigma_{AA}/\delta A\delta A$. Similarly, we note $\cL_{\gamma f}\equiv\cL_{\mu\nu,\alpha\bar\alpha}$, $\cL_{\gamma \gamma}\equiv\cL_{\mu\nu,\rho\sigma}$ etc. and similarly for the components of $\cM$, with the idea that the indices $\gamma$ and $f$ denote respectively the two-photon and two-fermion end of a given four-point function. With this notation, \Eqn{eq:eqeq2} reads, for $V_{\gamma f}$ and $V_{\gamma\gamma}$
\bea
\label{appeq:laprems}
 V_{\gamma f}&=&A_{\gamma f}+\frac{1}{2}V_{\gamma\gamma}\cM_{\gamma\gamma}\cL_{\gamma f}+V_{\gamma f}\cM_{ff}\cL_{ff}\,,\\
\label{appeq:ladeuz}
 V_{\gamma\gamma}&=&A_{\gamma\gamma}+\frac{1}{2}V_{\gamma\gamma}\cM_{\gamma\gamma}\cL_{\gamma\gamma}+V_{\gamma f}\cM_{ff}\cL_{f\gamma}\,,
\eea
where $A_{\gamma f}$ and $A_{\gamma\gamma}$ refer to the relevant component of the sum of the first two terms on the RHS of Eqs.~\eqn{eq:eqeq2}. For the present discussion, the only important feature is their two-photon-irreducible character. For later use, we also note that the function $V_{\gamma f}$ can be eliminated by means of the function $\cK$ introduced in \Eqn{appeq:Kfunction}, and that $V_{\gamma\gamma}$ can consequently be expressed in terms of the function $\cV$ introduced in \Eqn{eq:Vmnrs}: 
\beq\label{appeq:Vgg}
 V_{\gamma\gamma}=
B_{\gamma\gamma}+\frac{1}{2}V_{\gamma\gamma}\cM_{\gamma\gamma}{\cal K}_{\gamma\gamma}=
B_{\gamma\gamma}+\frac{1}{2}B_{\gamma\gamma}\cM_{\gamma\gamma}{\cal V}_{\gamma\gamma}
\eeq
with ($\cJ$ is defined in Eq.~\eqn{appeq:Jfunction})
\beq
 B_{\gamma\gamma}=A_{\gamma\gamma}+A_{\gamma f}\cM_{ff}\cL_{f\gamma}+A_{\gamma f}\cM_{ff}\cJ_{ff}\cM_{ff}\cL_{f\gamma}
\eeq
Note that $B$ is two-photon-irreducible. Note also that 
\beq\label{appeq:equality}
 V_{\gamma\gamma}\cM_{\gamma\gamma}{\cal K}_{\gamma\gamma}=B_{\gamma\gamma}\cM_{\gamma\gamma}{\cal V}_{\gamma\gamma}\,.
\eeq
Finally, a key ingredient of the argument below is the transversality property -- in momentum space -- of the function $V_{\gamma\gamma}\equiv\delta^2\bSigma_{AA}/\delta A\delta A$, see \Eqn{eq:tata}. We write this formally as:
\beq\label{appeq:transversal}
 k\cdot V_{\gamma\gamma}=0\,.
\eeq

We now employ a recurrence argument based on a formal coupling expansion, e.g.
\beq
 V_{\gamma f}=\sum_{n}e^{2n}\,V_{\gamma f}^{(2n)}\,,
\eeq
and similarly for all other functions above. Note that the functions $A$, $B$, $V$, $\cL_{\gamma\gamma}$, $\cK_{\gamma\gamma}$ and $\cV_{\gamma\gamma}$ start at order $e^4$, whereas the functions $\cL_{\gamma f}$, $\cL_{f\gamma}$ and $\cL_{ff}$ start at order $e^2$. Equation \eqn{appeq:transversal} yields
\beq\label{appeq:transversality}
 k\cdot V_{\gamma\gamma}^{(2n)}=0\quad{\rm for}\quad n\ge2\,.
\eeq
As a starting point of our recurence, we note that the one-loop diagrams $V_{\gamma f}^{(4)}=A_{\gamma f}^{(4)}$ and $V_{\gamma\gamma}^{(4)}=B_{\gamma\gamma}^{(4)}=A_{\gamma\gamma}^{(4)}$, being trivially free of subdivergences, are finite. This is clear for the former since it has negative superficial degree of divergence. For the latter, this follows from the transversality property \eqn{appeq:transversality} for $n=2$ and the general result of \App{appsec:four_photon}. We now assume that all $V_{\gamma f}^{(2p)}$ and $V_{\gamma \gamma}^{(2p)}$ are finite for $2\le p\le n-1$. \Eqn{appeq:laprems} yields, for $n\ge3$,
\beq
 V_{\gamma f}^{(2n)}=A_{\gamma f}^{(2n)}+\frac{1}{2}\sum_{p=2}^{n-1}V_{\gamma\gamma}^{(2p)}\left[\cM_{\gamma\gamma}\cL_{\gamma f}\right]^{(2n-2p)}+\frac{1}{2}\sum_{p=2}^{n-1}V_{\gamma f}^{(2p)}\left[\cM_{ff}\cL_{ff}\right]^{(2n-2p)}
\eeq
Using the two-particle-irreducibility of the kernels $\cL$, one sees that the only possible four-photon subgraphs are those of $V_{\gamma\gamma}^{(2p)}$ and $V_{\gamma f}^{(2p)}$ for $p\le n-1$, which are finite by hypothesis. The function $V_{\gamma f}^{(2n)}$ is thus void of subdivergences and consequently finite since it has negative superficial degree of divergence. As for $V_{\gamma\gamma}^{(2n)}$, equation \eqn{appeq:Vgg} yields, for $n\ge3$,
\beq
 V_{\gamma\gamma}^{(2n)}=B_{\gamma\gamma}^{(2n)}+\frac{1}{2}\sum_{p=2}^{n-1}V_{\gamma\gamma}^{(2p)}\left[\cM_{\gamma\gamma}{\cal K}_{\gamma\gamma}\right]^{(2n-2p)}\,.
\eeq
One can form four-photon subgraphs involving the two external legs of $\cK_{\gamma\gamma}$ and two internal photon lines of $V_{\gamma\gamma}^{(2p)}$. Using \Eqn{appeq:equality} one easily checks that these are in fact subgraphs of the function $\cV_{\gamma\gamma}$, which we have shown above to be finite. Using the two-photon-irreducibility of the kernels $\cK$ one sees that the only other possibility for four-photon subgraphs are either a subgraph of $V_{\gamma\gamma}^{(2p)}$ for $p\le n-1$, finite by hypothesis, or the full graphs $B_{\gamma\gamma}^{(2n)}$ and $V_{\gamma\gamma}^{(2p)}\left[\cM_{\gamma\gamma}{\cal K}_{\gamma\gamma}\right]^{(2n-2p)}$ for $2\le p\le n-1$. We conclude that $V_{\gamma\gamma}^{(2n)}$ is void of subdivergences. It follows from the 2PI Ward-Takahashi identity \Eqn{appeq:transversality} and the general result of \App{appsec:four_photon} that it is finite.\\

A similar analysis applies to higher 2PI vertex functions $\delta^n\bSigma/\delta\varphi_n\cdots\delta\varphi_1$. The latter satisfy self-consistent equations, similar to \eqn{eq:eqeq2}, which can be solved by means of the function $\cV$. This results in an expression for $\delta^n\bSigma/\delta\varphi_n\cdots\delta\varphi_1$ -- generalizing Eq.~(\ref{eq:eq3}) -- in terms of a finite number of building blocks, namely lower order 2PI vertex functions $\delta^p\bSigma/\delta\varphi_p\cdots\delta\varphi_1$ with $\smash{p<n}$, $\bar\cG$, $\cV$ and 2PI kernels $\delta^p\Gammaint^R/\delta\cG^p|_\bcG$ with $\smash{3\leq p <n}$. Assuming that 2PI vertex functions of order $\smash{p<n}$ have already been renormalized, the building blocks all are finite. Thus subdivergences in $\delta^n\bSigma/\delta\varphi_n\cdots\delta\varphi_1$ can only originate from subgraphs combining lines of different building blocks. An analysis similar\footnote{It is even simpler for four-photon subgraphs of the type depicted in Fig.~\ref{fig:2PI4_3_div2} do not appear.} to that performed for $\delta^2\bSigma/\delta\varphi_2\delta\varphi_1$ then shows that all such subgraphs are UV convergent.

\subsection{2PI-resummed vertex functions}
A similar method can be applied to higher 2PI-resummed vertex functions $\delta^n\Gamma/\delta\varphi_n\cdots\delta\varphi_1$. Contrary to 2PI vertex functions, these are not defined self-consistently but directly given in terms of the propagator $\bar\cG$, the 2PI kernel $\delta^2\Gammaint^R/\delta\cG\delta\varphi|_\bcG$ and lower 2PI vertex functions $\delta^p\bSigma/\delta\varphi_p\cdots\delta\varphi_1$ with $\smash{p<n}$, see Eq.~(\ref{appeq:gamma_n}). This equation is however not suited to show that $\delta^n\Gamma/\delta\varphi_n\cdots\delta\varphi_1$ is void of subdivergences. The reason for this is two-fold. First of all the 2PI-kernel $\delta^2\Gammaint^R/\delta\cG\delta\varphi|_\bcG$ is not UV convergent since it is proportional to either $\tilde Z_1$ or $\bar Z_1$. Moreover, there is always a contribution to $\delta^n\Gamma/\delta\varphi_n\cdots\delta\varphi_1$ involving only one 2PI vertex. This contribution reads
\beq\label{appeq:co}
\left(\bar\cG\,\frac{\delta^{n-1}\bSigma}{\delta\varphi_n\cdots\delta\varphi_2}\,\bar\cG\right)_{mn}\left.\frac{\delta^2\Gammaint^R}{\delta\cG_{mn}\delta\varphi_1}\right|_\bcG\,.
\eeq
Subdivergences of this contribution are not easy to analyze for they depend on the precise content of $\delta^{n-1}\bSigma/\delta\varphi_n\cdots\delta\varphi_2$. A strategy is then to replace the latter for its explicit expression in terms of lower 2PI vertex functions and $\cV$. The net result of this is an expression for $\delta^n\Gamma/\delta\varphi_n\cdots\delta\varphi_1$ in which no $\delta^2\Gammaint^R/\delta\cG\delta\varphi|_\bcG$ appears and in which the contribution \eqn{appeq:co} is also absent. Similarly to higher 2PI vertex functions, one can show that the building blocks appearing in this alternative expression for $\delta^n\Gamma/\delta\varphi_n\cdots\delta\varphi_1$ are UV convergent and so are also subgraphs which combine lines of different building blocks. It follows that $\delta^n\Gamma/\delta\varphi_n\cdots\delta\varphi_1$ is void of subdivergences.


\begin{thebibliography}{00}

\bibitem{Luttinger:1960ua}
J.~M. Luttinger, J.~C. Ward,  Phys.\ Rev.\ {\bf 118} (1960) 1417;

G. Baym, Phys.\ Rev.\ {\bf 127} (1962) 1391;

C. De Dominicis, P.~C. Martin, J.\ Math.\ Phys.\ {\bf 5} (1964) 14.

\bibitem{Cornwall:1974vz}
J.~M. Cornwall, R. Jackiw, E. Tomboulis, Phys.\ Rev.\ D {\bf 10} (1974) 2428;

  R.~E.~Norton, J.~M.~Cornwall,
  Annals Phys.\  {\bf 91} (1975) 106.


\bibitem{Blaizot:1999ip}
J.~P. Blaizot, E. Iancu, A. Rebhan, Phys.\ Rev.\ Lett.\ {\bf 83} (1999) 2906;
Phys.\ Lett.\ B {\bf 470} (1999) 181;
Phys.\ Rev.\ D {\bf 63} (2001) 065003;

  E.~Braaten, E.~Petitgirard,
  Phys.\ Rev.\  D {\bf 65} (2002) 041701;
  {\it ibid} 085039.

\bibitem{Andersen:2004re}
  J.~O.~Andersen, M.~Strickland,
  Phys.\ Rev.\ D {\bf 71} (2005) 025011.

\bibitem{Berges:2004hn}
J. Berges, Sz. Bors{\'a}nyi, U. Reinosa, J. Serreau, Phys.\ Rev.\ D {\bf 71} (2005) 105004.

\bibitem{Blaizot:2005wr}
J.~P. Blaizot, A. Ipp, A. Rebhan, U. Reinosa, Phys.\ Rev.\ D {\bf 72} (2005) 125005.

\bibitem{Aarts:2003bk}
G.~Aarts, J.~M.~Martinez Resco, Phys.\ Rev.\ D {\bf 68} (2003) 085009;
JHEP {\bf 0402} (2004) 061;
JHEP {\bf 0503} (2005) 074;

  M.~E.~Carrington, E.~Kovalchuk,
  Phys.\ Rev.\  D {\bf 76} (2007) 045019.

\bibitem{Alford:2004jj}
  M.~Alford, J.~Berges, J.~M.~Cheyne,
  Phys.\ Rev.\  D {\bf 70} (2004) 125002.

\bibitem{Arrizabalaga:2006hj}
A.~Arrizabalaga, U.~Reinosa, Nucl.\ Phys.\  A {\bf 785} (2007) 234.




\bibitem{Berges:2004vw}
For a short review, see: J.~Berges, J.~Serreau, hep-ph/0410330.

\bibitem{Berges:2000ur}
  J.~Berges, J.~Cox,
  Phys.\ Lett.\  B {\bf 517} (2001) 369;

  J.~Berges,
  Nucl.\ Phys.\ A {\bf 699} (2002) 847;

  F.~Cooper, J.~F.~Dawson, B.~Mihaila,
  Phys.\ Rev.\ D {\bf 67} (2003) 056003;

  S.~Juchem, W.~Cassing, C.~Greiner,
  Phys.\ Rev.\ D {\bf 69} (2004) 025006;

  A.~Rajantie, A.~Tranberg,
  JHEP {\bf 0611} (2006) 020.

\bibitem{Aarts:2002dj}
  G.~Aarts, D.~Ahrensmeier, R.~Baier, J.~Berges, J.~Serreau,
  Phys.\ Rev.\ D {\bf 66} (2002) 045008.

\bibitem{Berges:2002cz}
  J.~Berges, J.~Serreau,
  Phys.\ Rev.\ Lett.\ {\bf 91} (2003) 111601;

  A.~Arrizabalaga, J.~Smit, A.~Tranberg,
  JHEP {\bf 0410} (2004) 017;

  G.~Aarts, A.~Tranberg,
  Phys.\ Lett.\  B {\bf 650} (2007) 65;
  Phys.\ Rev.\  D {\bf 77} (2008) 123521;


  A.~Tranberg,
  JHEP {\bf 0811} (2008) 037.



\bibitem{Berges:2002wr}
  J.~Berges, Sz.~Bors\'anyi, J.~Serreau,
  Nucl.\ Phys.\ B {\bf 660}( 2003) 51;

  J.~Berges, Sz.~Bors\'anyi, C.~Wetterich,
  Nucl.\ Phys.\  B {\bf 727} (2005) 244;

  M.~Lindner, M.~M.~Muller,
  Phys.\ Rev.\  D {\bf 77} (2008) 025027.



\bibitem{Gasenzer:2005ze}
  T.~Gasenzer, J.~Berges, M.~G.~Schmidt, M.~Seco,
  Phys.\ Rev.\  A {\bf 72} (2005) 063604.


\bibitem{Reinosa:2007vi}
U.~Reinosa, J.~Serreau, JHEP {\bf 0711} (2007) 097.

\bibitem{Reinosa:2006cm}
U.~Reinosa, J.~Serreau, JHEP {\bf 0607} (2006) 028.

\bibitem{vanHees:2002bv}
H. van Hees, J. Knoll, 
Phys.\ Rev.\ D {\bf 66} (2002) 025028;

\bibitem{Arrizabalaga:2002hn}
A. Arrizabalaga, J. Smit, Phys.\ Rev.\ D {\bf 66} (2002) 065014;

M.E. Carrington, G. Kunstatter, H. Zaraket, Eur.\ Phys.\ J.\  C {\bf 42} (2005) 253.

\bibitem{Mottola:2003vx}
  E.~Mottola,
  arXiv:hep-ph/0304279.

\bibitem{Borsanyi:2007bf}
  Sz.~Bors\'anyi, U.~Reinosa,
  Phys.\ Lett.\  B {\bf 661} (2008) 88.

\bibitem{VanHees:2001pf}
  H.~Van Hees, J.~Knoll,
  Phys.\ Rev.\ D {\bf 65} (2002) 105005;
{\it ibid.} 025010;

J.-P.~Blaizot, E.~Iancu, U.~Reinosa, Phys.\ Lett.\ B {\bf 568} (2003) 160;
Nucl.\ Phys.\ A {\bf 736} (2002) 149.

\bibitem{Cooper:2004rs}
  F.~Cooper, B.~Mihaila, J.~F.~Dawson,
  Phys.\ Rev.\ D {\bf 70} (2004) 105008;

  A.~Jakovac,
  Phys.\ Rev.\  D {\bf 76} (2007) 125004;

  A.~Patk\'os, Z.~Sz\'ep,
  Nucl.\ Phys.\  A {\bf 811} (2008) 329;

  Sz.~Bors\'anyi, U.~Reinosa,
  arXiv:0809.0496 [hep-th].


\bibitem{Berges:2005hc}
J. Berges, Sz. Bors{\'a}nyi, U. Reinosa, J. Serreau,  Ann.\ Phys.\ {\bf 320} (2005) 344.



\bibitem{Reinosa:2005pj}
U.~Reinosa, Nucl.\ Phys.\ A {\bf 772} (2006) 138.

\bibitem{Calzetta:2004sh}
  E.~A.~Calzetta,
  Int.\ J.\ Theor.\ Phys.\  {\bf 43} (2004) 767.

\bibitem{Fejos:2009dm}
  G.~Fejos, A.~Patkos and Z.~Szep,
  arXiv:0902.0473 [hep-ph].

\bibitem{Gies:2004hy}
 H.~Gies, J.~Jaeckel,
 Phys.\ Rev.\ Lett.\  {\bf 93} (2004) 110405.

\bibitem{Weinberg} 
S. Weinberg,
{\it The quantum Theory of Fields}, Vol. I, Cambridge University Press (1995)

\end{thebibliography}
\end{document}